\documentclass[12pt]{elsarticle}

\usepackage{comment}
\usepackage{csquotes}

\usepackage{pgfplots}  
\usepgfplotslibrary{groupplots}
\usepackage{tikz}
\usetikzlibrary{decorations, arrows.meta, calligraphy}
\pgfplotsset{compat=newest}
\usepackage{amssymb}
\usepackage{bm}
\usepackage{MnSymbol}
\usepackage{siunitx}
\usepackage{xcolor}
\usepackage{lineno}

\usepackage{glossaries}
\newacronym{ad}{AD}{automatic differentiation}
\newacronym{adam}{ADAM}{adaptive moment estimation}
\newacronym{bvp}{BVP}{boundary value problem}
\newacronym{cnn}{CNN}{convolutional neural network}
\newacronym{dof}{dof}{degrees of freedom}
\newacronym{dl}{DL}{deep learning}

\newacronym{rnn}{RNN}{recurrent neural network}
\newacronym{fe}{FE}{finite element}
\newacronym{fv}{FV}{finite volume}
\newacronym{lstm}{LSTM}{long-short term memory}
\newacronym{ml}{ML}{machine learning}
\newacronym{nn}{NN}{neural network}
\newacronym{ode}{ODE}{ordinary differential equation}
\newacronym{pinn}{PINN}{physics-informed neural network}
\newacronym{pde}{PDE}{partial differential equation}
\newacronym{relu}{ReLU}{rectified linear units}

\DeclareMathOperator*{\argmin}{argmin}

\definecolor{TUDa-0d}{cmyk/RGB/HTML}{0,0,0,.8/83,83,83/535353}
\definecolor{TUDa-0c}{cmyk/RGB/HTML}{0,0,0,.6/137,137,137/898989}
\definecolor{TUDa-0b}{cmyk/RGB/HTML}{0,0,0,.4/181,181,181/B5B5B5}
\definecolor{TUDa-0a}{cmyk/RGB/HTML}{0,0,0,.2/220,220,220/DCDCDC}
\definecolor{TUDa-10c}{cmyk/RGB/HTML}{.5,1,.3,0/149,17,105/951169}

\journal{-}

\begin{document}

\begin{frontmatter}
\title{Error Approximation and Bias Correction in Dynamic Problems using a Recurrent Neural Network/Finite Element Hybrid Model}

\author[aff1]{Moritz von Tresckow\corref{cor1}}
\author[aff1]{Herbert De Gersem}
\author[aff1]{Dimitrios Loukrezis}

\affiliation[aff1]{
organization={Technische Universität Darmstadt, Institute for Accelerator Science and Electromagnetic Fields (TEMF)},
addressline={Schlossgartenstr. 8}, 
city={64289 Darmstadt},
country={Germany}
}

\cortext[cor1]{Corresponding author: moritz.von{\textunderscore}tresckow@tu-darmstadt.de}

\begin{abstract}
This work proposes a hybrid modeling framework based on recurrent neural networks (RNNs) and the finite element (FE) method to approximate model discrepancies in time-dependent, multi-fidelity problems, and use the trained hybrid models to perform bias correction of the low-fidelity models.
The hybrid model uses FE basis functions as a spatial basis and RNNs for the approximation of the time dependencies of the FE basis' degrees of freedom.
The training data sets consist of sparse, non-uniformly sampled snapshots of the discrepancy function, pre-computed from trajectory data of low- and high-fidelity dynamic FE models.
To account for data sparsity and prevent overfitting, data upsampling and local weighting factors are employed, to instigate a trade-off between physically conforming model behavior and neural network regression.
The proposed hybrid modeling methodology is showcased in three highly non-trivial engineering test-cases, all featuring transient FE models, namely, heat diffusion out of a heat sink, eddy-currents in a quadrupole magnet, and sound wave propagation in a cavity. 
The results show that the proposed hybrid model is capable of approximating model discrepancies to a high degree of accuracy and accordingly correct low-fidelity models.
\end{abstract}


\end{frontmatter}


\section{Introduction}
Over the last decade, \gls{ml} has established itself as one of the prime research subjects in the contemporary sciences, finding applications in biology \cite{alber2019integrating}, engineering \cite{mason2019review, johnson2020invited, vadyala2022review}, physics \cite{carleo2019machine, ion2021local}, and medicine \cite{yadav2019deep}, to name only a few areas.
Even though this development can to some extent be attributed to hype \cite{shi2023role}, there are domains in which \gls{ml} methods have become the state-of-the-art.
Explicit examples include the development of \glspl{cnn} and graph neural networks for image classification \cite{mallat2016understanding, chen2021review, wu2021comprehensive}, large language models for natural language processing \cite{openai2023gpt4}, and deep reinforcement learning for applications in autonomous driving \cite{sallab2017deep}.
However, \gls{ml} models are subjected to the curse of dimensionality and training often involves solving nonlinear optimization problems which causes training times to be a significant concern.
Nevertheless, with processing units such as CPUs and GPUs becoming more performant each year \cite{burg2021moore}, the impact of \gls{ml} will most likely increase even further.

On the basis of these recent successes, \gls{ml} quickly penetrated traditional scientific computing methods \cite{kashinath2021physics}.
Recent advances in this area lie in the emergence of \glspl{pinn} \cite{han2018solving, raissi2019physics, cai2021physics,  karniadakis2021physics}, hybrid modeling \cite{kurz2022hybrid, maier2022known}, but also in the field of data-driven computing \cite{kirchdoerfer2016data, galetzka2021data}.
\Glspl{pinn} incorporate physics-based loss functions to train \glspl{nn} as to exhibit physically conforming behavior, hybrid modeling approaches seek to combine ``the best of both worlds", and data-driven computing methods extend traditional numerical approximation methods for \glspl{pde} to directly incorporate measurement data.
Worthy of note, however, is that \gls{ml} does not always improve upon established methods and its success or lack thereof significantly depends on the problem setting to which it is applied \cite{markidis2021old}.
For instance, deep learning is inherently based on nonlinear optimization, therefore, it is at natural disadvantage when competing against traditional solvers in low dimensional, linear problem settings. 
Consequently,  deep learning methods should preferably be applied to resolve nonlinear and/or high-dimensional problems in order to achieve a computational benefit \cite{dumitru2013advantages}.

In the authors' opinion, data-based bias correction and model validation, a framework discussed in detail in the recent work of Levine and Stuart \cite{levine2022framework}, provides exactly such a setting.
In particular, let us consider the case of  multi-fidelity modeling, where computationally inexpensive, low-fidelity models are employed to approximate time- and resource-demanding high-fidelity models, resulting in a model hierarchy according to accuracy and computational cost \cite{park2017remarks, peherstorfer2018survey, alber2019integrating}.
Therein, bias correction and model validation occur naturally.
In bias correction, a corrective term is computed, which accounts for the discrepancy between models of different fidelity \cite{peherstorfer2018survey}. 
In model validation, the accuracy of the low-fidelity model is validated against a reference \cite{roache1998verification, oberkampf2010verification, roy2010complete}.
In both cases, approximating the systematic modeling error, also referred to as model bias or model discrepancy, is necessary. 
This systematic error commonly takes the form of a so-called discrepancy function. 
Primary approaches for learning a discrepancy function include Bayesian inference \cite{chen2008design, wang2009bayesian, li2016integrating, conn2018guide} and Gaussian Process regression \cite{kennedy2000predicting, arendt2012quantification, li2016integrating, parussini2017multi, gardner2021learning}.
In more recent years, bias correction by means of neural networks has also been attempted \cite{lary2009machine, moghim2017bias, wang2022deep}.

In this work, we propose a hybrid modeling framework to approximate discrepancy functions between low- and high-fidelity \gls{fe} simulations of dynamical systems. 
The hybrid model is based on an \gls{rnn} and on a low-fidelity \gls{fe} model, and is trained on sparse, non-uniformly sampled data from high-fidelity solution trajectories.
The low- and high-fidelity simulations differ by the choice of the \gls{fe} mesh, as well as due to modeling assumptions, where the former induces discretization errors and the latter modeling errors.
Accordingly, the \gls{rnn}/\gls{fe}-based hybrid model is constructed by splitting the spatial and temporal dependencies, such that the \gls{rnn} approximates the discrepancy function dynamics, while the low-fidelity \gls{fe} model accounts for spatial discrepancy effects.
To account for the sparsity of the training data and prevent overfitting, the training data is upsampled artificially using linear interpolation operators and localized Gaussian priors at missing time steps to provide noise stabilization, an approach commonly used in game design and image processing \cite{hung2011fast, kavan2011physics, pons2021upsampling}. 
Local weighting factors on the training data are also employed to control the interpolation behavior of the model.
The hybrid model is applied to three non-trivial engineering test cases, namely, heat diffusion out of a heat sink, eddy-currents in a quadrupole magnet, and sound wave propagation in a cavity.
The results show that the proposed hybrid model and training regimen is capable of approximating model discrepancies to a high degree of accuracy. 

The main contribution of this work lies in the application of \glspl{rnn} for discrepancy function approximation in the setting of multi-fidelity \gls{fe}-based modeling and simulation. 
In essence, the present paper extends the framework suggested by Levine and Stuart \cite{levine2022framework} to the case of time-dependent \glspl{pde} solved by means of the \gls{fe} method, which has not been previously considered.
Additionally, the present work addresses the challenges posed by sparsely sampled training data and applies the suggested hybrid modeling framework to highly non-trivial engineering applications.
\textcolor{black}{
More importantly, the methodology developed in this work addresses the need to characterize and correct model discrepancy over time, which is highly practically relevant for numerous real-world applications. 
One exemplary application concerns the use of simplified models to accelerate simulation-based predictions, e.g., symmetry-based 2D models \cite{thuillier2020modeling} or data-driven surrogate models \cite{koziel2011surrogate, trehan2017error}. These reduced order models can significantly accelerate industrial design procedures or even allow for online model-based estimation in real time.
At the same time, the employed simplifications often neglect important physical constraints and phenomena, thus inducing significant model-form errors. 
Using the present methodology allows to derive an operator that captures and corrects the omissions of the reduced order model, at the cost of only a few evaluations of the full order model.
}

The present work contributes to the increasing volume of literature on hybrid numerical methods that enhance traditional solvers by means of \gls{ml}. 
Therein, a number of related works have been identified.
Um et al. \cite{um2020solver} and Kochkov et al. \cite{kochkov2021machine} learn correction operators for fluid flow simulation by integrating a differentiable (finite-difference or finite-volume) solver into the training loop of a \gls{cnn}.   
A similar approach utilizes adjoint-based gradients instead of a differentiable solver \cite{sirignano2020dpm}. 
Other approaches learn the correction operator directly, by training a feedforward \gls{nn} \cite{baiges2020finite} or a \gls{cnn} \cite{chen2022multi} with data generated from low- and high-fidelity models.
Oishi and Yagawa \cite{oishi2021finite} combine discrepancy approximation and correction via feedforward \glspl{nn} with a posteriori \gls{fe} error analysis, in particular for stress estimation in static mechanical simulations.
Zhou and Tang \cite{zhou2021efficient} train a composite \gls{nn} with multi-fidelity data, to be used as a surrogate model for frequency-response estimation under uncertainty. 
A related method by Chahar and Mukhopadhyay for multi-fidelity surrogate modeling under uncertainty employs a Gaussian process instead \cite{chahar2023multi}.
In the two latter works, the surrogate model essentially approximates the high-fidelity model directly, that is, these two approaches concern neither discrepancy approximation nor bias correction, at least not explicitly.

In contrast to the aforementioned related works, the methodology developed in this paper distinguishes itself in various ways: It specifically concerns dynamic problems governed by time-dependent \glspl{pde} and compared to \cite{um2020solver}, resolved with the \gls{fe} method; it does not require differentiable or adjoint solvers; it addresses discrepancies due to differences not only in mesh resolution, but also in the underlying modeling assumptions.
Compared to works focussing on the interaction of \gls{ml} and the \gls{fe} method such as \cite{chahar2023multi, zhou2021efficient}, our approach focusses on approximating the discrepancy function between low and high fidelity model instead of providing a surrogate for the high fidelity model directly.
Furthermore, we seek to operate in a sparse data regimen.
Last, an important distinction is that the hybrid modeling framework suggested in this work employs an \gls{rnn}, which is crucial to address the time dependency of the discrepancy function.

The remaining of this paper provides, first, a theoretical overview of data-driven discrepancy approximation for model validation and bias correction (Section~\ref{sec:preliminaries}), followed by a presentation of the suggested hybrid modeling approach (Section~\ref{sec:hybrid_modeling}), which highlights the hybrid model architecture and discusses the treatment of sparse training data. 
Numerical experiments showcasing the benefits of the proposed hybrid modeling approach 
are presented in Section~\ref{sec:numerical_examples}. 
Last, concluding remarks and considerations for further methodological developments are available in Section~\ref{sec:conclusion}.

\section{Preliminaries and notation}
\label{sec:preliminaries}

\subsection{Dynamical system simulation}
\label{subsec:dynamical-system}

The simulation of dynamical systems commonly requires the spatial and temporal discretization of an underlying \gls{bvp}, defined on $\Omega \times [0,T]$, where $\Omega \subset\mathbb{R}^d$ denotes the spatial domain at dimension $d \leq 3$ and $T \in \mathbb{R}_{\geq 0}$ the time range.
Spatial discretization refers to the approximation of the system states by a finite dimensional basis representation and temporal discretization refers to state propagation along a discrete time axis.
Therein, implicit time stepping schemes are frequently used, due to their  advantages in numerical stability.
The discrete dynamical system can be described by
\begin{align}
\label{eq:dyn_sys}
    \mathbf{x}_{t_{k+1}} = \mathbf{\Psi}(\mathbf{x}_{t_{k+1}},\mathbf{x}_{t_k},t_{k+1}, t_k), 
\end{align}
where $\mathbf{\Psi}:\mathbb{R}^{d}\rightarrow \mathbb{R}^{d}$ is a time propagation operator and $\mathbf{x}_{t_k}:\Omega \rightarrow \mathbb{R}^{d}$ with $\mathbf{r}\mapsto\mathbf{x}_{t_k}(\mathbf{r})$, $\mathbf{r}\in\Omega$, are the system states at time $t=t_k$ where $k$ is the time step indexing. 
In the case where the spatial discretization does not change over time, $\mathbf{\Psi}$ only operates on the \gls{dof} of the states' basis coefficients.
Accordingly, we can express \eqref{eq:dyn_sys} as $\mathbf{\hat{x}}_{t_{k+1}} = \mathbf{\Psi}(\mathbf{\hat{x}}_{t_k})$ with $\mathbf{\Psi}:\mathbb{R}^{N_{\text{dof}}} \rightarrow \mathbb{R}^{N_{\text{dof}}}$ and $\mathbf{\hat{x}}\in \mathbb{R}^{N_{\text{dof}}}$, where $N_{\text{dof}} \in \mathbb{N}$ denotes the number of spatial \gls{dof}.
Given an initial state $\mathbf{x}_{t_0}$, we can solve \eqref{eq:dyn_sys} for $N_T \in \mathbb{N}$ time steps of size $\Delta t\in \mathbb{R}$. 
The solution is an approximation of the states' trajectories, the accuracy of which is dictated by the values of $N_{\text{dof}}$ and $N_T$ and the time stepping scheme. 
In this work, we focus on the \gls{fe} method for spatial discretization and an implicit Euler scheme for time-stepping.

\subsection{Model fidelity and discrepancy}
Real-world scientific and engineering applications governed by dynamical problems are inherently complex, often requiring a large number of \gls{dof} and time steps to be resolved to sufficient accuracy. 
Quite often, solution accuracy must be balanced against limitations in computational resources.
A predominant factor in this trade-off is model fidelity, that is, the model's capability to precisely represent the physical system it approximates \cite{fernandez2016review}.
In the following, high- and low-fidelity models are denoted with $\mathbf{\Psi}_{\text{hifi}}$ and $\mathbf{\Psi}_{\text{lofi}}$, respectively, as they will refer to dynamical system models, as they were introduced in Section~\ref{subsec:dynamical-system}.
Accordingly, low- and high-fidelity system states at time $t_k$ are respectively denoted with $\mathbf{x}_{t_k}^{\text{lofi}}$ and $\mathbf{x}_{t_k}^{\text{hifi}}$.

The difference in the results between a low- and a high-fidelity model can be predominantly attributed to a systematic error, which is commonly called model discrepancy or prediction bias.
Model discrepancy can be quantified using a so-called discrepancy function $\delta_{t} :\mathbb{R}^d \rightarrow \mathbb{R}^d$ with $\delta_{t}(\mathbf{r}) = \mathbf{x}_{t}^{\text{hifi}}(\mathbf{r}) - \mathbf{x}_{t}^{\text{lofi}}(\mathbf{r})$, which maps the system state onto its corresponding systematic error \cite{arendt2012quantification, mahadevan2006inclusion}.
Non-systematic errors such as noise can similarly be expressed using an error function $\varepsilon_t:\mathbb{R}^d\rightarrow\mathbb{R}^d$, where a common assumption is $\varepsilon_t \sim\mathcal{N}(\mathbf{0},\mathbf{\Sigma}), \,\mathbf{\Sigma} \in \mathbb{R}^{d\times d}$.
Under standard assumptions regarding the nature of the error and bias terms, for example, additivity or multiplicativity, the relationship between $\mathbf{\Psi}_{\text{hifi}}$ and $\mathbf{\Psi}_{\text{lofi}}$ can be expressed explicitly. 
For instance, under the additive bias and noise assumption, it holds that
\begin{align} 
\label{eq:correction_op}
\mathbf{x}_{t_k}^{\text{hifi}} = \mathbf{x}_{t_k}^{\text{lofi}}  + \delta_{t_k}  + \varepsilon_{t_k},
\end{align}
for $k = 1,...,N_T$. 
Multiplicative or more complex relations are also possible.
For the remaining of this work, we neglect the noise term in \eqref{eq:correction_op} and focus solely on the discrepancy function.

\subsection{Data-driven discrepancy function approximation for model validation and correction}
In the field of model validation, one is interested in whether the discrepancy function, $\delta_t$, fulfills certain conditions, as to determine whether the use of $\mathbf{\Psi}_{\text{lofi}}$ is justifiable \cite{langemann2022model}.
In this context, a very general validation condition reads
\begin{align}
	\label{eq:error_bound}
	\frac{1}{T}\int_{[0,T]}\left\|\delta_{t}\right\|^2_2\,\mathrm{d}t < C,
\end{align}
where $C>0$ and $\left\| \cdot\right\|_2$ the $L^2\left(\Omega\right)$ norm.
If $\delta_{t}$ fulfills \eqref{eq:error_bound}, the use of $\mathbf{\Psi}_{\text{lofi}}$ is tenable \cite{langemann2022model}. 
To that end, the functional form of $\delta_{t_k}$, which essentially quantifies the accuracy of $\mathbf{\Psi}_{\text{lofi}}$ with respect to $\mathbf{\Psi}_{\text{hifi}}$, must be inferred.
While this could be accomplished by sufficiently sampling the high- and the low-fidelity models, the evaluation of the high-fidelity model must in many cases be avoided, typically due to its prohibitive computational cost. 
Instead, observed snapshot data $\mathbf{X}:=\{\mathbf{x}^{\mathrm{o}}_{t_k}\}_{t_k\in T_{\mathrm{o}}}$, $T_{\mathrm{o}} = \{t_k\}_{k=1}^{N_T}$, of its state trajectory may be available \cite{levine2022framework}, for instance, originating from measurement setups or auxiliary simulations.
To that end the superscript ``${\mathrm{o}}$'' is chosen to denote an observation.

Then, the validation of the low-fidelity model requires a parametric model $\delta_{\theta}$, which approximates the discrepancy based on the snapshots $\mathbf{X}$. 
This approximation can be accomplished with supervised learning algorithms.
However, using data as a reference increases substantially the complexity of the validation problem, since $\mathbf{X}$ might contain sparse, non-uniformly sampled, and noisy data.
Possible remedies to this issue can be sought in novel physics-informed \gls{ml} approaches, which integrate physics-inspired objective functions into the approximation process to induce physics conforming behavior in sparse data regimes \cite{karniadakis2021physics}.
Irrespective of the choice of the parametric model $\delta_{\theta}$ and the multi-fidelity models, the minimization problem to be solved in order to determine an optimal parameter set $\theta^*$ given the training data set $\mathcal{D}:=\left\{\mathbf{x}_{t_k}^{\mathrm{o}} - \mathbf{x}_{t_k}^{\text{lofi}} \right\}_{t_k\in T_{\mathrm{o}}}$, reads
\begin{align}
\label{eq:obj_func}
   \theta^* = \argmin_\theta \mathcal{J}_\theta, \: \text{where} \:\: \mathcal{J}_\theta= \frac{1}{T} \int_{0}^T\left\|\delta_{t}
    -\delta_{\theta}\right\|_2 \,\mathrm{d}t,
\end{align}
for $\delta_t \in \mathcal{D}$.
The parametric model $\delta_\theta:\mathbb{R}^d \rightarrow \mathbb{R}^d$ for $\theta \in \mathbb{R}^{N_\theta}$, where $N_\theta$ denotes the number of parametric \gls{dof}, can be chosen to map the low-fidelity states $\mathbf{x}^{\text{lofi}}_{t_k}\mapsto \delta_{\theta}$ onto the respective systematic error for all time instances $t_k \in T_{\mathrm{o}}$ and points in the spatial domain $\mathbf{r}\in \Omega$ \cite{um2020solver}.
We note however, that this is one example and that other mappings are also possible.

Assuming a successful solution of the minimization problem \eqref{eq:obj_func}, there are numerous uses for the function $\delta_{\theta^*}$.
On the one hand, $\delta_{\theta^*}$ can be used to approximate the validation condition \eqref{eq:error_bound}, ultimately verifying the low-fidelity model.
On the other hand, it can be used for bias correction, where we define a \emph{corrected} system state
\begin{align}
\label{eq:corr_model}
\mathbf{x}_{t_k}^{\text{corr}}  = \mathbf{x}_{t_k}^{\text{lofi}}  + \delta_{\theta^*}\left(\mathbf{x}_{t_k}^{\text{lofi}} \right),
\end{align}
where $\delta_{\theta^*}$ models the discrepancy between the low-fidelity model and the reference data.
The state trajectory resulting from \eqref{eq:corr_model} is the bias-corrected trajectory of the low-fidelity model.
In any case, model validation or correction depend significantly on the accuracy of the parametric model.

\subsection{Modeling and discretization errors in transient finite element models}

Multi-fidelity modeling and simulation introduces different types of systematic errors. 
In the particular case of \gls{fe} analysis, these errors can be categorized into three different classes, namely, discretization errors, numerical errors, and modeling errors \cite{braack2003posteriori, strang1972variational}.
Discretization errors occur when a function of a continuous variable is approximated by a finite dimensional basis representation, and primarily depend on the mesh resolution and the choice of the finite dimensional ansatz space.
Reducing the discretization error requires a finer mesh resolution, thus increasing the number of \gls{dof}.
Numerical errors occur due to the finite precision of computation hardware, e.g., in the representation of real numbers as floating data types and the finite precision of iterative solvers. 
Truncation errors fall in the same category.
Modeling errors occur due to assumptions and simplifications with respect to the problem itself. 
Common examples include misspecification of boundary conditions, incorrect definition of loading terms, linear approximations of otherwise nonlinear material responses, and 2D approximations of 3D geometries.

For the approximation of the discrepancy function between \gls{fe} models of varying fidelity, a suitable representation basis must be chosen.
Assuming solely the discretization error $\delta^{\text{d}}$, a natural choice would be to use the \gls{fe} basis functions.
In the context of this work, we use the \gls{fe} basis of the low-fidelity model $\mathbf{\Psi}_{\text{lofi}}$, as we focus on bias correction.
Consequently, a low-fidelity representation of the high-fidelity system states must be made available. 
Thus, it is necessary to define a projection operator $\mathcal{T}:\mathbb{R}^{N_\text{dof}^\text{hifi}}\rightarrow \mathbb{R}^{N_\text{dof}^{\text{lofi}}}$, where $N_\text{dof}^\text{hifi}$ and $N_\text{dof}^\text{lofi}$ denote the number of \gls{dof} of the high- and the low-fidelity model, respectively.  
In the context of the \gls{fe} method, $\mathcal{T}$ is the Galerkin projection from the high-fidelity state onto the low-fidelity basis.
\footnote{Note that, when considering model validation, a projection  $\mathcal{T}:\mathbb{R}^{N_\text{dof}^\text{lofi}}\rightarrow\mathbb{R}^{N_\text{dof}^{\text{hifi}}}$ is required, to get a high-fidelity representation of the low-fidelity states.}
Assuming $\mathbf{\Psi}_{\text{hifi}}$ and $\mathbf{\Psi}_{\text{lofi}}$ to be defined on the same time grid, the discretization error at time step $t_{k}$ is given as 
\begin{align}
\label{eq:disc_error}
    \delta^{\textrm{d}}_{t_{k}} = \left\|\mathcal{T}\left( \mathbf{x}_{t_{k}}^{\text{hifi}}\right) - \mathbf{x}_{t_{k}}^{\text{lofi}}\right\|_2,
\end{align}
where $\mathbf{x}_{t_k}^{\text{hifi}}=\mathbf{\Psi}_{\text{hifi}}\left(\mathbf{x}^{\text{hifi}}_{t_{k-1}} \right)$ and $\mathbf{x}_{t_k}^{\text{lofi}}=\mathbf{\Psi}_{\text{lofi}}\left(\mathbf{x}^{\text{lofi}}_{t_{k-1}} \right)$ are the high-fidelity and low-fidelity states at time $t$.

On the other hand, modeling errors  $\delta^{\text{m}}$ occur in the assumptions that define the \gls{bvp}, ultimately affecting the definition of the time propagators.
This fact is illustrated most clearly by assuming that $\mathbf{\Psi}_{\text{hifi}}$ and $\mathbf{\Psi}_{\text{lofi}}$ are parametrized by parameter sets $\pmb{\lambda}:=\left(\lambda_1,...,\lambda_l\right)$, such that $\mathbf{x}_{t_{k}} = \mathbf{\Psi}(\mathbf{x}_{t_{k-1}}\,|\,\pmb{\lambda})$, and defined on the same computational mesh.
Consequently, if $\pmb{\lambda}_{\text{hifi}} \neq \pmb{\lambda}_{\text{lofi}}$, the high-fidelity model reads $\mathbf{x}^{\text{hifi}}_{t_k}=\mathbf{\Psi} (\mathbf{x}^{\text{hifi}}_{t_{k-1}}\,|\,\pmb{\lambda}_{\text{hifi}})$ and the low-fidelity model $\mathbf{x}^{\text{lofi}}_{t_k} = \mathbf{\Psi} (\mathbf{x}^{\text{lofi}}_{t_{k-1}}\,|\,\pmb{\lambda}_{\text{lofi}})$ with $\mathbf{x}^{\text{hifi}}_{t_0}=\mathbf{x}^{\text{lofi}}_{t_0}=\mathbf{x}_{t_0}$.
At each time step, this error can be quantified as 
\begin{align}
\label{eq:modelling_error}
    \delta^{\textrm{m}}_{t_k} = \left\| \mathbf{x}_{t_{k}}^{\text{hifi}} - \mathbf{x}_{t_{k}}^{\text{lofi}}\right\|_2.
    \end{align}

In most multi-fidelity modeling settings, discretization and modeling errors are present simultaneously.
Thus, a combined error term must be considered, which captures both.
Considering \eqref{eq:disc_error} and \eqref{eq:modelling_error}, the combined error term can be quantified as
\begin{align}
\label{eq:comb_error}
    \delta_{t_k} = \left\|\mathcal{T}\left( \mathbf{x}_{t_{k}}^{\text{hifi}}\right) - \mathbf{x}_{t_{k}}^{\text{lofi}}\right\|_2,
\end{align}
where $ \mathbf{x}_{t_{k}}^{\text{hifi}}=\mathbf{\Psi}_{\text{hifi}}\left(\mathbf{x}^{\text{hifi}}_{t_{k-1}} \,|\,\pmb{\lambda}_{\text{hifi}}\right)$ and $ \mathbf{x}_{t_{k}}^{\text{lofi}}=\mathbf{\Psi}_{\text{lofi}}\left(\mathbf{x}^{\text{lofi}}_{t_{k-1}}\,|\,\pmb{\lambda}_{\text{lofi}}\right)$ with $\mathcal{T}\left(\mathbf{x}_{t_{0}}^{\text{hifi}}\right)=\mathbf{x}_{t_{0}}^{\text{lofi}}$.
In the following sections, \eqref{eq:comb_error} provides the basis for preprocessing the training data.

\section{Hybrid modeling methodology}
\label{sec:hybrid_modeling}


\subsection{\textcolor{black}{Hybrid model: Structure and building blocks}}
\label{sec:building-blocks}

In this section we discuss the nature of the discrepancy function $\delta_{t}$ and our subsequent choice for the parametric model $\delta_\theta \approx \delta_{t}$.
Discrepancy functions often display complex dynamics, exhibiting piecewise smooth and non-smooth behavior in the problem domain.
These dynamics occur because discrepancy functions are required to capture harmonic propagating behavior, but also phase transitions, material interfaces, and interpolation errors.
This phenomenon can be attributed to a variety of reasons, arising partially due to the underlying physics, but also due to the spatial and temporal discretization schemes, e.g., because of mesh inconsistencies or time stepping errors.

As we seek to calculate a bias correction term for the low-fidelity model, we assume that the discrepancy function has the form
\begin{align}
\label{eq:ansatz}
	\delta_{t_k}(\mathbf{r}) = \sum_{i=1}^{N_{\text{dof}}^{\text{lofi}}} \hat{\delta}_{i,t_k}  \, \phi_i(\mathbf{r}),
\end{align}
where $\hat{\delta}_{i,t_k}$ are the time-dependent \gls{dof} of the finite dimensional basis at time $t_k$ and the $i$-th spatial \gls{dof}, $\{\phi_i\}_{i=1}^{N^{\text{lofi}}_{\text{dof}}}$ are the basis functions of the low-fidelity model, and $\mathbf{r}\in \Omega$.
We denote with $\bm{\hat{\delta}}_{t_k}:=\left(\hat{\delta}_{1,t_k},...,\hat{\delta}_{N_{\text{dof}}^{\text{lofi}},t_k}\right)$ the vector containing the coefficients of \eqref{eq:ansatz}.

\textcolor{black}{The suggested hybrid modeling approach is based on splitting the spatial and temporal dependencies of the discrepancy function.
To that end, an \gls{rnn} is employed to learn the time-dependent \gls{dof} $\bm{\hat{\delta}}_{t_k}$ at each time step, while the \gls{fe} shape functions provide a spatial basis.
The \gls{rnn}/\gls{fe} hybrid model is then defined as
\begin{align}
	\delta_{\theta}\left(\mathbf{r}\right) = \sum^{N_{\text{dof}}^{\text{lofi}}}_{i=1}\delta^{\text{\tiny NN}}_{i,t_k} \,\phi_i\left(\mathbf{r}\right),
\end{align}
where $\bm{\delta}^{\text{NN}}_{t_k} \approx \bm{\hat{\delta}}_{t_k}$ denotes the \gls{rnn}-based approximation of the time-dependent \gls{dof}.
Due to this spatial-temporal split, the \gls{rnn} needs only to handle the variation of the coefficients in time, a task for which it is ideally suited. On the other hand, the \gls{fe} basis functions provide the basis for the spatial discretization of the computational domain, as is their intended function.
Combining these two modeling methods, each of which targets a separate approximation task, results in significantly more efficient training and higher prediction accuracy.
}

\textcolor{black}{
The \gls{rnn} can be written in the form of a map  $\delta^{\text{\tiny NN}} :\mathbb{R}^{N^\text{lofi}_{\text{dof}}\times\{t_k,..., t_{k+N_p-1}\}} \rightarrow \mathbb{R}^{N^\text{lofi}_{\text{dof}}\times\{t_k,..., t_{k+N_p-1}\}}$, such that
\begin{align}
\label{eq:approx_func}
\delta^{\text{\tiny NN}}\left(\mathbf{\hat{x}}^{\textcolor{black}{\text{lofi}}}_{t_k},...,\mathbf{\hat{x}}^{\textcolor{black}{\text{lofi}}}_{t_{k+N_p-1}}\right) 
= \left\{\bm{\delta}^{\text{\tiny NN}}_{t_k},...,\bm{\delta}^{\text{\tiny NN}}_{t_{k+N_p-1}}\right\}
\approx \left\{\bm{\hat{\delta}}_{t_k},...,\bm{\hat{\delta}}_{t_{k+N_p-1}}\right\},
\end{align}
where $\bm{\delta}^{\text{\tiny NN}}_{t_k}=\left(\delta^{\text{\tiny NN}}_{1,t_k},...,\delta^{\text{\tiny NN}}_{N_{\text{dof}}^{\text{lofi}},t_k}\right)$ denotes the component-wise output of the \gls{rnn} and $N_p$ the number of consecutive time steps considered in the input data.
Hence, the input and output dimension of the \gls{rnn} is $\mathbb{R}^{N^{\text{lofi}}_{\text{dof}}\times N_p}$.
Calculating \eqref{eq:approx_func} with inputs $\{\mathbf{\hat{x}}^{\textcolor{black}{\text{lofi}}}_{t_k}\}_{k=l \cdot N_p}^{(l+1)N_p-1}$ for $l=0,1,...,\frac{N^{\text{lofi}}_T}{N_p}-1$ yields an approximation of $\hat{\delta}_{i,t_k}$ on the whole trajectory.
Note that, the \gls{dof} $\bm{\hat{\delta}}_{t_k}$ corresponding to the high fidelity trajectory data $\mathbf{x}^{\text{hifi}}_{t_k}$ are used to train the \gls{rnn}, as explained in detail in Section \ref{sec:hybrid_model}.
}

\begin{figure}[t!]
	\centering
	\includegraphics[width = \linewidth]{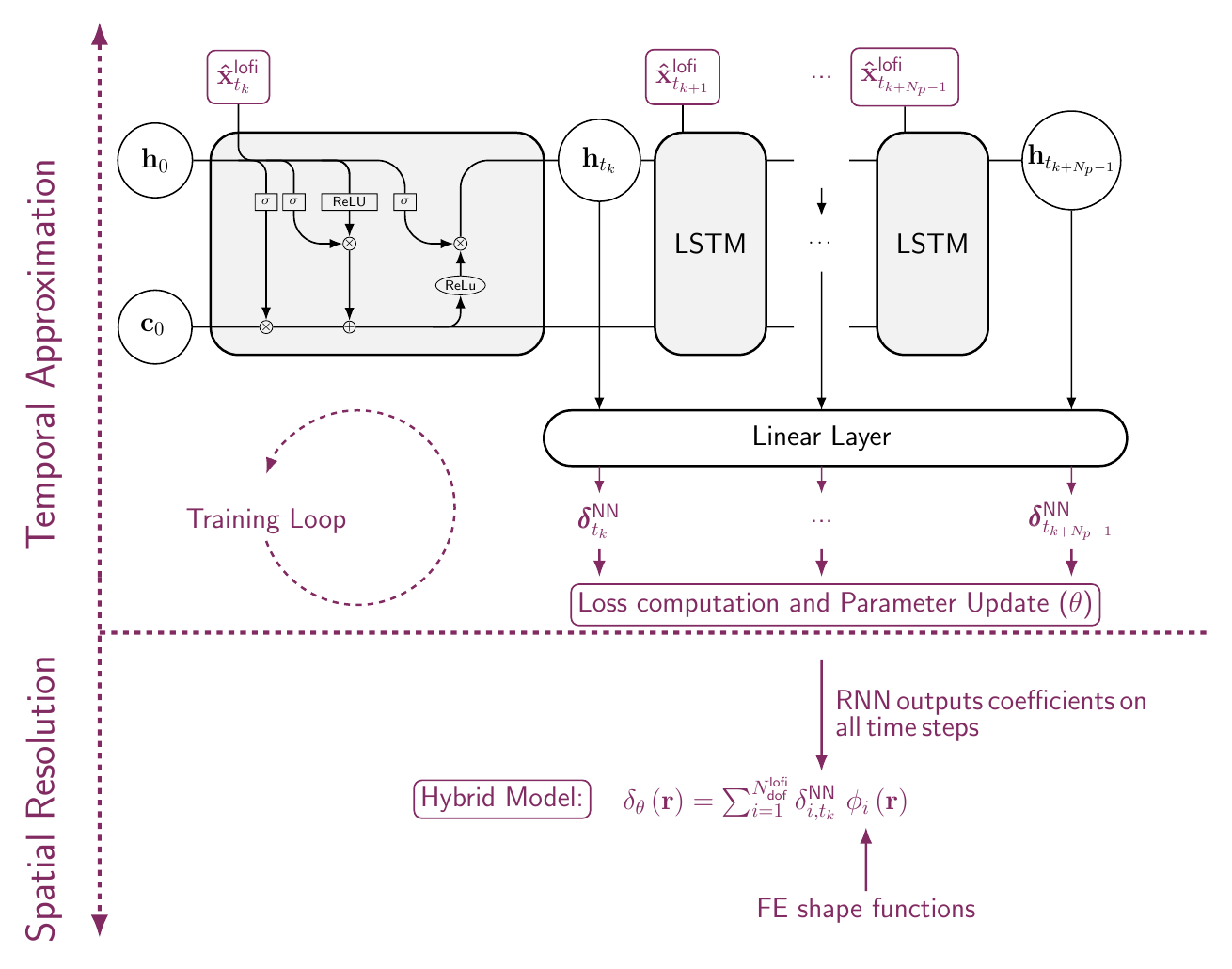}
	\caption{\textcolor{black}{Hybrid model architecture. The \gls{rnn} comprises $N_p$ \gls{lstm} units and a linear output layer and approximates the coefficients of the discrepancy function. The latter are subsequently combined with the \gls{fe} shape functions in the resulting hybrid model. The vectors $\mathbf{c}_{t}$ and $\mathbf{h}_t$, $t = t_k,...,t_{k+N_p-1}$, denote the cell states and hidden states of the \gls{lstm} units. The \gls{rnn}'s inputs and outputs are highlighted in purple.}}
	\label{fig:lstm_nn_error}
\end{figure}

\begin{figure}[t!]
	\centering
	\includegraphics[width = \linewidth]{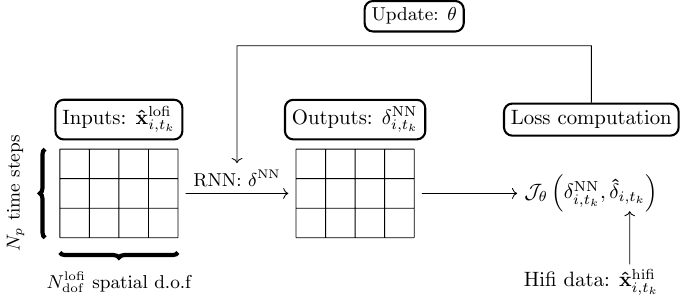}
	\caption{\textcolor{black}{Overview of the data exchange in the hybrid model's training process. Details regarding loss computation and pre-processing of the training data $\hat{\delta}_{t_k}$ are given in Sections \ref{sec:hybrid_model}, \ref{sec:upsampling}, and \ref{sec:non_dimensionalisation}.}}
	\label{fig:in_out}
\end{figure}

\textcolor{black}{
The structure of the hybrid model is shown in Figure~\ref{fig:lstm_nn_error}, which also highlights the separation into temporal and spatial approximation of the discrepancy function.
The data exchange within the \gls{rnn}/\gls{fe} hybrid model during the training procedure is depicted in Figure~\ref{fig:in_out}. 
Specific details on the architecture of the \gls{rnn} model and of the training procedure are provided in Sections \ref{sec:rnn-architecture}-\ref{sec:non_dimensionalisation}.
}

\subsection{\textcolor{black}{\gls{rnn} architecture}}
\label{sec:rnn-architecture}
\textcolor{black}{Due to their universal approximation property, as well as their use of residual connections to account for time-dependencies in the training data, \glspl{rnn} offer sufficient flexibility to approximate the discrepancy dynamics.
In this work, the \gls{rnn} architecture comprises two building blocks, namely, a concatenation of \gls{lstm} cells and a linear output layer. 
For the \gls{rnn} to approximate the underlying dynamics in the time-dependent data, $N_p>1$ consecutive time steps are considered simultaneously, where the coefficients of the low-fidelity system states, $\mathbf{\hat{x}}^{\text{lofi}}_{t_k}$, serve as inputs to the corresponding \gls{lstm} cells, while the outputs of the \gls{lstm} cells  approximate the discrepancy function's coefficients, that is, $\bm{\delta}^{\text{NN}}_{t_k} \approx \bm{\hat{\delta}}_{t_k}$.
The choice of $N_p$ is in fact a hyperparameter of the \gls{rnn} architecture.
}

\textcolor{black}{The hyperparameters of the \gls{rnn} model are chosen as follows.
As already mentioned, the number of \gls{lstm} cells is equal to the number of considered consecutive time steps $N_p$, which is chosen heuristically for each computational example.	
The values of $N_p = 2,3$ have been proven to be suitable choices. 
The number of neurons of each \gls{lstm} cell and of the linear output layer is equal to $N_{\text{dof}}^{\text{lofi}}$, such that each \gls{dof} of the low-fidelity \gls{fe} model to one neuron. 
This choice is made in order to avoid a more complicated encoder-decoder architecture.
To account for the piecewise and non-differentiable behavior occurring in phase transitions, it is necessary to employ \gls{relu} activation functions.
Last, with respect to the learning rates employed in the training process, step-based learning rate decay with a starting rate of $\alpha = 1 \cdot 10^{-3}$ is used. 
Explicit values for all hyperparameters are given in the computational examples in Section \ref{sec:numerical_examples}.
}

\subsection{Hybrid model training}
\label{sec:hybrid_model}

\begin{figure}[t!]
	\centering
	\begin{tikzpicture}[ArrowC2/.style={
			rounded corners=.5cm,
			very thick,
		}]
		\node (hifi) at (-5.5,2.5){\Large \textcolor{TUDa-10c}{$\mathbf{x}^{\text{hifi}}_{t_k}$}};

		\node[draw =TUDa-10c, rounded corners] (t0) at (-5.5,0.7){\textcolor{TUDa-10c}{\Large $t_0$}};
		\node[draw =TUDa-10c, rounded corners] (t1) at (-3.5,0.7){\textcolor{TUDa-10c}{\Large $t_1$}};
		\node[draw =TUDa-10c, rounded corners] (t2) at (0.5,0.7){\textcolor{TUDa-10c}{\Large $t_2$}};
		\node[draw =TUDa-10c, rounded corners] (t3) at (2.5,0.7){\textcolor{TUDa-10c}{\Large $t_3$}};
		\node[draw =TUDa-10c, rounded corners] (t4) at (5.5,0.7){\textcolor{TUDa-10c}{\Large $t_4$}};
		
		\node (t01) at (-4.5,-0.7){\Large $t_{0_1}$};
		\node (t11) at (-2.5,-0.7){\Large $t_{1_1}$};
		\node (t12) at (-1.5,-0.7){\Large $t_{1_2}$};
		\node (t13) at (-0.5,-0.7){\Large $t_{1_3}$};
		
		\node (t21) at (1.5,-0.7){\Large $t_{2_1}$};
		\node (t31) at (3.5,-0.7){\Large $t_{3_1}$};
		\node (t32) at (4.5,-0.7){\Large $t_{3_2}$};
		
		\draw[TUDa-10c, very thick, ->, arrows = {-Stealth[scale=1]}] (hifi) -- (t0);
		\draw[TUDa-10c,very thick, ->, arrows = {-Stealth[scale=1]}, rounded corners] (hifi) -- (-3.5,2.5)--(t1);
		\draw[TUDa-10c,very thick, ->, arrows = {-Stealth[scale=1]}, rounded corners] (hifi) -- (0.5,2.5)--(t2);
		\draw[TUDa-10c,very thick, ->, arrows = {-Stealth[scale=1]}, rounded corners] (hifi) -- (2.5,2.5)--(t3);
		\draw[TUDa-10c,very thick, ->, arrows = {-Stealth[scale=1]}, rounded corners] (hifi) -- (5.5,2.5)--(t4);
		
		\node (pic) at (0,0){\includegraphics[scale = 4]{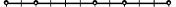}};
		\draw [very thick, decorate, decoration = {calligraphic brace, mirror}] (-2.8,-1.2) --  (-0.15,-1.2);
		\node (n) at (-1.4,-1.75){\large $N_{I_1}=3$};
	\end{tikzpicture}
	\caption{The high-fidelity solution $\mathbf{x}^{\text{hifi}}_{t_k}$ is defined on $T_{\text{hifi}}$. The intermediary time steps of the low-fidelity model are on the time axis. $N_{I_1}=3$ denotes the number of intermediary steps for the interval $I_1$.}
	\label{fig:time_axis}
\end{figure}
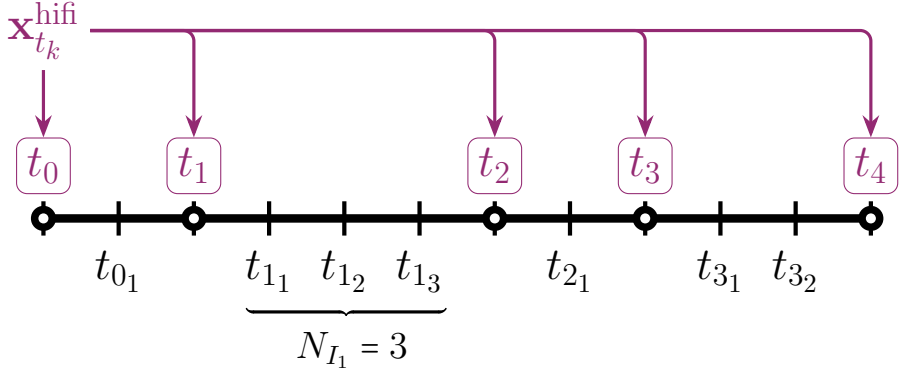

In this section, we discuss the training of the hybrid model, in particular of its \gls{rnn} part, including details on the choice of training data and how we deal with data sparsity.
The training data set consists of snapshot data of the discrepancy function, which are calculated from the down-projected high-fidelity solution and the low-fidelity solution.
Assuming $N^{\text{hifi}}_T$ trajectory samples of the high-fidelity solution $\{ \mathbf{x}_{t_k}^{\text{hifi}} \}_{t_k\leq N^{\text{hifi}}_T}$, we denote with $T_{\text{hifi}}:=\{t_0,t_1,...,t_{N^{\text{hifi}}_T} \}$ the respective time instances on which they are defined.
Then, we can partition the time axis of the \gls{bvp} into seperate intervals $I_k:= [t_k, t_{k+1}]$ such that $[0,T] = \bigcup_{k=0}^{N^{\text{hifi}}_T}\,I_k$ holds for $t_0=0$ and $t_{N^{\text{hifi}}_T}=T$.
As we can evaluate the low-fidelity model at more time instances than those included in the high-fidelity data, low-fidelity states $\mathbf{x}^{\text{lofi}}_{t_k}$ can be evaluated for $t_k \in T_{\text{hifi}}$, but also on intermediate time instances $t_k<t_{k_j}<t_{k+1}$ with $j=1,...,N_{I_k}$, where $N_{I_k}$ denotes the number of intermediate time steps in the interval $I_k$.
Consequently, the low-fidelity states $\mathbf{x}^{\text{lofi}}_{t_k}$ are defined on more time steps, namely $T_{\text{lofi}}:=T_{\text{hifi}}\cup\left\{\bigcup_{k=1}^{N^{\text{hifi}}_T}\bigcup_{j=1}^{N_{I_k}}t_{k_j}\right\}$.
We construct $T_{\text{lofi}}$ by choice of the low-fidelity model, such that $|T_{\text{hifi}}|\leq |T_{\text{lofi}}|$ holds and the time instances are uniform, i.e., $t_{k+1} - t_k = \Delta t$, $\forall \, t_k, t_{k+1}\in T_{\text{lofi}}$.
Especially the latter point is important for the inputs of \glspl{rnn}.
For a visualization of the difference between the sets $T_{\text{hifi}}$ and $T_{\text{lofi}}$, see \textcolor{black}{Figure \ref{fig:time_axis}}.
Given theses considerations, the training data consisting of discrepancy function snapshots is given as the set
\begin{align}
	\mathcal{D}_{\text{d}}:=\left\{\mathcal{T}\left(\mathbf{x}^{\text{hifi}}_{t_k}\right) - \mathbf{x}^{\text{lofi}}_{t_k} \right\}_{t_k\in T_{\text{hifi}}},
\end{align}
where $\mathcal{T}$ is a linear projection operator and $t_k \in T_{\text{hifi}}$.
Note that the sampled instances of the high-fidelity data are not chosen randomly. Instead, they depend on the dynamics of the underlying physical system.
In areas where the high-fidelity systems is very dynamic, for instance, when it is heavily excited, the trajectory is sampled more frequently than in areas where the system approaches steady state.
In this work, we sample heuristically and note that there are more sophisticated methods to perform data sampling, for instance, using active learning or optimal experimental design techniques.

Whilst sparse data sets are the norm for most practical scenarios, for example in measurement setups, \glspl{nn} typically show bad interpolation behavior when trained solely with sparse data sets. This is partially due to the fact that no knowledge of the objective function is provided in sparsely sampled areas, wherein the \gls{rnn}'s interpolation accuracy decreases dramatically.
In these scenarios, one has to ``inform'' the \gls{nn} on the correct mode of behavior.
One way to achieve this would be to add physically motivated constraints to the loss function, an approach frequently pursued in so-called physics-informed machine learning \cite{karniadakis2021physics}. 
Another approach is to upsample the available training data as to reflect correct physical behavior, an approach commonly used in game design and image processing \cite{hung2011fast, kavan2011physics, pons2021upsampling}.
The main difference between these two approaches is on how the information regarding the correct interpolation behavior is encoded. 
In the former case, it is encoded in the formulation of the optimization problem. 
In the latter case, in the artificial (upsampled) data points the model is trained with. 
The latter approach is explained in more detail in Section \ref{sec:upsampling}.

\subsection{Localized data upsampling and noise stabilization}
\label{sec:upsampling}
In this work, we resort to the approach of data upsampling, such that model training is performed within a standard supervised learning context.
Introducing physics-inspired loss functions, whilst also a promising avenue, has the drawback of significantly complicating the optimization problem, often resulting in elongated training times.

We propose an upsampling scheme based on a combination of localized, linear interpolation to calculate intermediate artificial system states for $t_k\in T_{\text{up}}:=T_{\text{hifi}} \backslash T_{\text{lofi}}$, and a Gaussian prior for noise stabilization to prevent the overfitting phenomenon.
Noise stabilization is a common approach in Gaussian Processes to improve numerical stability \cite{bernardo1998regression, williams2006gaussian} and can be used to prevent \glspl{nn} from overfitting.
The linear interpolation aspect controls the \gls{nn}'s behavior in the sparse data regime, whereas the prior distribution prevents the \gls{nn} from overfitting.
In each training epoch, the prior distribution is sampled to generate new artificial states that are locally bounded by the variance of the Gaussian prior.
We choose a linear interpolation approach due to its simplicity and ease of use, as well as its applicability to a large category of problem types.
We denote the locally linear interpolation function  $\overline{\delta}_{I_k}:I_k \rightarrow \mathbb{R}^{N^{\text{lofi}}_{\text{dof}}}$ as

\begin{align}
\label{eq:lin_up}
	\overline{\delta}_{I_k}(t) = \frac{\delta_{t_k} \,t_{k+1} - \delta_{t_{k+1}} \,t_{k}}{t_{k+1}-t_k} + t \left(\frac{\delta_{t_{k+1}} - \delta_{t_k}}{t_{k+1}-t_k}\right).
\end{align}
Essentially, \eqref{eq:lin_up} interpolates linearly in intervals of sparse data, based on the boundary values found in the training data.
We apply noise stabilization by assuming a Gaussian prior on the artificial intermediate states.
Let $\overline{\delta}_{i,t_{k_j}} = \overline{\delta}_{I_{k}}(t_{k_j})\big|_i$ be evaluated in $t_{k_j}\in I_k$ for $j=1,...,N_{I_k}$ and restricted to the $i$-th spatial \gls{dof}.
Then, we define the Gaussian prior
\begin{align}
\label{eq:prior}
	p\left(\overline{\delta}_{i,t_{k_j}}\right) = \overline{\delta}_{i,t_{k_j}} + \mathcal{N}\left(0,\alpha \left\| \overline{\delta}_{i,t_{k_j}}\right\|_2\right),
\end{align}
for $i = 1,...,N^{\text{lofi}}_{\text{dof}}$ and $t_{k_j} \in T_{\text{up}}$, where $\alpha\in\mathbb{R}$ is a weighting factor controlling the variance.
The distribution \eqref{eq:prior} is defined locally for the individual \gls{dof}, where a lower variance is assumed for \gls{dof} close to zero, thus preserving known boundary conditions, and a larger variance is allowed in domains where the discrepancy is non-zero.
An upsampled data set $\mathcal{D}_{\text{up}}$ for $t_{k_j}\in T_{\text{up}}$ based on \eqref{eq:prior} is then given as
\begin{align}
	\mathcal{D}_{\text{up}}:=\left\{ \bm{\delta}_{t_{k_j}}\in \mathbb{R}^{N^\text{lofi}_{\text{dof}}}\,\big|\quad \delta_{i,t_{k_j}}\sim p\left(\overline{\delta}_{i,t_{k_j}}\right), \quad i = 1,...,N^{\text{lofi}}_{\text{dof}}\right\}_{t_{k_j}\in T_{\text{up}}}.
\end{align}
In each training epoch of the \gls{rnn}, $\mathcal{D}_{\text{up}}$ is generated by sampling \eqref{eq:prior} for all spatial \gls{dof}. 
Thus, the sparse time series is upsampled to a complete trajectory using the varying artificial system states in $\mathcal{D}_{\text{up}}$ and the elements of $\mathcal{D}_{\text{d}}$ which remain fixed as part of the ground truth.

The training of the \gls{rnn} is based on a locally weighted loss function, given by
\begin{align}
	\mathcal{J}_{\theta} = \frac{1}{N^{\text{lofi}}_{\text{dof}}} \left( \overbrace{\sum_{\substack{t_k\in T_{\text{hifi}}}}\beta_{t_{k}}\,|\bm{\delta}_{t_{k}} - \pmb{\delta}^{\text{\tiny NN}}_{t_{k}} |}^{\text{ground truth}} 
	+ \overbrace{\sum_{
	\substack{
	t_{k_j}\in T_{\text{up}}
	}}|\bm{\delta}^{*}_{t_{k_j}}
	 - \bm{\delta}^{\text{\tiny NN}}_{t_{k_j}} |}^{\text{artificial states}}\right),
\end{align}
where $\bm{\delta}_{t_{k}}\in \mathcal{D_{\text{d}}}$ are part of the ground truth states, $\bm{\delta^{*}}_{t_{k_j}}\in \mathcal{D}_{\text{up}}$ the sampled artificial states, $\beta_{t_{k}}\in\mathbb{R}$ a local, time-dependent weighting factors, and $|\cdot|$ the Euclidean norm. 
In the initial training stages, we choose $\beta_{t_{k}}=1$, $\forall t_{k} \in T_{\text{hifi}}$, until the \gls{rnn} has a coarse fit on the data.
In the later training stages, we change the local weighting factors on $\mathcal{D}_{\text{d}}$ to $\beta_{t_{k}}=1+\| \delta_{t_{k}} - \delta_{\theta} \|_2$, which are calculated by numerical quadrature.
Note that $\beta_{t_{k}} \geq 1$. 
In that way, we ensure that the data set $\mathcal{D}_{\text{d}}$ is always weighted more heavily than $\mathcal{D}_{\text{up}}$ during \gls{rnn} training. 
For the optimization, we use the \gls{adam} algorithm with learning rate decay.

\subsection{Normalization by non-dimensionalization}
\label{sec:non_dimensionalisation}
A significant drawback of \glspl{nn} is their inability to adequately represent data with small values, a scenario which comes up when considering error functions.
Thus, normalization procedures are in order, which scale the input data and significantly improve the performance of \glspl{nn}.
Especially in physical systems, the quantities of interest can become very small as they are expressed relative to very small physical constants.
Examples of such constants are the magnetic permeability $\mu \sim 10^{-7}\,\si{\frac{Vs}{Am}}$ or the thermal conductivity $\kappa \sim 10^{-3} \,\si{\frac{W}{mK}}$.
Small physical geometries induce similar problems.
The resulting error functions of such systems consist of small, fluctuating values, thus requiring a normalization procedure which scales the dynamical system appropriately, while remaining physically conforming at the same time.

An example of such a scaling procedure is the non-dimensionalization of the physical system.
In essence, non-dimensionalization removes the physical dimensions from underlying differential equations by suitable variable substitutions.
The resulting differential equations have their physical dimensions partially or even completely removed. 
As an illustrative example, we present how the temporal and spatial differential operators transform under change of variables.
Let $\mathbf{r} = \left(r_x,r_y\right)$ and $\tau=t^{-1}_ct$, $\underline{r_x}=r^{-1}_{x,c} \,r_x$, $\underline{r_y}=r_{y,c}^{-1} \,r_y$ for $t_c\,[\mathrm{s}],\, r_{x,c}\,[\mathrm{m}],\, r^{-1}_{y,c} \,[\mathrm{m}] \in \mathbb{R}$, be a coordinate transformation for which physical dimensions have been removed.
The non-dimensionalised $n$-th time derivative as well as the Laplace operator transformation is given by
\begin{align}
	\frac{\partial^n}{\partial t^n} = \frac{1}{t^n_c}\frac{\partial^n}{\partial \tau^n} \quad \text{and}\quad \Delta = \frac{1}{\sqrt{|g|}}\sum_{i=1}^2\partial_i\left(\sqrt{|g|}g_{ii}\partial_i \right) \text{, with} \quad
	g = 
	\begin{pmatrix}
		r^{-2}_{x,c} & 0 \\
		0 & r^{-2}_{y,c}
	\end{pmatrix},
\end{align}
where $g$ is the metric tensor and $\Delta$ the Laplace-Beltrami operator.
The transformed wave equation then reads
\begin{align}
	\left[\frac{1}{t^{2}_c} \frac{\partial^2}{\partial \tau^2} + c\left(\frac{1}{r^{2}_{x,c}} \frac{\partial^2}{\partial \underline{r_x}^2} + \frac{1}{r^{2}_{y,c}}\frac{\partial^2}{\partial \underline{r_y}^2} \right)\right]u(\tau,\underline{r_x},\underline{r_y}) = f(\tau,\underline{r_x},\underline{r_y}),
\end{align}
where the solution to the original system can then be recovered by the backwards transformation \textcolor{black}{$t=t_c\tau$, $r_x=r_{x,c} \,\underline{r_x}$ and $r_y=r_{y,c}\,\underline{r_y}$}.
The resulting system can then be scaled such that \glspl{nn} can optimally fit the data, as well as conform to the physical equations.

\section{Numerical Examples}
\label{sec:numerical_examples}
In the following numerical investigations, we employ the proposed hybrid model to approximate discrepancy functions in three engineering test cases governed by transient \glspl{pde}, namely, heat diffusion on a heat sink, eddy-currents in a quadrupole magnet, and sound wave propagation inside a cavity.
For each test case, we consider a high-fidelity and a low-fidelity \gls{fe} representation of the \gls{bvp}. 
The difference between the two lies in the \gls{fe} mesh refinement, as well as in modeling errors in the material laws, excitation, and domain geometry, depending on the test case. 
To highlight the necessity of the upsampling approach proposed in Section \ref{sec:upsampling}, we consider hybrid models with and without data upsampling and observe the impact of overfitting in the latter case.
Both simulations are then compared to reference data of the discrepancy function, which is calculated from densely sampled high-fidelity data.
The trained hybrid models are then employed for bias correction of the low-fidelity models using \eqref{eq:corr_model}, leading to significantly more accurate results.

\textcolor{black}{
To assess the accuracy of the discrepancy function approximation,
we use the relative	 $L^2(\Omega)$ error
\begin{align}
	\Delta_{L^2} \, \delta_{\theta} = \frac{\int_{[0,T]}\| \delta_\theta - \delta_t \|_2 \,\mathrm{d}t}{\int_{[0,T]} \|\delta_t \|_2 \,\mathrm{d}t},
\end{align}
where $\|\cdot \|_2$ is approximated via numerical integration over the computational mesh and $\int_{[0,T]}\cdot \, \mathrm{d}t$ via Riemannian sums.
Accordingly, the accuracy of the corresponding bias-corrected model is evaluated using the relative $L^2(\Omega)$ error
\begin{align}
	\Delta_{L^2} \, \mathbf{x} = \frac{\int_{[0,T]}\| \mathbf{x}_t - \mathcal{T}\left(\mathbf{x}^{\text{hifi}}_t\right) \|_2 \,\mathrm{d}t}{\int_{[0,T]} \|\mathcal{T}\left(\mathbf{x}^{\text{hifi}}_t\right) \|_2 \,\mathrm{d}t}.
\end{align}
}

Considering the \gls{fe} simulations, we restrict ourselves to the 2D case without loss of generality.
Consequently, for the derivation of the \gls{fe} formulation, we restrict the space of test functions to
\begin{align}
	H_0(\textrm{grad};\,\Omega) := \left\{u\in L^2(\Omega) \text{ with } \nabla(u)\in L^2(\Omega) \text{ and } u|_{\partial \Omega}=0 \right\},
\end{align}
in all following test cases.
Thus, $w \in H_0(\textrm{grad};\,\Omega)$ are defined by the mesh nodes of the triangulation of $\Omega$.
For all test cases, we choose first order shape functions.
For the \gls{fe} implementation we use \texttt{gmsh} \cite{geuzaine2009gmsh} and \texttt{FEniCSx} \cite{AlnaesEtal2014, LoggEtal2012, BasixJoss, ScroggsEtal2022}. \glspl{rnn} are implemented using \texttt{jax}/\texttt{flax} \cite{jax2018github, flax2020github}.

\subsection{Heat diffusion on a heat sink}
\begin{figure}[t!]
\begin{tikzpicture}
\node at (-2,0){
\includegraphics[width = 8cm,
		height = 7.5cm]{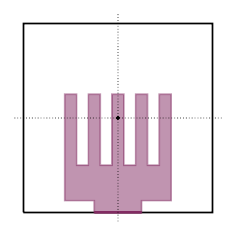}
};
\node at (4.5,-0.1){
\begin{tikzpicture}
\begin{axis}[scatter/classes={%
    a={mark=x,draw=green}, b={mark=x,draw=blue}},
		grid=both,
		width = 5.5cm,
		height = 7.5cm,
		xlabel={$r_y$},
		ylabel={$\kappa(0,r_y)$},
		legend columns = 2,
		legend style={at ={(0.975,1.1)}, fill=none, column sep = 1ex}
		]
		\addplot [TUDa-10c, very thick] table [col sep=comma, x=epochs, y = data] {./data/heat_sink/kappa.csv};
		\addlegendentry{$\kappa_{\text{hifi}}$};
		\addplot [TUDa-0d, densely dotted, very thick] table [col sep=comma, x=epochs, y = data] {./data/heat_sink/kappa_hom.csv};
		\addlegendentry{$\kappa_{\text{lofi}}$};
		\end{axis}
		\end{tikzpicture}
		};
\draw[TUDa-10c] node (a1) at (-2,-3.45) {$u|_{\partial\Omega_{\text{nd}}}=c$};
\draw node (a2) at (-4,3.35) {$u|_{\partial\Omega_{\text{d}}}=0$};	
\draw[fill = white, opacity = 0.5] node (a3) at (-1,-2) {$\Omega_{\text{con}}$};	
\draw node (a4) at (0,2) {$\Omega_{\text{air}}$};	
\end{tikzpicture}
\caption{\textbf{Left:} Schematic of a heat sink cross-section, where $\Omega_{\text{con}}$ is the thermally conductive domain and $\Omega_{\text{air}}$ the non-conductive domain. Non-homogeneous Dirichlet boundary condition are applied at the heat sink base at $\partial \Omega_{\text{nd}} = \partial \Omega \cap \partial \Omega_{\text{con}}$. \textbf{Right:} Plot of the thermal conductivity of the material along the middle fin, where $\kappa_{\text{hifi}}$ is the conductivity with defects and $\kappa_{\text{lofi}}$ without. The conductivity in $\Omega_{\text{air}}$ is $\kappa = 0.5 \si{W\left(mK\right)^{-1}}$ for both cases.}
\label{fig:heat_sink_geo}
\end{figure}
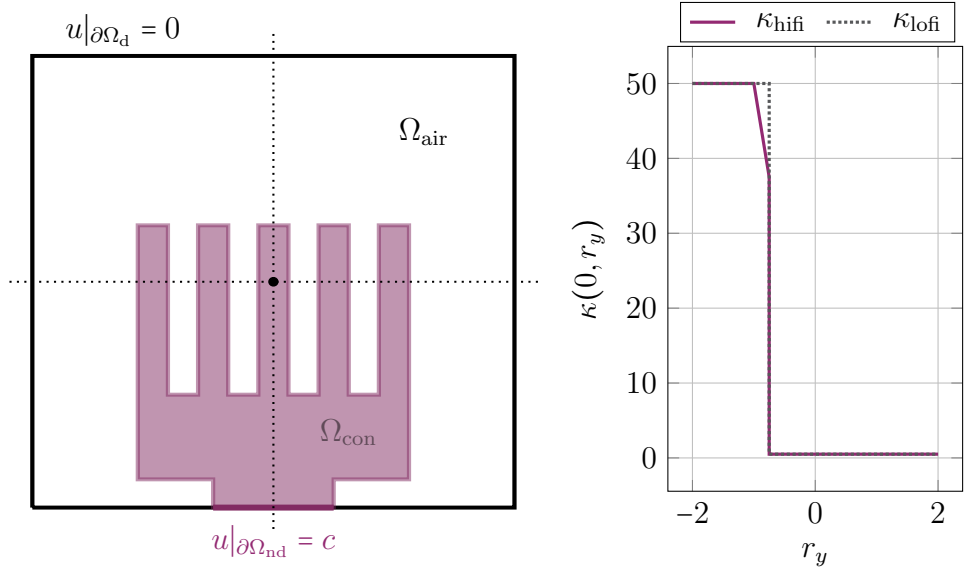

As first test case, we consider the heat diffusion problem on a 2D cross-section of a heat sink, see Figure~\ref{fig:heat_sink_geo} (left).
The heat sink geometry is defined on the domain $\Omega=[-l,l]^2$, $l\in\mathbb{R}$, which consists of a thermally conductive region $\Omega_{\text{con}}$ and a less thermally conductive air region $\Omega_{\text{air}}$.
The boundary of the geometry consists of $\partial \Omega_{\text{d}}= \partial \Omega \cap \partial \Omega_{\text{air}}$, to which we apply homogeneous Dirichlet boundary conditions and $\partial \Omega_{\text{nd}} = \partial \Omega \cap \partial \Omega_\text{con}$, to which we apply non-homogeneous Dirichlet boundary conditions.
The \gls{bvp} reads
\begin{align}
\label{eq:heat_eq}
    \rho c_v \frac{\partial u}{\partial t} - \nabla \cdot(\kappa \, \nabla u)&=0 \text{ on } \Omega,  \nonumber \\
    u|_{\partial \Omega_{\text{nd}}} &= c,        	  \\
    u|_{\partial \Omega_{\text{d}}} &= 0, \nonumber
\end{align}
where $u$ is the temperature, $\rho c_v$ the heat capacity, $\kappa$ the thermal conductivity, and $c$ a fixed temperature.
The non-homogeneous Dirichlet boundary condition signifies a heat source underneath the heat sink, while the homogeneous Dirichlet boundary conditions assume constant temperature on the boundary.
On the fins of the heat sink, we assume a heat conductivity of $\kappa = 50 \, \si{W\left(mK\right)^{-1}}$, following a linear deterioration of the heat conductivity by $25\%$ close to the proximity of the tips.
This deteroration can be attributed to material aging or other defects. 
\textcolor{black}{Figure~\ref{fig:heat_sink_geo}} (right) shows the thermal conductivity including material defects, $\kappa_{\text{hifi}}$, plotted against the thermal conductivity without material defects, $\kappa_{\text{lofi}}$, along the middle fin of the heat sink.
In $\Omega_{\text{air}}$ we assume $\kappa = 0.5 \,\si{W\left(mK\right)^{-1}}$.

\subsubsection{Finite element modeling}
To solve the heat diffusion problem using the \gls{fe} method, we introduce the corresponding variational form by multiplying \eqref{eq:heat_eq} with a test function $w_i\in H_0(\textrm{grad};\,\Omega)$ and integrate over the domain. 
The variational form reads
\begin{align}
    \int_{\Omega} \rho c_v \frac{\partial u}{\partial t}\, w_i \, \mathrm{d} \Omega - 
     \int_{\Omega} \nabla \cdot(\kappa \nabla u)\, w_i \, \mathrm{d}\Omega = 0,
\end{align}
for $i=1,...,N_{\text{dof}}^{\text{lofi}}$.
Applying integration by parts and first order difference quotients $\frac{\partial u}{\partial t} = \frac{u_{t_{k+1}} - u_{t_{k}}}{\Delta t}$ results in the implicit Euler time stepping scheme
\begin{align}
    \int_{\Omega} \rho c_v \,u_{t_{k+1}} \, w_i \,\mathrm{d} \Omega + 
     \Delta t \int_{\Omega} \kappa \nabla u_{t_{k+1}} \cdot \nabla w_i \,\mathrm{d}\Omega = \int_{\Omega} \rho c_v \, u_{t_{k}}\, w_i \,\mathrm{d} \Omega.
\end{align}
We discretize the temperature $u=\sum^{N^{\text{lofi}}_{\text{dof}}+N^{\text{lofi}}_{\text{bdry}}}_{j=1} \hat{u}_j\,w_j$. 
The first $N^{\text{lofi}}_{\text{dof}}$ shape functions $w_j\in H_0(\textrm{grad};\,\Omega)$ for $j=1,...,N^{\text{lofi}}_{\text{dof}}$ are the trial functions and equal to the test functions, adopting the Ritz-Galerkin approach.
$\left\{ \hat{u}_j\right\}_{j=1}^{N^{\text{lofi}}_{\text{dof}}}$ are the \gls{dof}.
The additional $N^{\text{lofi}}_{\text{bdry}}$ shape functions $w_j$ for $j=N^{\text{lofi}}_{\text{dof}}+1,...,N^{\text{lofi}}_{\text{dof}}+N^{\text{lofi}}_{\text{bdry}}$ together with the coefficients $\{\hat{u}_j\}_{j=N^{\text{lofi}}_{\text{dof}}+1}^{N^{\text{lofi}}_{\text{dof}}+N^{\text{lofi}}_{\text{bdry}}}$ discretize the non-homogeneous Dirichlet data.
The resulting system reads
\begin{align}
\label{eq:update_scheme}
  \left(\Delta t \,\mathbf{A} + \mathbf{M}\right)\, \mathbf{\hat{u}}_{t_{k+1}} = \mathbf{M} \, \mathbf{\hat{u}}_{t_{k}},
\end{align}
where $\mathbf{A}$ is the stiffness matrix and $\mathbf{M}$ the mass matrix and the individual entries of the respective matrices are given as
\begin{align}
\label{eq:heat_discr}
    \mathbf{M}_{ij} =
    \int_{\Omega} \rho c_v w_i\, w_j  \,\mathrm{d}\Omega \quad \text{ and }\quad 
    \mathbf{A}_{ij} =
    \int_{\Omega}  \kappa  \nabla w_i \cdot \nabla w_j\,\mathrm{d}\Omega.
\end{align}
The columns of $\mathbf{A}$ and $\mathbf{M}$ for $j=N^{\text{lofi}}_{\text{dof}}+1,...,N^{\text{lofi}}_{\text{dof}}+N^{\text{lofi}}_{\text{bdry}}$ are shifted from the left to right-hand side of \eqref{eq:update_scheme} by applying a Dirichlet lift \cite{quarteroni2008numerical}.

For the simulation with either model, i.e, low- or high-fidelity, we assume $\rho c_v=1 \,\si{JK^{-1}m^{-3}}$, $c=10\, \si{K}$, $\Delta t = 2\cdot 10^{-2}\, \si{s}$, and $N_T=100$.
The low-fidelity model $\mathbf{\Psi}_{\text{lofi}} \left(\mathbf{\hat{u}}_{t_k}^{\text{lofi}}\,|\,\kappa_{\text{lofi}}\right)$ has thermal conductivity $\kappa_{\text{lofi}}$ with $N_{\text{dof}}^{\text{lofi}}=278$, while the high-fidelity model $\mathbf{\Psi}_{\text{hifi}}\left(\mathbf{\hat{u}}^{\text{hifi}}_{t_ k}\,|\,\kappa_{\text{hifi}}\right)$ has thermal conductivity $\kappa_{\text{hifi}}$ and $N_{\text{dof}}^{\text{hifi}}= 1408$. 
The different material choices and mesh discretizations induce modeling and discretization errors between low-fidelity and high-fidelity model.

\subsubsection{Dicrepancy function approximation}
To approximate the discrepancy function, we employ 19 out of 100 trajectory samples as training data, where more samples are chosen in the initial stages of excitation and a reduced number of samples when the system reaches steady-state.
The training data instances are depicted as crosses at the respective time steps in Figure \ref{fig:error_heat_sink}.
The \gls{rnn} is trained according to the parameters in \textcolor{black}{Table \ref{table:heat_sink_par}}, from which we observe that $1000$ training epochs are required to reduce $\Delta_{L^2}\,\delta_{\theta}$ of the upsampled model to $0.27\,\%$ and equivalently, the non-upsampled model to $8.187\,\%$.
\textcolor{black}{The \gls{rnn} has $N_p=2$ \gls{lstm} cells and $N_{\text{dof}}^{\text{lofi}}=278$ neurons per layer, amounting to $1\,316\,330$ trainable parameters.}

\begin{table}[b!]
	\centering
	\begin{tabular}{l l c c}
		\hline\hline
		Description & Symbol & 0-500 & 500-1000\\ [0.5ex] 
		\hline
		Learning rate    & $\eta$    &  $1\cdot 10^{-3}$ & $1\cdot 10^{-4}$ \\
		Local weighting factors & $\beta$ & $\times$ & $\checkmark$ \\
		Variance weighting factor            & $\alpha$  &  $\frac{1}{25}$  & $\frac{1}{25}$  \\
		Error with upsampling & $\Delta_{L^2} \delta_{\theta^{\text{up}}}$  &  $2.502\,\%$  & $0.270\,\%$       \\
		Error without upsampling & $\Delta_{L^2} \delta_{\theta}$  &  $10.014\,\%$  & $8.187\,\%$    \\[1ex]
		\hline
	\end{tabular}
	\caption{\textcolor{black}{Training parameters and hybrid model errors for the heat sink test case. The \gls{rnn} has $N_p=2$ \gls{lstm} cells, equivalently, $N_p=2$ consecutive time steps are taken into account in each training epoch. Each layer of the \gls{rnn} has $N_{\text{dof}}^{\text{lofi}}=278$ neurons.}}
	\label{table:heat_sink_par}
\end{table}

In Figure \ref{fig:error_heat_sink}, the spatially integrated discrepancy function $\|\delta_{t} \|_{2}$ is displayed at each time step, once for the hybrid model, with and without upsampling, and once for reference data.
In case the hybrid model is trained solely on the sparse data set, we observe a good agreement on the training data, however, large errors appear for previously unseen data, due to overfitting.
This phenomenon is especially pronounced in areas where larger gaps exist in the training data, for example between the time steps $t_{30}=0.6\,\si{s}$ and $t_{100}=2 \,\si{s}$.
However, we also note that upsampling is not necessary in all regions of the trajectory for the hybrid model to exhibit correct interpolation behavior, as can be observed between time steps $t_5=0.1 \,\si{s}$ and $t_{15} = 0.3\,\si{s}$ for the non-upsampled model.

\begin{figure}[t!]
	\begin{tikzpicture}[scale = 1]
		\pgfplotsset{
			every axis/.style={xmin=0, xmax = 100}
		}
		\begin{axis}[scatter/classes={%
				a={mark=x,draw=TUDa-10c}},
			grid = both,
			width = \textwidth,
			height = 7 cm,
			ymax = 0.75,
			xmax = 2,
			xlabel={$t$},
			ylabel={$\|\delta_{t}\|_{2}$},
			legend columns = 4,
			legend style={at={(1,1.125)}, fill=none, column sep = 1ex}
			]
			\addplot [TUDa-0d, thick] table [col sep=comma, x=epochs, y=error, x expr = 0.02*\thisrow{epochs}] {./data/heat_sink/d_func_ref.csv};
			\addlegendentry{\footnotesize reference};
			\addplot [scatter, only marks, draw = TUDa-10c, mark = x, mark size = 4pt, very thick] table [col sep=comma, x=epochs, y=error, x expr = 0.02*\thisrow{epochs}] {./data/heat_sink/d_func_training_data.csv};
			\addlegendentry{\footnotesize training data};
			\addplot [TUDa-0c, densely dotted, thick] table [col sep=comma, x=epochs, y=error, x expr = 0.02*\thisrow{epochs}] {./data/heat_sink/d_func_approx_regular.csv};
			\addlegendentry{\footnotesize hybrid model};
			\addplot [TUDa-10c, densely dashed, thick] table [col sep=comma, x=epochs, y=error, x expr = 0.02*\thisrow{epochs}] {./data/heat_sink/d_func_approx_upsampling.csv};
			\addlegendentry{\footnotesize hybrid model + up.};
		\end{axis}
	\end{tikzpicture}
	\label{fig:error_heat_sink}
	\caption{Spatially integrated discrepancy function $\|\delta_{t}\|_{2}$ for $t_0=0 \,\si{s}$ until $t_{100}=2\,\si{s}$. The reference and the hybrid model with and without artificial upsampling are depicted. The training data is indicated at the respective time steps with \textcolor{TUDa-10c}{$\mathbf{\times}$}.}
	\label{fig:error_heat_sink}
\end{figure}
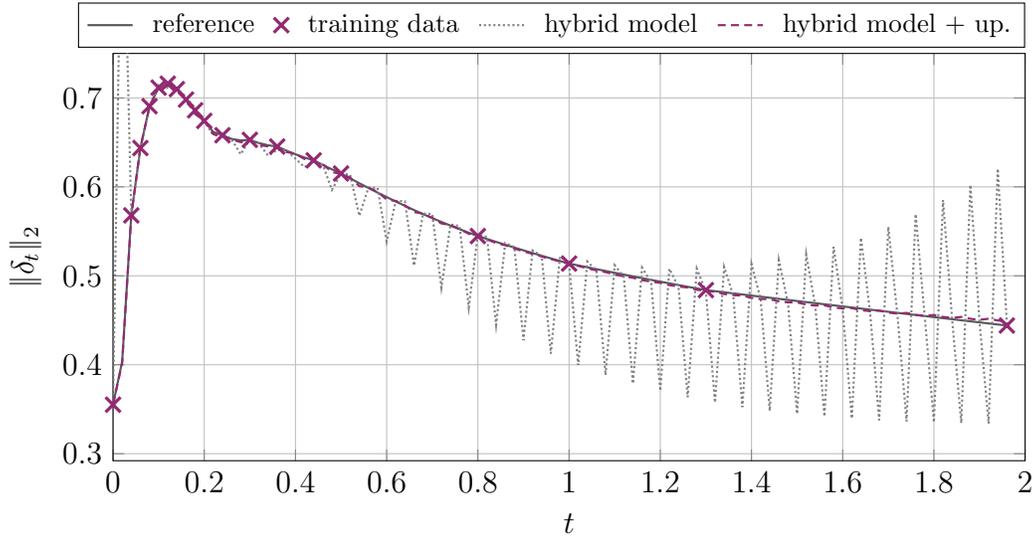

\subsubsection{Bias correction}
\begin{figure}[t!]
	\vspace{-5em}
	\centering	
	\begin{tabular}{cccc}
		$u^{\text{\tiny hifi}}_{t_1}$ & 
		$u^{\text{\tiny hifi}}_{t_{10}}$ & 
		$u^{\text{\tiny hifi}}_{t_{20}}$ &  
		$u^{\text{\tiny hifi}}_{t_{50}}$
		\\
		\includegraphics[width=0.25 \linewidth]{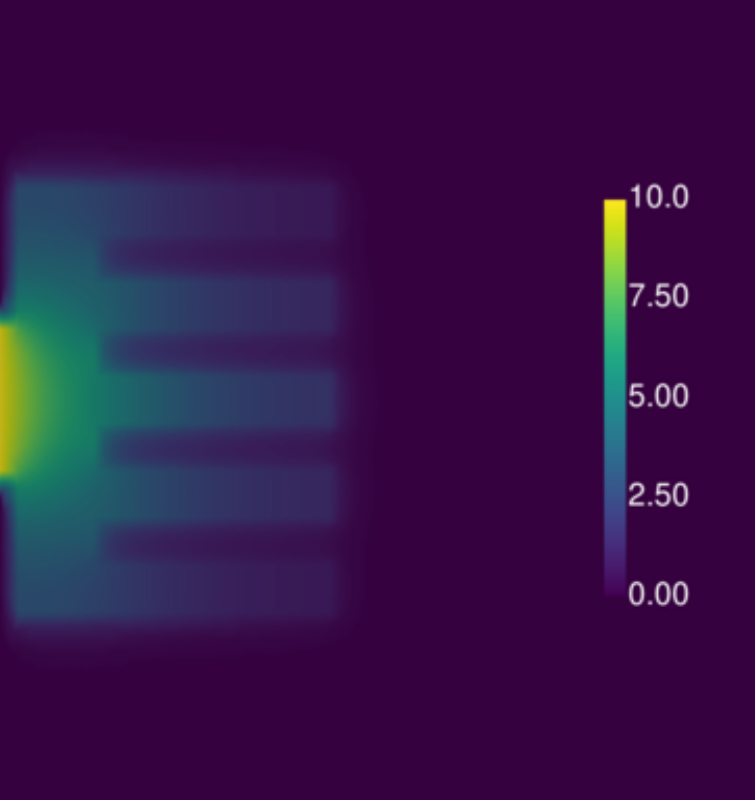}&
		\includegraphics[width=0.25 \linewidth]{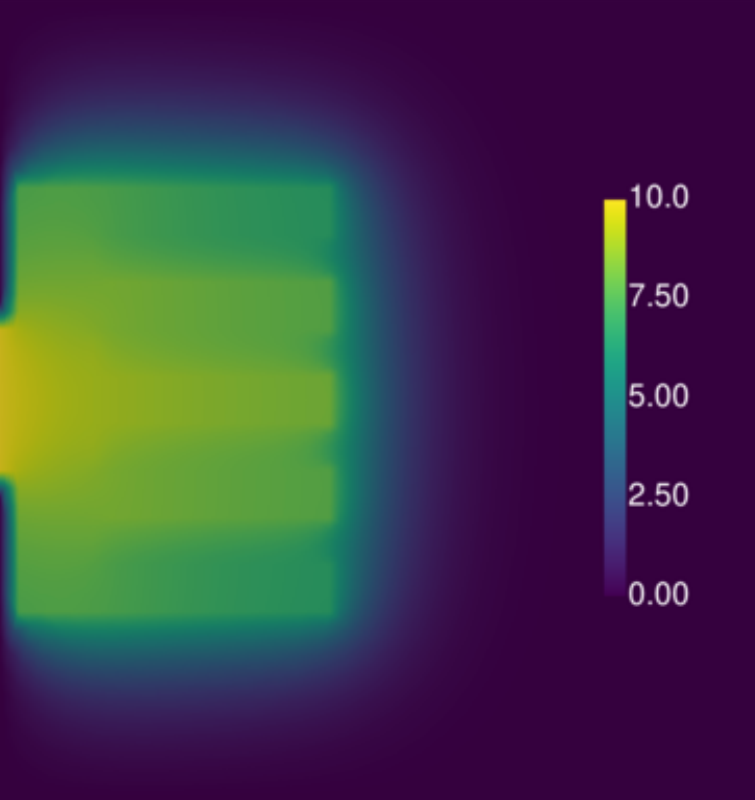}&
		\includegraphics[width=0.25 \linewidth]{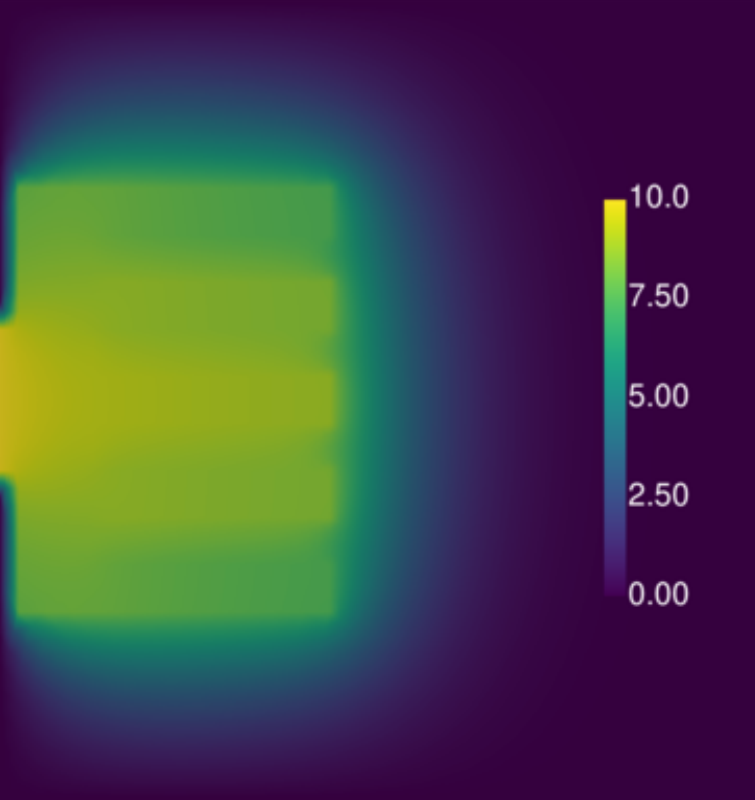}&
		\includegraphics[width=0.25 \linewidth]{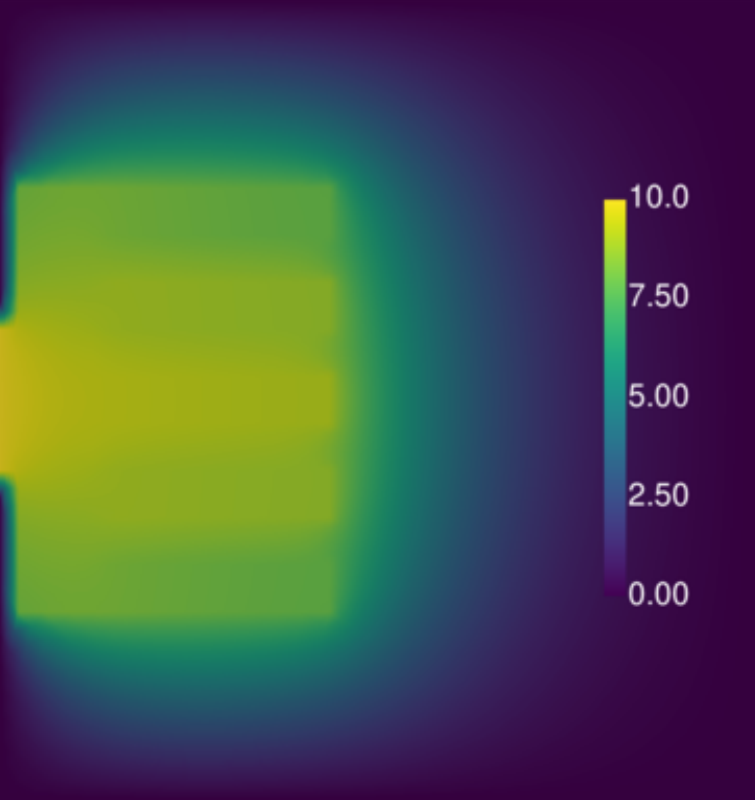} 
		\\
		$u^{\text{\tiny lofi}}_{t_1}$ & 
		$u^{\text{\tiny lofi}}_{t_{10}}$ & 
		$u^{\text{\tiny lofi}}_{t_{20}}$ &  
		$u^{\text{\tiny lofi}}_{t_{50}}$
		\\
		\includegraphics[width=0.25 \linewidth]{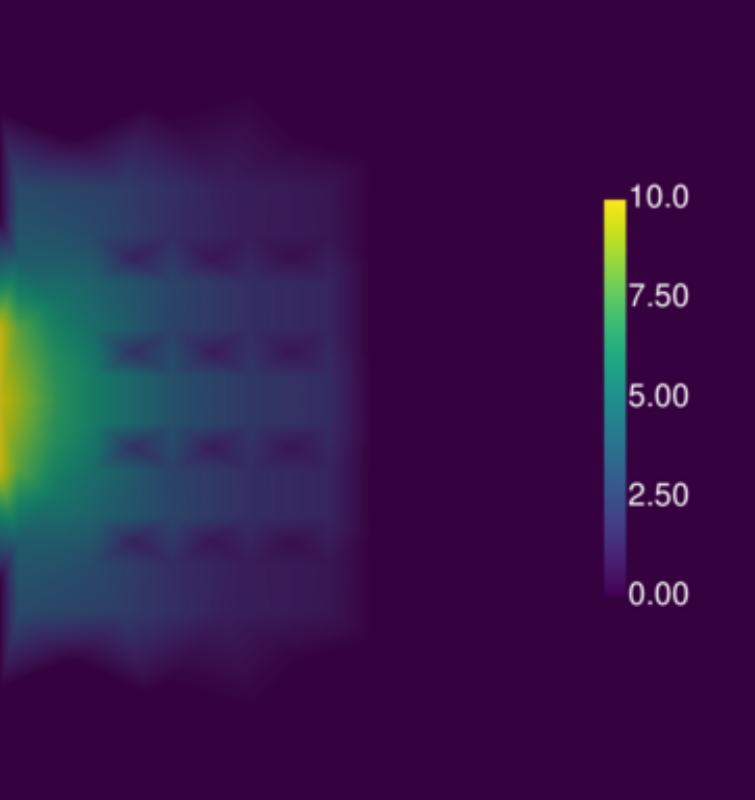}&
		\includegraphics[width=0.25 \linewidth]{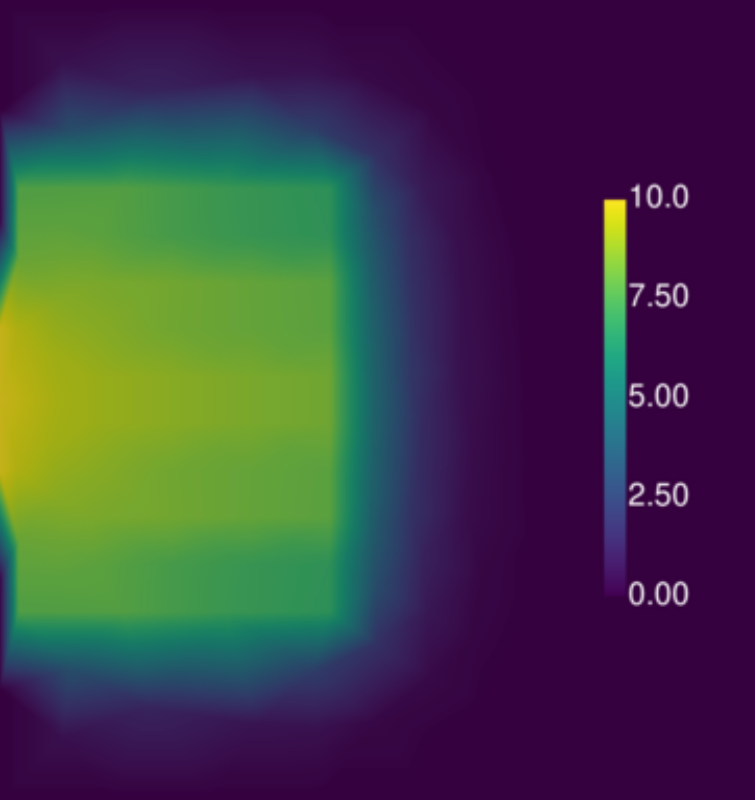}&
		\includegraphics[width=0.25 \linewidth]{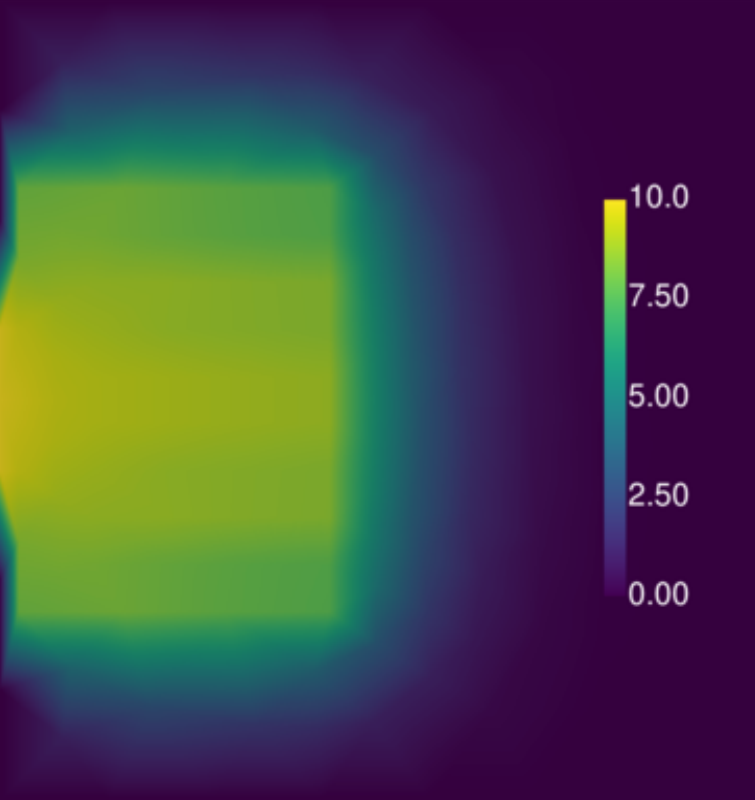}&
		\includegraphics[width=0.25 \linewidth]{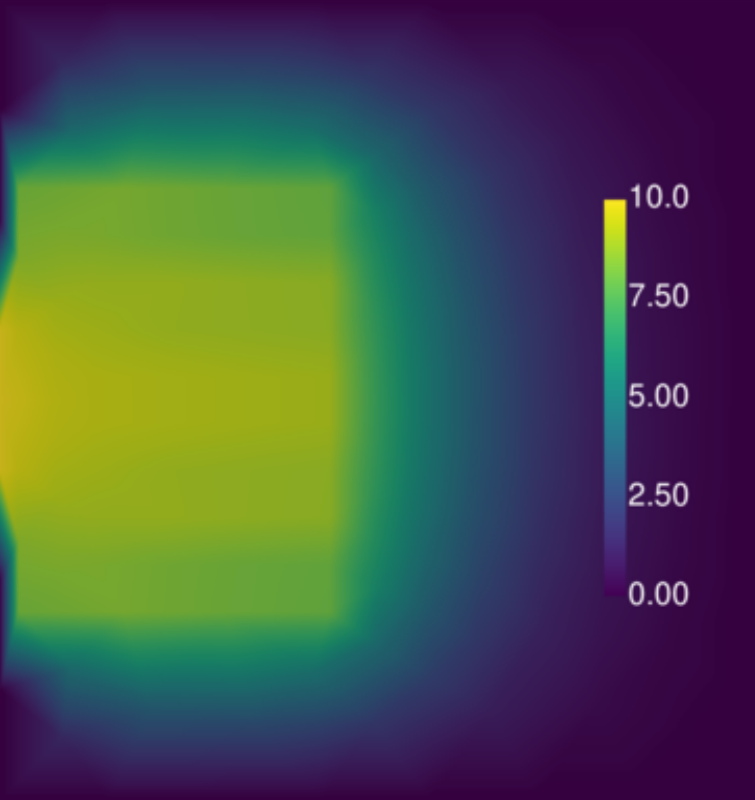} 
		\\
		$u^{\text{\tiny corr}}_{t_1}$ & 
		$u^{\text{\tiny corr}}_{t_{10}}$ & 
		$u^{\text{\tiny corr}}_{t_{20}}$ &  
		$u^{\text{\tiny corr}}_{t_{50}}$ 
		\\
		\includegraphics[width=0.25 \linewidth]{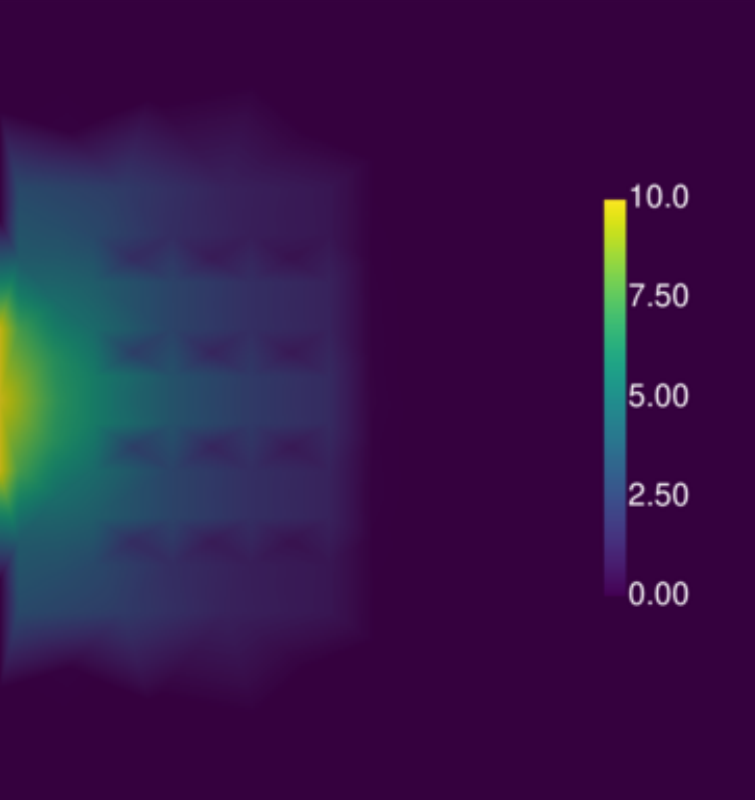}&
		\includegraphics[width=0.25 \linewidth]{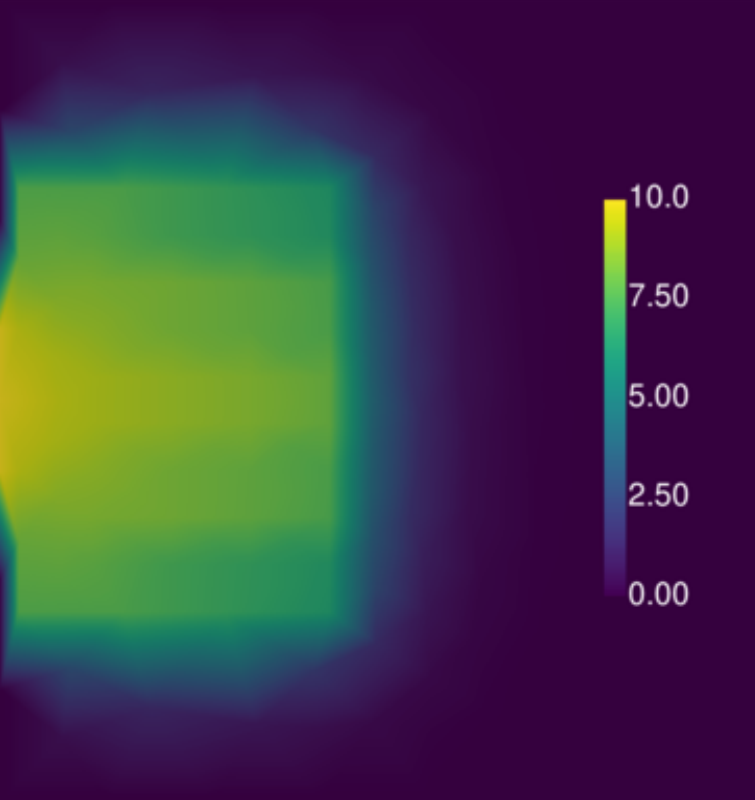}&
		\includegraphics[width=0.25 \linewidth]{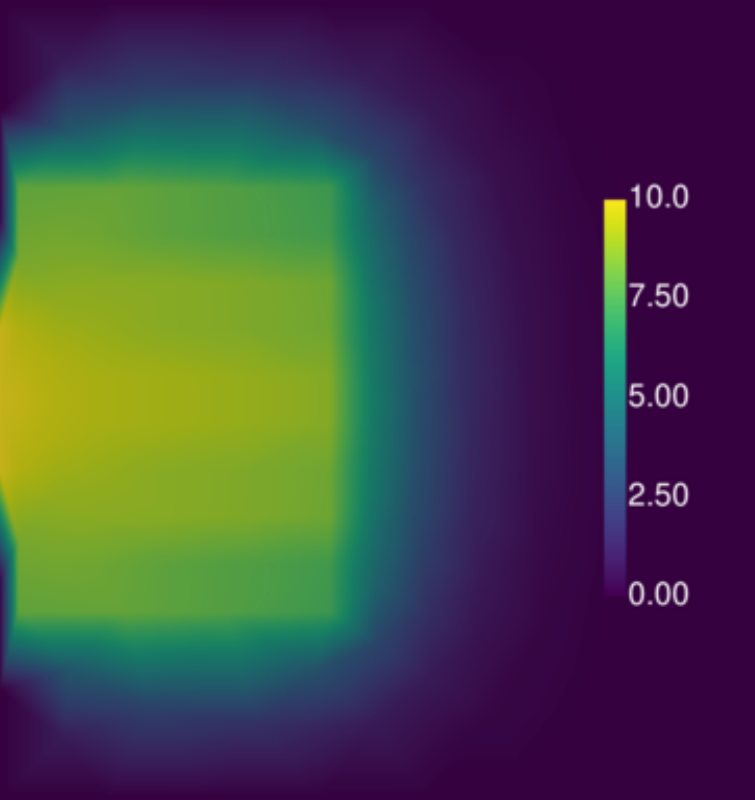}&
		\includegraphics[width=0.25 \linewidth]{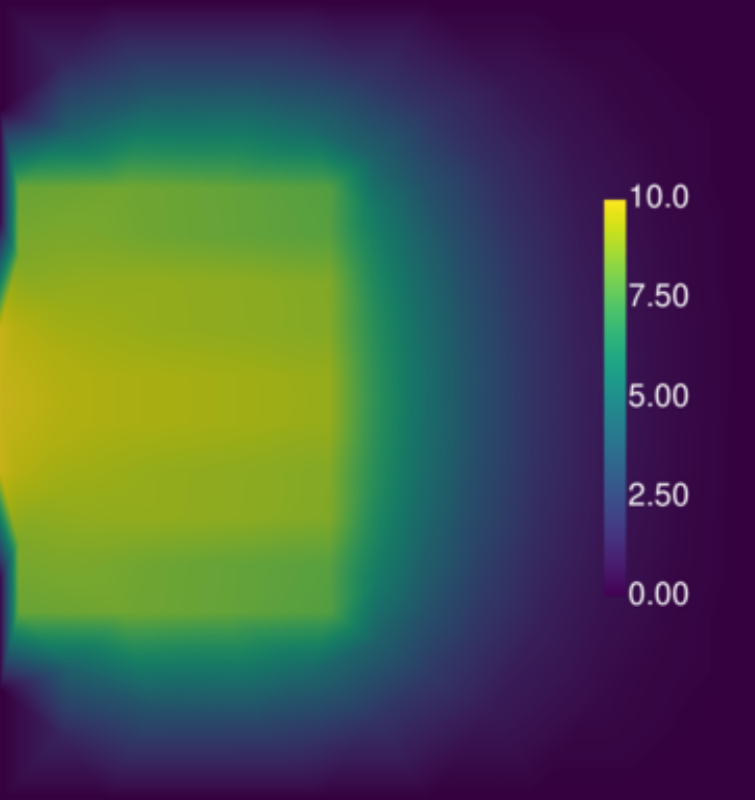} 
		\\
		\scriptsize $\sum_{i=1}^{N^\text{lofi}_{\text{dof}}}\,\left|\delta^{\text{\tiny NN}}_{i,t_1}\,\right|\phi_i(\mathbf{r})$ & 
		\scriptsize $\sum_{i=1}^{N^\text{lofi}_{\text{dof}}}\,\left|\delta^{\text{\tiny NN}}_{i,t_{10}}\,\right|\phi_i(\mathbf{r})$ & 
		\scriptsize $\sum_{i=1}^{N^\text{lofi}_{\text{dof}}}\,\left|\delta^{\text{\tiny NN}}_{i,t_{20}}\,\right|\phi_i(\mathbf{r})$ &  
		\scriptsize $\sum_{i=1}^{N^\text{lofi}_{\text{dof}}}\,\left|\delta^{\text{\tiny NN}}_{i,t_{50}}\,\right|\phi_i(\mathbf{r})$
		\\
		\includegraphics[width=0.25 \linewidth]{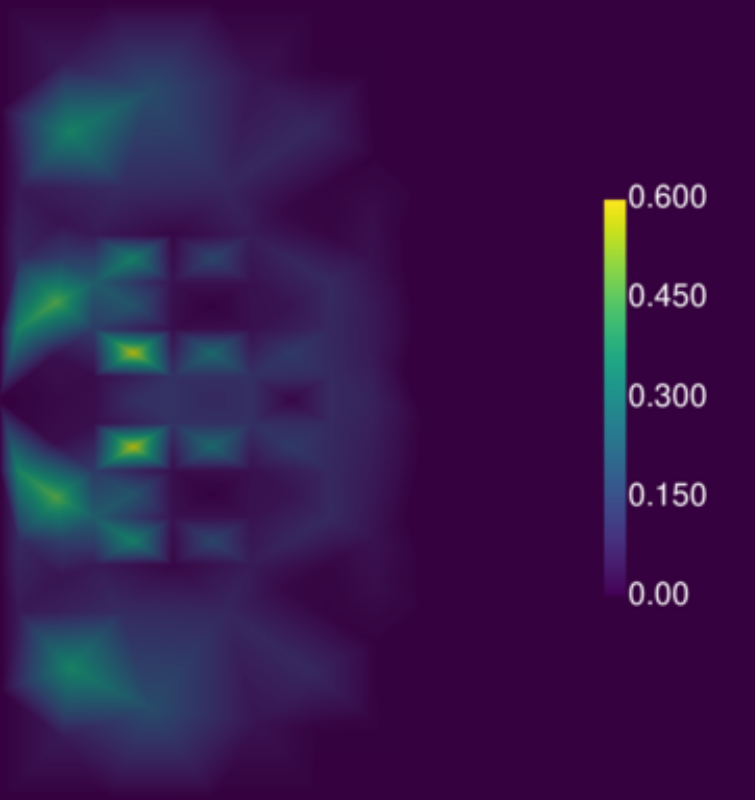} &
		\includegraphics[width=0.25 \linewidth]{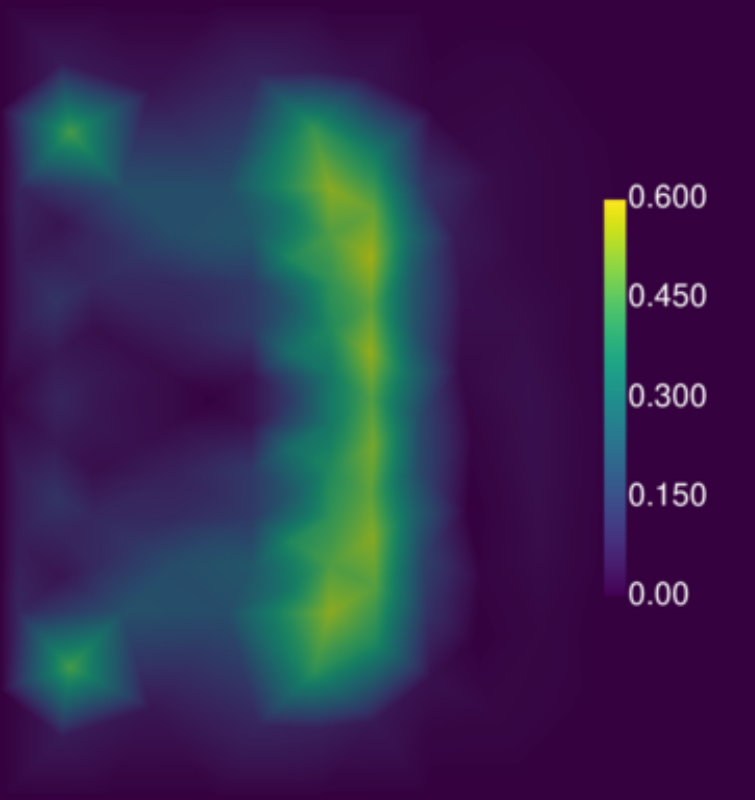}&
		\includegraphics[width=0.25 \linewidth]{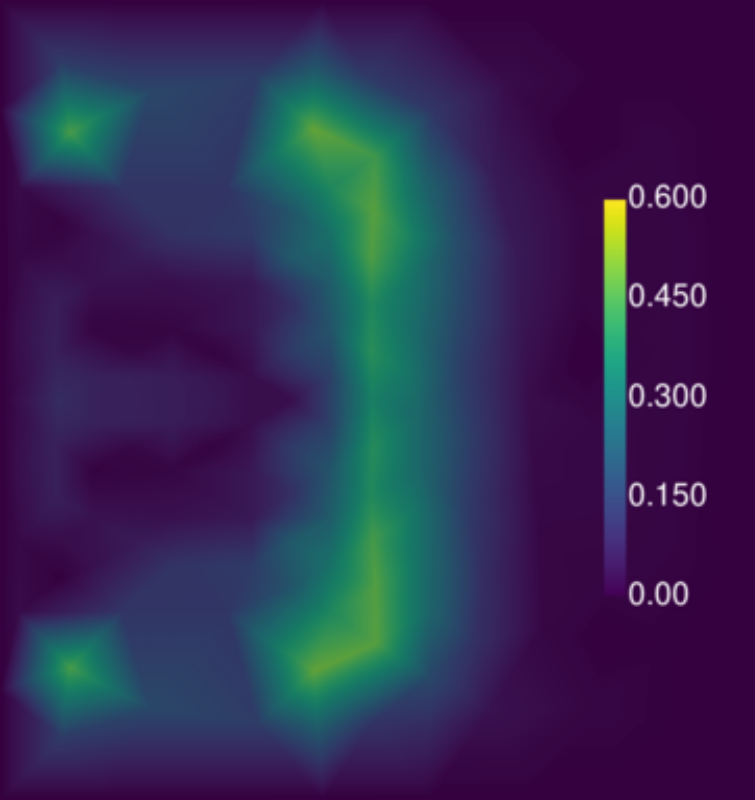}&
		\includegraphics[width=0.25 \linewidth]{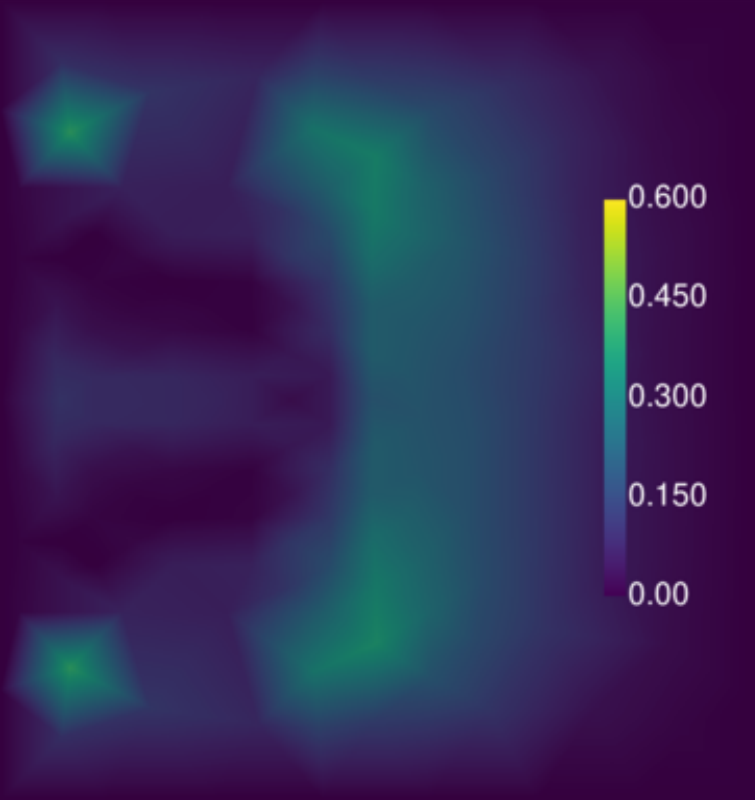}
	\end{tabular}
	\caption{Transient heat diffusion for $k=1,10,20,50$. 
		\\ \indent \textbf{Row 1}: High-fidelity model, parametrized by $\kappa_{\text{hifi}}$. 
		\\ \indent \textbf{Row 2}: Low-fidelity model, parametrized by $\kappa_{\text{lofi}}$ $\left(\Delta_{L^2}\,u^{\text{lofi}}=2.73\,\%\right)$.
		\\ \indent \textbf{Row 3}: Bias-corrected model $\left(\Delta_{L^2}\,u^{\text{corr}}=7.59\cdot 10^{-3}\,\%\right)$.  
		\\ \indent \textbf{Row 4}: Absolute values of the discrepancy function coefficients.}
	\label{fig:trans_flow_pointwise}
\end{figure}

\textcolor{black}{
The heat distribution obtained with the high- and low-fidelity heat sink models is depicted in Figure \ref{fig:trans_flow_pointwise} (rows 1 and 2, respectively).
For both models, we can observe that the heat sink rapidly transports heat from the excited base to its surroundings, eventually reaching a steady-state.
Relative to the high-fidelity model, the error of the low-fidelity model is $\Delta_{L^2}\,u^{\text{lofi}}=2.73\,\%$.
Figure \ref{fig:trans_flow_pointwise} (row 3) shows the heat distribution given by the bias-corrected model, the relative error of which is $\Delta_{L^2}\,u^{\text{corr}}=7.59\cdot 10^{-3}\,\%$. 
This is a significant improvement to the low-fidelity model, although not easily discernible with bear eye.
Figure \ref{fig:trans_flow_pointwise} (row 4) shows the absolute value of the discrepancy function between the low- and high-fidelity model.
As would be expected, the maximum discrepancy is observed at the tips of the heat sink fins, exactly where the material defect occurs.
}

\clearpage

\subsection{Transient eddy-current problem in a quadrupole magnet}
\begin{figure}[b!]
\centering
\begin{tikzpicture}	
\node at (0,0){};	
\node at (-3.5,0){
\includegraphics[scale = 2.5]{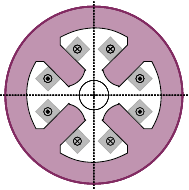}
};
\node at (-0.85,3.7){\textcolor{TUDa-10c}{$\mathbf{a}_z|_{\partial \Omega} = 0$}};
\node at (-0.75,1.9)[fill = TUDa-10c, opacity = 0.01]{$\Omega_{\text{Fe}}$};
\node at (-3,0)[fill = white, opacity = 1]{$\Omega_{\text{p}}$};
\node at (-3.5,2.45) [fill = white, opacity = 1]{$\Omega_{\text{s}}$};

\node at (3.5,0){
\begin{tikzpicture}
\begin{axis}[scatter/classes={%
    a={mark=x,draw=green}, b={mark=x,draw=blue}},
		grid=both,
		width = 5cm,
		height = 7.5cm,
		xlabel={$t$},
		ylabel={$I(t)$},
		legend columns = 2,
		legend style={at ={(1,1.125)}, fill=none, column sep = 1ex}
		]
		\addplot [TUDa-10c, thick] table [col sep=comma, x=epochs, y = current] {./data/quadrupole/current_hf.csv};
		\addlegendentry{\footnotesize $I_{\text{hifi}}(t)$};
		\addplot [TUDa-0d, thick] table [col sep=comma,x=epochs, y=current] {./data/quadrupole/current_lf.csv};
		\addlegendentry{\footnotesize $I_{\text{lofi}}(t)$};
		\end{axis}
		\end{tikzpicture}
		};
\end{tikzpicture}
\caption{\textbf{Left:} Schematic of the quadrupole magnet, where $\Omega_{\text{Fe}}$ denotes the domain of the iron yoke, $\Omega_{\text{s}}$ the domain of current exitation and $\Omega_{\text{p}}$ the air domain. \textbf{Right:} Current excitation $I_{\text{hifi}}$ for the high-fidelity model and $I_{\text{lofi}}$ for the low-fidelity model.}
\label{fig:quadrupole_geo}
\end{figure}
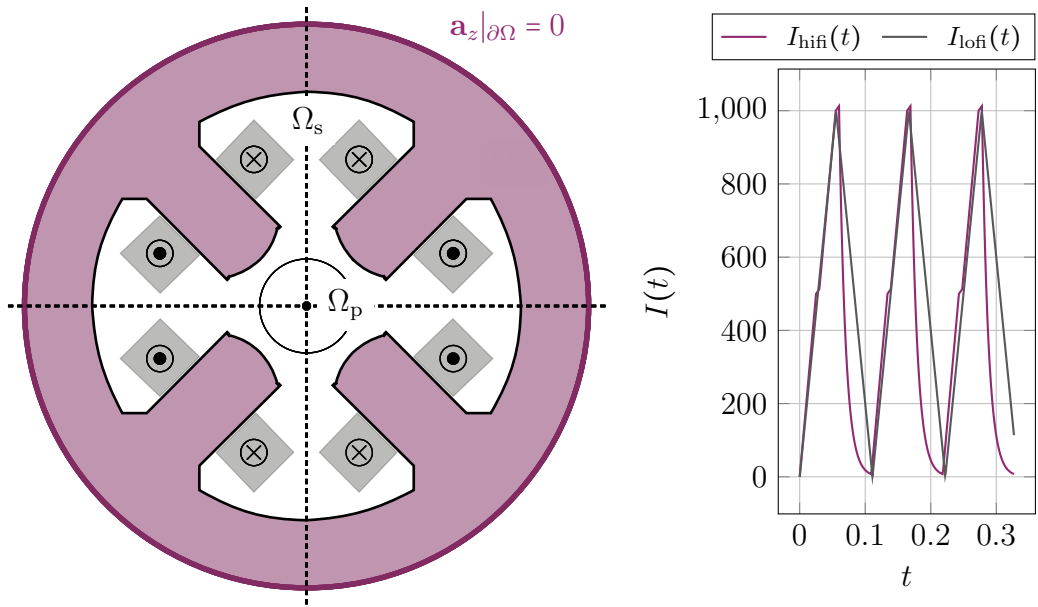

As second test case, we examine the transient eddy-current problem on a $2$D cross-section of a quadrupole magnet. The geometry is depicted in Figure~\ref{fig:quadrupole_geo} (left) \cite{diehl2022quadrupole}.
The quadrupole is defined on a circular domain $\Omega$, which consists of an iron yoke $\Omega_{\text{Fe}}$, coils for current excitation $\Omega_{\text{s}}$, and an aperture $\Omega_{\text{p}}$.
The domain boundary $\partial \Omega$ consists of the outer boundary of the iron domain $\partial \Omega_{\text{Fe}}$.
The \gls{bvp} describing the dynamical system is given by the magneto-quasistatic Maxwell equations.
Choosing the vector potential ansatz $\mathbf{b}=\nabla \times \mathbf{a}$, where $\mathbf{a}$ is the magnetic vector potential and $\mathbf{b}$ the magnetic flux density, the eddy-current problem can be given as the time-dependent \gls{bvp}
\begin{align}
\label{eq:eddy_curr}
    \nabla \times  \left(\nu \nabla \times \mathbf{a} \right) + \sigma \frac{\partial \mathbf{a}}{\partial t} &= \mathbf{j}_s, \\
    \mathbf{a}|_{\partial \Omega} &= 0, \nonumber
\end{align}
where $\mathbf{j}_s$ the source current density.
For the current excitation, we assume the current $I_{\text{hifi}}(t)$ depicted in Figure \ref{fig:quadrupole_geo} (right).
This choice of current excitation is motivated qualitatively by the ramping procedure that provides a linear field increase during acceleration with current plateaus for beam insertion and extraction and a current decay after switch-off.
An approximate current density $I_{\text{lofi}}(t)$ is additionally considered, also depicted in Figure \ref{fig:quadrupole_geo}, which consists of a linear ramp and de-ramp.
In both cases, the currents are distributed in the excitation domain, yielding a current density that can be calculated by $\mathbf{j}_s(\mathbf{r},t) = |\Omega_{\text{s}}|^{-1}\, I(t)\, \mathbf{e}_z$, where $\mathbf{e}_z$ is the unit vector in $z$-direction.
To confine the magnetic quadrupole field inside the magnet, we choose homogeneous boundary conditions on $\partial \Omega$.

\subsubsection{Finite element modeling}
We express \eqref{eq:eddy_curr} in its variational form and spatially discretize with vectorial first-order shape and test functions over a triangulation of $\Omega$. 
Due to geometrical considerations, we can safely neglect transverse effects of the vector potential, which results in a single vectorial component $\mathbf{w}_i = w_i \, \mathbf{e}_z$, with $w_i\in H_0(\text{grad} ; \Omega)$.
The magnetic vector potential is approximated via the ansatz function $\mathbf{a} = \sum_{j=1}^{N_{\text{dof}}^{\text{\text{lofi}}}} \hat{a}_j\,\mathbf{w}_j$, where the \gls{dof} $\{\hat{a}_j\}_{j\leq N_{\text{dof}}^{\text{lofi}}}$ lie on the mesh nodes.
In matrix-vector notation, the \gls{fe} formulation reads
\begin{align}
\label{eq:update_scheme}
   (\Delta t \,\mathbf{A} + \mathbf{M})\, \mathbf{\hat{a}}_{t_{k+1}} = \Delta t\, \mathbf{b} \left(t_{k+1} \right) + \mathbf{M} \, \mathbf{\hat{a}}_{t_k},
\end{align}
 where $\mathbf{A}$ and $\mathbf{M}$ are the stiffness and mass matrix, respectively, and $\mathbf{b}$ is the loading vector.
 The entries of $\mathbf{A}$ and $\mathbf{M}$ are given as
\begin{align}
\label{eq:eddy_curr_disc}
    \mathbf{A}_{ij} =
    \int_{\Omega}\left(\nu \nabla \times \mathbf{w}_i \right) \cdot (\nabla \times\mathbf{w}_j ) \,\mathrm{d}\Omega \quad \text{ and }\quad 
    \mathbf{M}_{ij} =
    \int_{\Omega} \sigma \, \mathbf{w}_i \cdot \mathbf{w}_j\,\mathrm{d}\Omega,
\end{align}
while the entries of the right hand side loading vector are
\begin{align}
\label{eq:eddy_curr_disc}
    \mathbf{b}_{i}\left( t_k\right) =
    \int_{\Omega} \mathbf{j}_{s}\left(t_k \right) \cdot \mathbf{w}_i \, \mathrm{d}\Omega.
\end{align}

For the simulation, we assume $\sigma_{\text{Fe}} = 1.04 \cdot 10^{7} \,\si{Sm^{-1}}$ and $\nu_{\text{Fe}} = 2 \cdot 10^{-3} \, \nu_0 $ in $\Omega_{\text{Fe}}$, and $\sigma = 1\,\si{Sm^{-1}}$ and $\nu = \nu_0$ in $\Omega_{\text{p}}$ and $\Omega_{\text{s}}$.
Furthermore, we assume a constant time step of $\Delta t = 1\cdot 10^{-2}\, \si{s}$ and $N_T=327$ time steps.
The low-fidelity model $\mathbf{\Psi}_{\text{lofi}}\left(\mathbf{\hat{a}}^{\text{lofi}}_{t_{k+1}}, \mathbf{\hat{a}}^{\text{lofi}}_{t_{k}},\Delta t \,|\,I_{\text{lofi}}(t_{k+1})\right)$ is parametrized with the current $I_{\text{lofi}}$ and mesh resolution $N_{\text{dof}}^{\text{lofi}} = 895$,
while the high-fidelity model $\mathbf{\Psi}_{\text{hifi}}\left(\mathbf{\hat{a}}^{\text{hifi}}_{t_{k+1}}, \mathbf{\hat{a}}^{\text{hifi}}_{t_k}, \Delta t\,|\,I_{\text{hifi}}(t_{k+1})\right)$ with current $I_{\text{hifi}}$ and $N_{\text{dof}}^{\text{hifi}}=277\,594$.
Both multi-fidelity models are non-dimensionalized as described in Section \ref{sec:non_dimensionalisation}, such that the magnetic vector potential is normalized to the unit square.

\subsubsection{Discrepancy function approximation}
\begin{figure}[t!]
	\begin{tikzpicture}[scale = 1]
		\begin{axis}[scatter/classes={%
				a={mark=x,draw=TUDa-10c, very thick}},
			grid = both,
			width = \textwidth,
			height = 8cm,
			xlabel={$t$},
			ylabel={$\|\delta_{t}\|_{2}$},
			xmax = 0.35,
			xtick = {0,0.05,...,0.4},
			legend columns = 4,
			legend style={at={(1,1.125)}, fill=none, column sep = 1ex}
			]
			\addplot [TUDa-0c, thick] table [col sep=comma, x=epochs, y=error, x expr = 0.001*\thisrow{epochs}] {./data/quadrupole/d_func_ref.csv};
			\addlegendentry{\footnotesize reference};
			\addplot [scatter, only marks, draw = TUDa-10c, mark = x, mark size = 4pt, very thick] table [col sep=comma, x=epochs, y=error, x expr = 0.001*\thisrow{epochs}] {./data/quadrupole/d_func_training_data.csv};
			\addlegendentry{\footnotesize training data};
			\addplot [TUDa-0d, densely dotted, thick] table [col sep=comma,x=epochs, y=error, x expr = 0.001*\thisrow{epochs}] {./data/quadrupole/d_func_approx_regular.csv};
			\addlegendentry{\footnotesize hybrid model};
			\addplot [TUDa-10c, very thick] table [col sep=comma,x=epochs, y=error, x expr = 0.001*\thisrow{epochs}] {./data/quadrupole/d_func_approx_upsampling.csv};
			\addlegendentry{\footnotesize hybrid model + up.};
		\end{axis}
	\end{tikzpicture}
	\caption{Spatially integrated discrepancy function $\|\delta_{t}\|_{2}$ for $t_0 = 0\,\si{s}$ until $t_{327}=0.327\,\si{s}$. The reference and the hybrid model with and without artificial upsampling are depicted. The training data is indicated at the respective time steps with \textcolor{TUDa-10c}{$\mathbf{\times}$.}}
	\label{fig:error_quadrupole}
\end{figure}
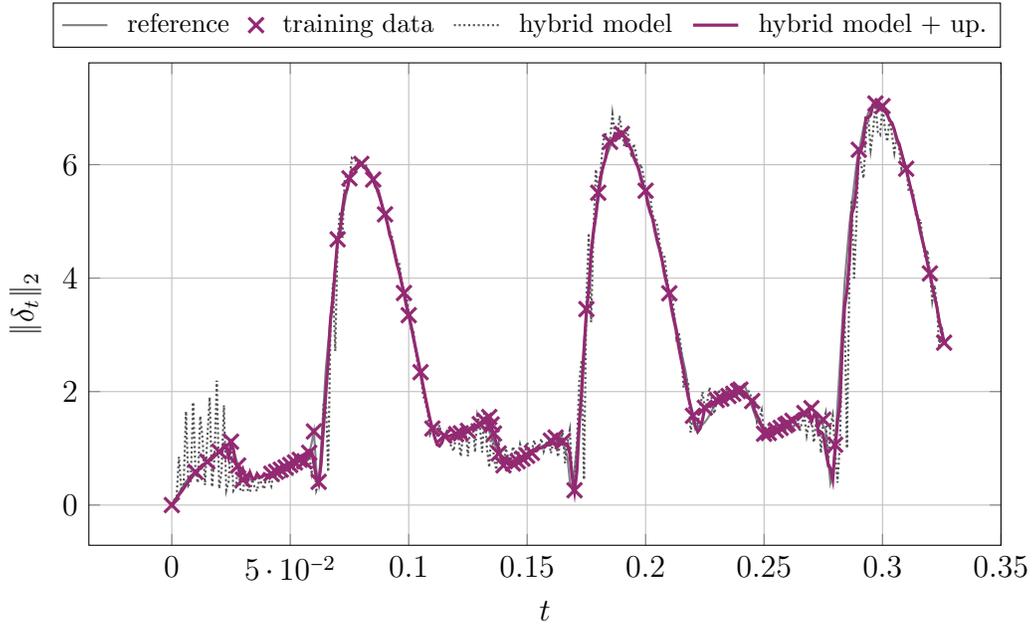

As training data, we use 79 out of 327 trajectory samples at the respective time steps, depicted as crosses in Figure \ref{fig:error_quadrupole}. 
In this test case, we increase the number of samples in the areas where the current excitation changes from an increasing to a decreasing flank.
We train the model according to the parameters given in Table \ref{table:quadrupole_par}, from which we observe that $1000$ training epochs are required to reduce $\Delta_{L^2} \delta_{\theta}$ to $1.538\%$ for the upsampled model and to $8.187\%$ for the non-upsampled model.
\textcolor{black}{The \gls{rnn} has $N_p=2$ \gls{lstm} cells and $N_{\text{dof}}^{\text{lofi}} = 895$ neurons per layer, amounting to $13\,625\,480$ trainable parameters.}

\begin{table}[b!]
	\centering
	\begin{tabular}{l l c c}
		\hline\hline
		Description & Symbol & 0-500 & 500-1000\\ [0.5ex] 
		\hline
		Learning rate    & $\eta$    &  $1\cdot 10^{-3}$ & $1\cdot 10^{-4}$ \\
		Local weighting factors & $\beta$ & $\times$ & $\checkmark$ \\
		Variance weighting factor            & $\alpha$  &  $\frac{1}{25}$  & $\frac{1}{25}$  \\
		Error with upsampling & $\Delta_{L^2} \delta_{\theta^{\text{up}}}$  &  $2.806\,\%$  & $1.538\,\%$       \\
		Error without upsampling & $\Delta_{L^2} \delta_{\theta}$   &  $13.094\,\%$  & $8.187\,\%$    \\[1ex]
		\hline
	\end{tabular}
	\caption{\textcolor{black}{Training parameters and hybrid model errors for the quadrupole magnet test case. The \gls{rnn} has $N_p=2$ \gls{lstm} cells, equivalently, $N_p=2$ consecutive time steps are taken into account in each training epoch. Each layer of the \gls{rnn} has $N_{\text{dof}}^{\text{lofi}}=895$ neurons.}}
	\label{table:quadrupole_par}
\end{table}

In Figure \ref{fig:error_quadrupole} we display the spatially integrated discrepancy function $\|\delta_{t}\|_{2}$ at each time step, for the hybrid model with and without upsampling and the reference data.
If the hybrid model is trained using solely the sparse data set, we observe a good agreement on the data points. However, the model overfits and performs poorly for previously unseen data, albeit not as strongly as in the heat sink test case.
This phenomenon is again prevalent in domains with sparse training data.
Similar to the previous section, the hybrid model with artificial upsampling generally yields good agreement on the complete data set.
However, there are certain areas in which the approximation is suboptimal, in particular when the current excitation changes from a rising to a falling flank.

\subsubsection{Bias correction}
\begin{figure}[t!]
	\vspace{-4em}
	\centering
	\noindent
	\begin{tabular}{cccc}
		$\mathbf{a}^{\text{\tiny hifi}}_{t_{105}}$ & 
		$\mathbf{a}^{\text{\tiny hifi}}_{t_{110}}$ & 
		$\mathbf{a}^{\text{\tiny hifi}}_{t_{115}}$ &  
		$\mathbf{a}^{\text{\tiny hifi}}_{t_{120}}$
		\\
		\includegraphics[width=0.25 \linewidth]{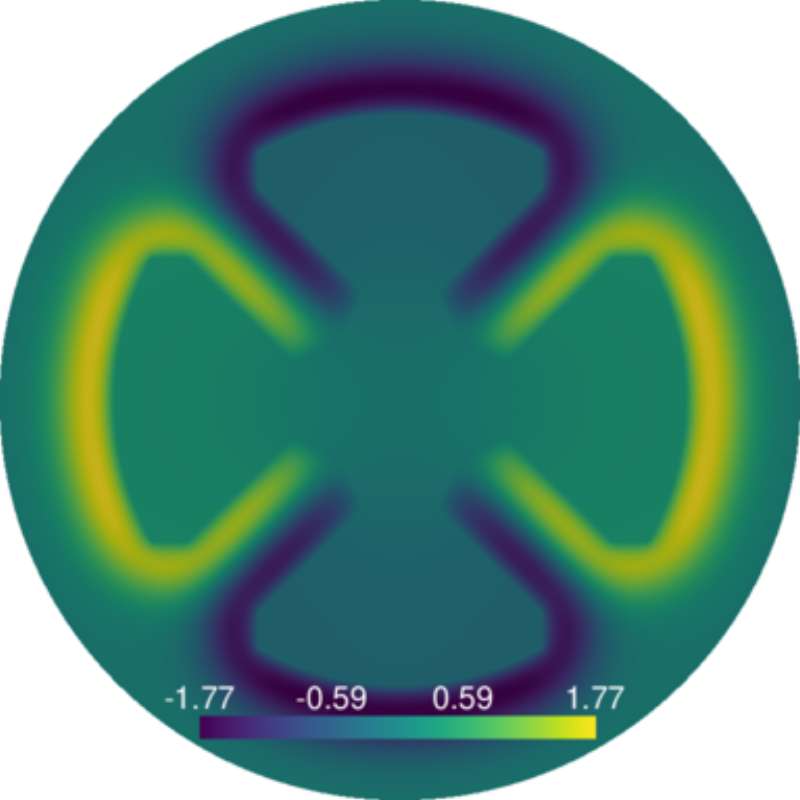}&
		\includegraphics[width=0.25 \linewidth]{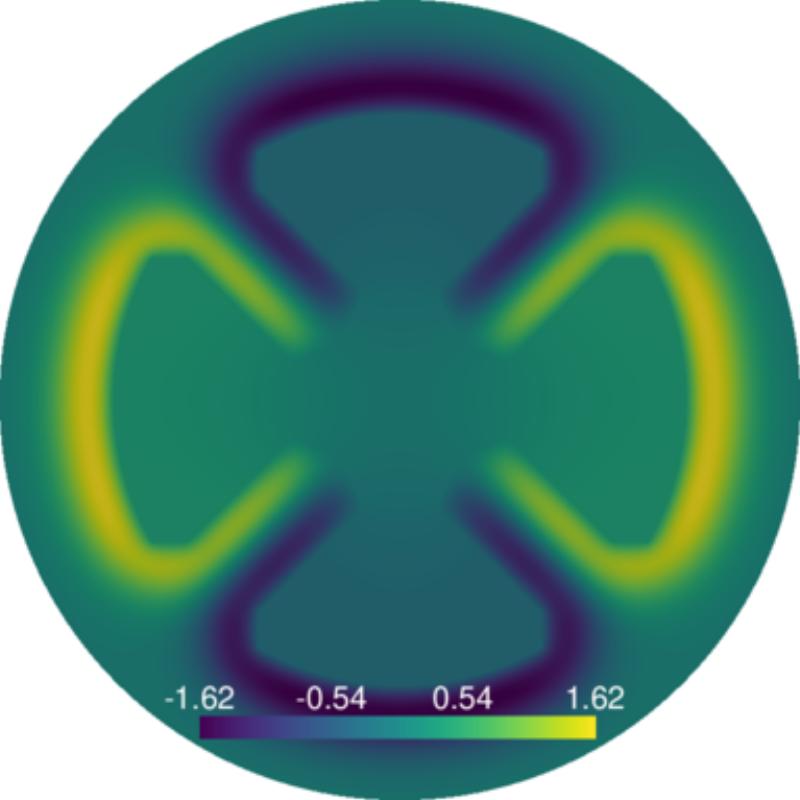}&
		\includegraphics[width=0.25 \linewidth]{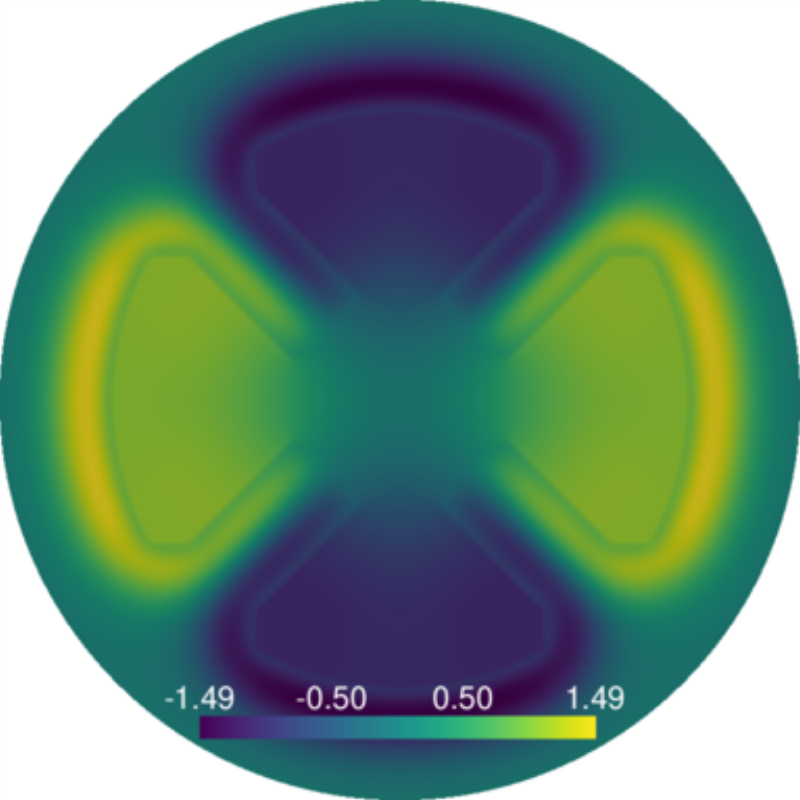}&
		\includegraphics[width=0.25 \linewidth]{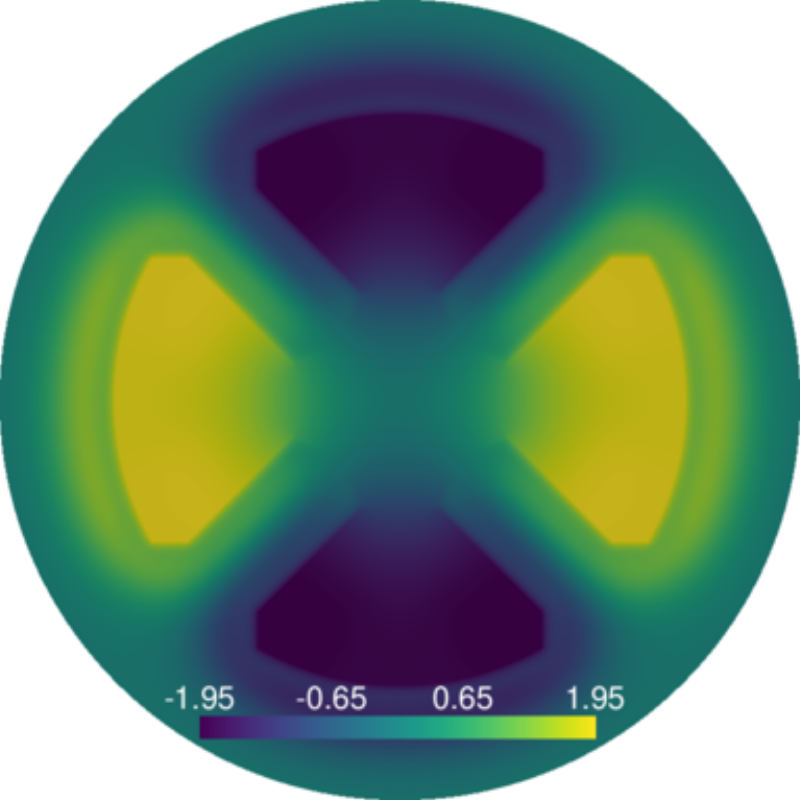}  
		\\
		$\mathbf{a}^{\text{\tiny lofi}}_{t_{100}}$ & 
		$\mathbf{a}^{\text{\tiny lofi}}_{t_{110}}$ & 
		$\mathbf{a}^{\text{\tiny lofi}}_{t_{115}}$ &  
		$\mathbf{a}^{\text{\tiny lofi}}_{t_{120}}$
		\\
		\includegraphics[width=0.25 \linewidth]{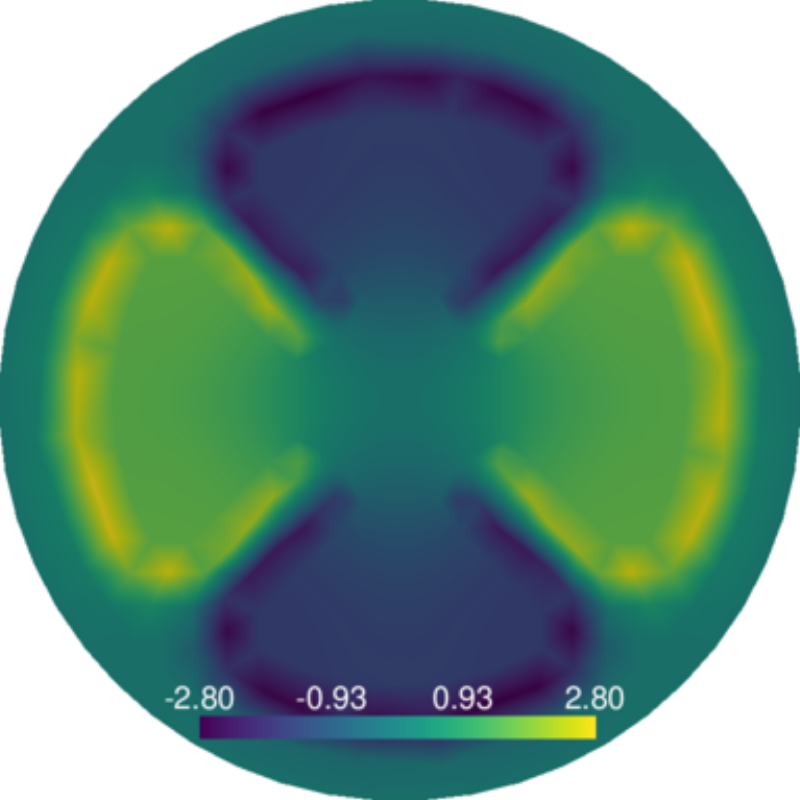}&
		\includegraphics[width=0.25 \linewidth]{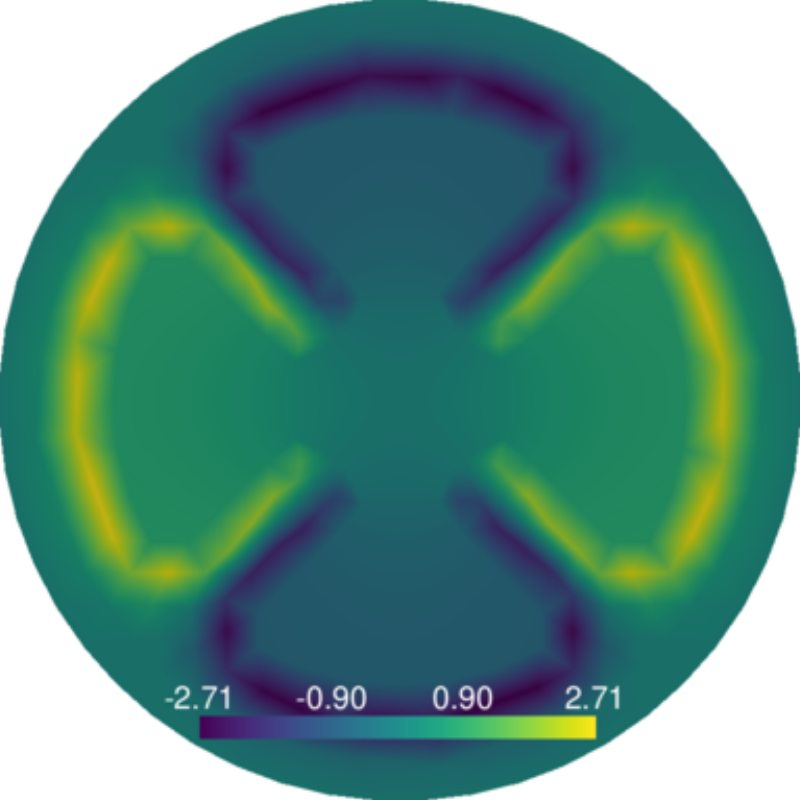}&
		\includegraphics[width=0.25 \linewidth]{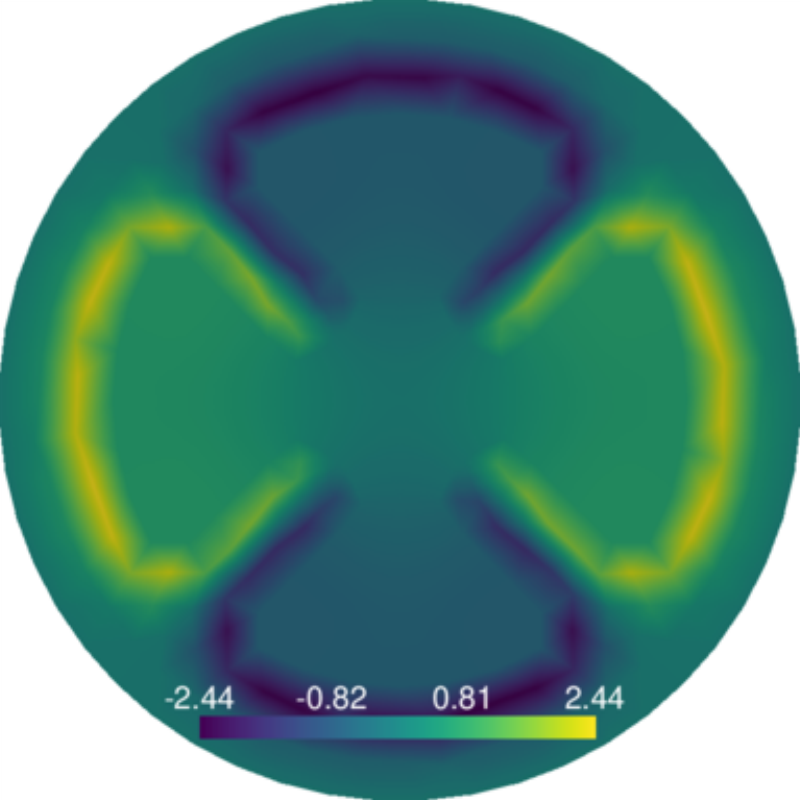}&
		\includegraphics[width=0.25 \linewidth]{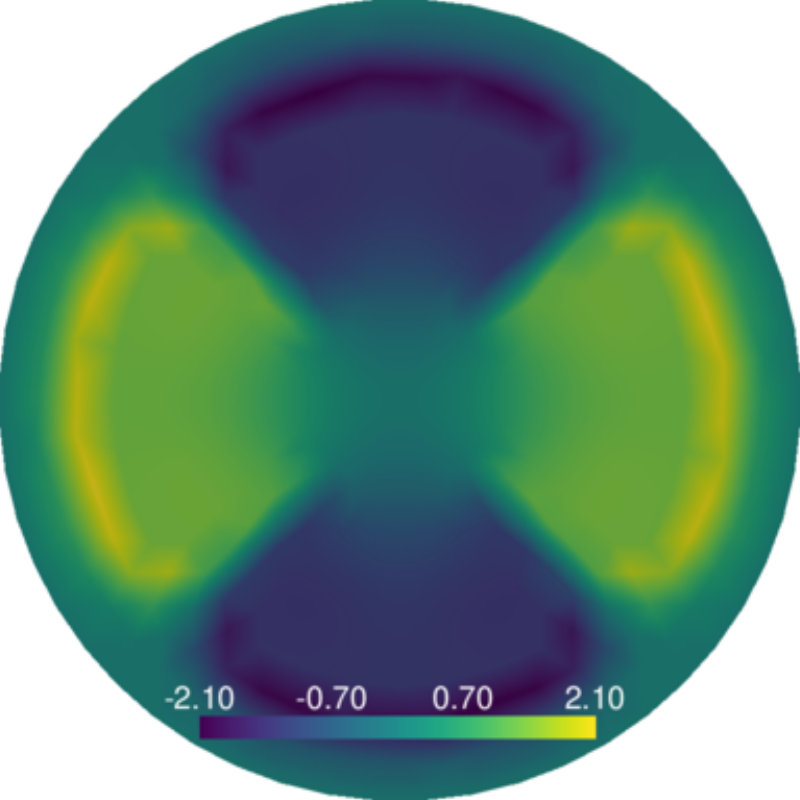} 
		\\
		$\mathbf{a}^{\text{\tiny corr}}_{t_{100}}$ & 
		$\mathbf{a}^{\text{\tiny corr}}_{t_{110}}$ & 
		$\mathbf{a}^{\text{\tiny corr}}_{t_{115}}$ &  
		$\mathbf{a}^{\text{\tiny corr}}_{t_{120}}$ 
		\\ 
		\includegraphics[width=0.25 \linewidth]{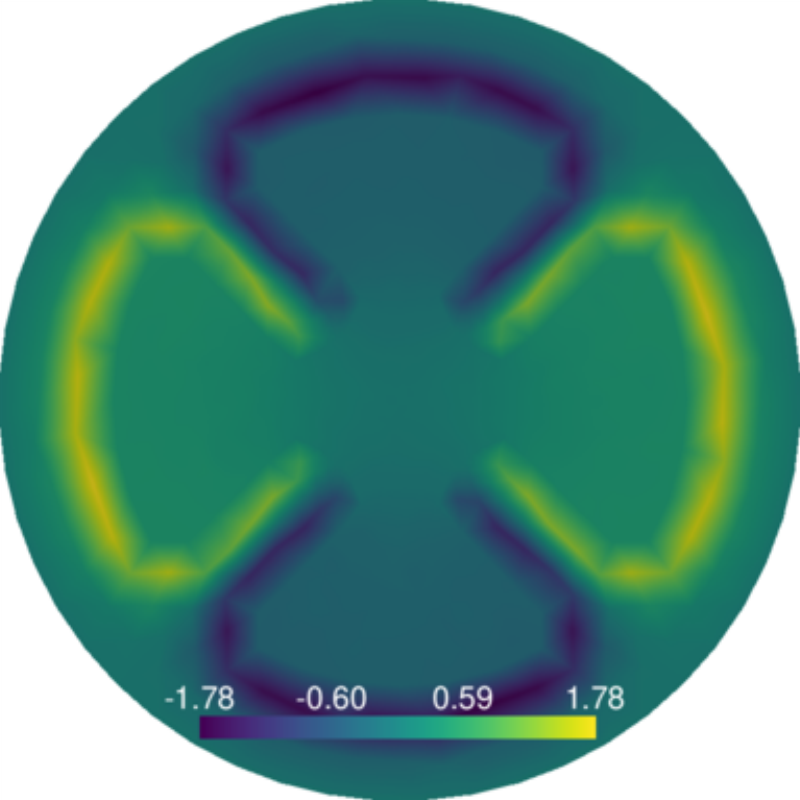}&
		\includegraphics[width=0.25 \linewidth]{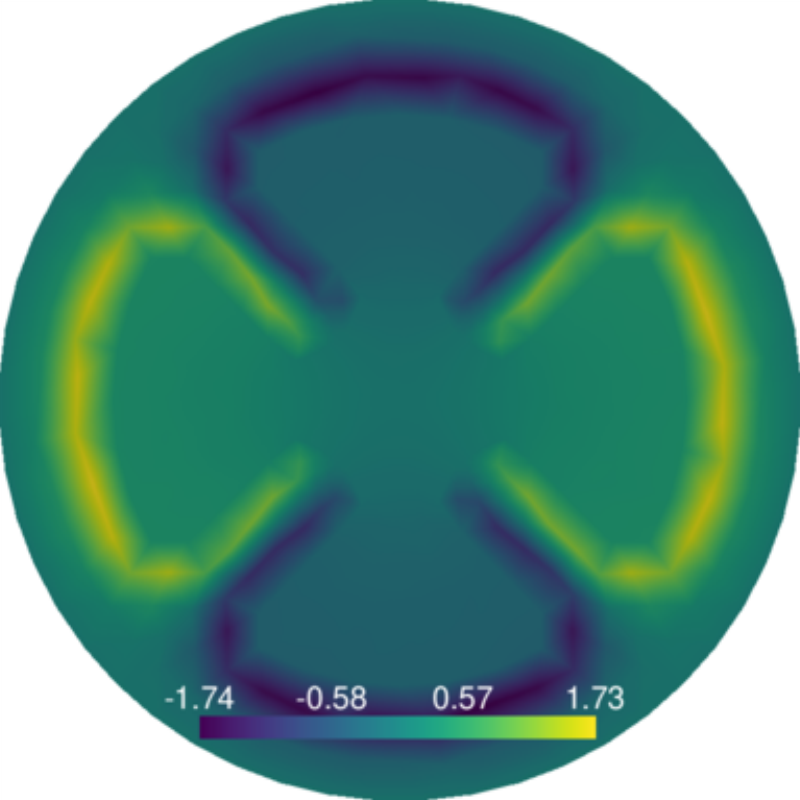}&
		\includegraphics[width=0.25 \linewidth]{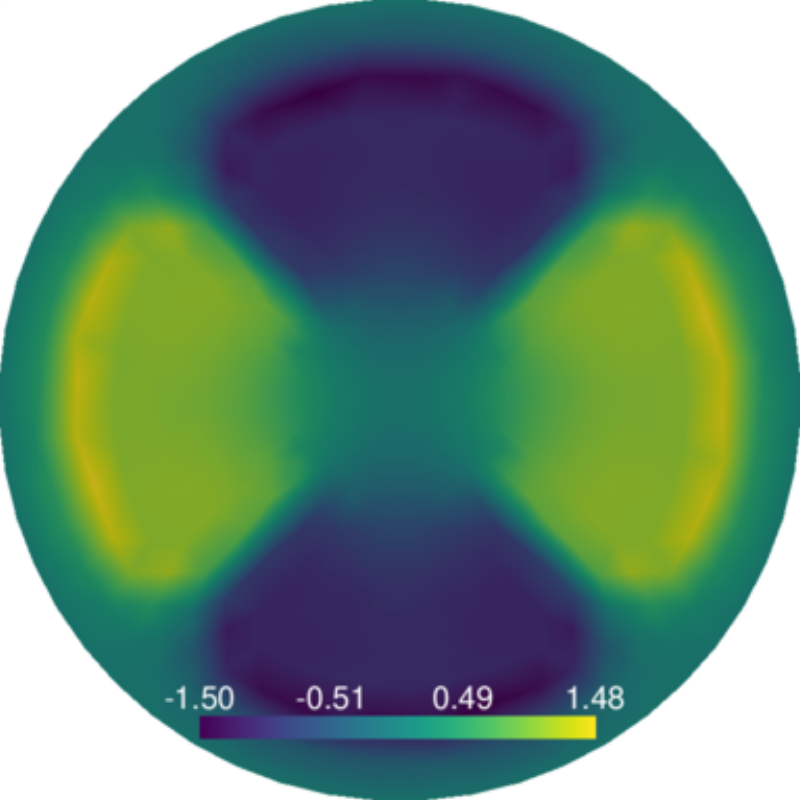}&
		\includegraphics[width=0.25 \linewidth]{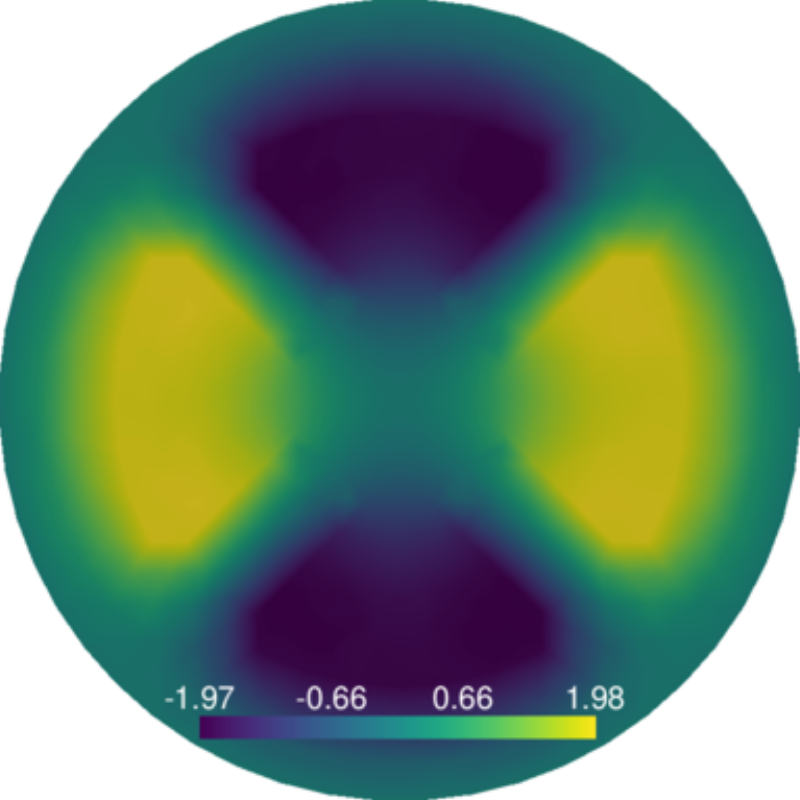} 
		\\
		\scriptsize $\sum_{i=1}^{N^\text{lofi}_{\text{dof}}} \,\left|\delta^{\text{\tiny NN}}_{i,t_{100}}\,\right|\phi_i(\mathbf{r})$ & 
		\scriptsize $\sum_{i=1}^{N^\text{lofi}_{\text{dof}}}\,\left|\delta^{\text{\tiny NN}}_{i,t_{110}}\,\right|\phi_i(\mathbf{r})$ & 
		\scriptsize $\sum_{i=1}^{N^\text{lofi}_{\text{dof}}}\,\left|\delta^{\text{\tiny NN}}_{i,t_{115}}\,\right|\phi_i(\mathbf{r})$ &  
		\scriptsize $\sum_{i=1}^{N^\text{lofi}_{\text{dof}}}\,\left|\delta^{\text{\tiny NN}}_{i,t_{120}}\,\right|\phi_i(\mathbf{r})$
		\\  
		\includegraphics[width=0.25 \linewidth]{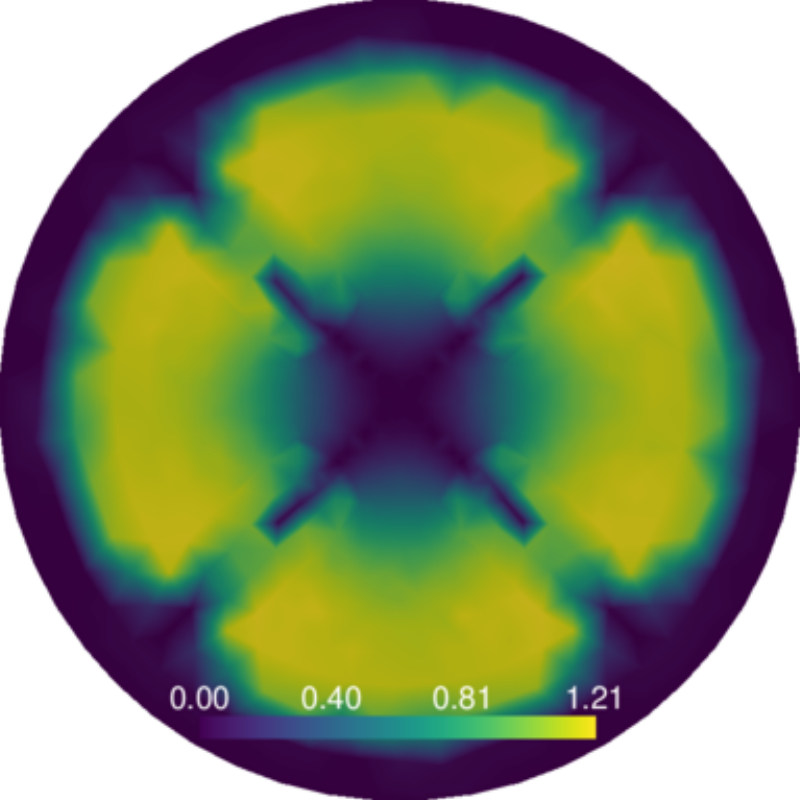}&
		\includegraphics[width=0.25 \linewidth]{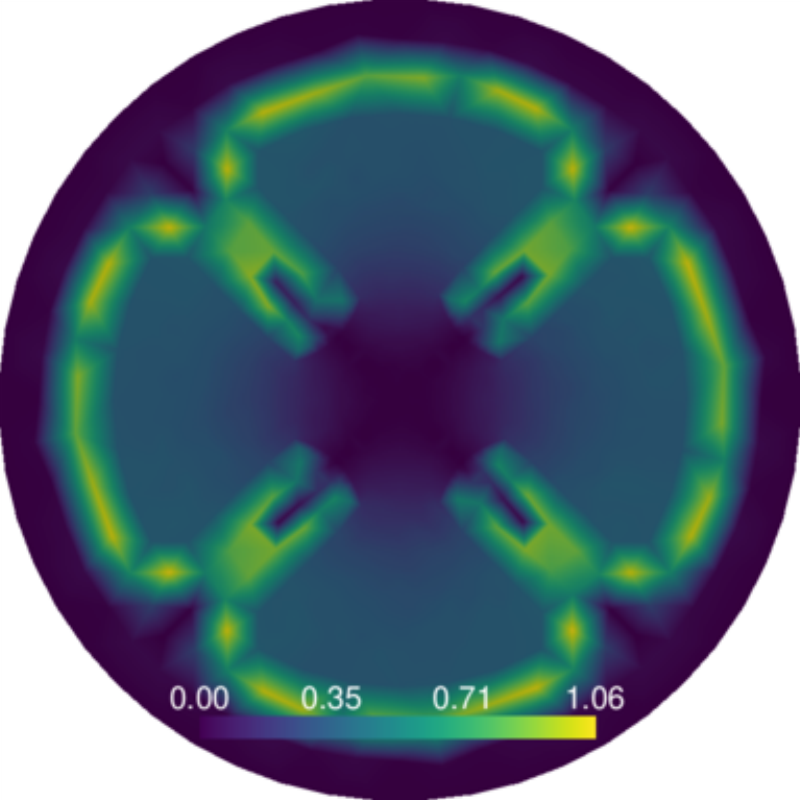}&
		\includegraphics[width=0.25 \linewidth]{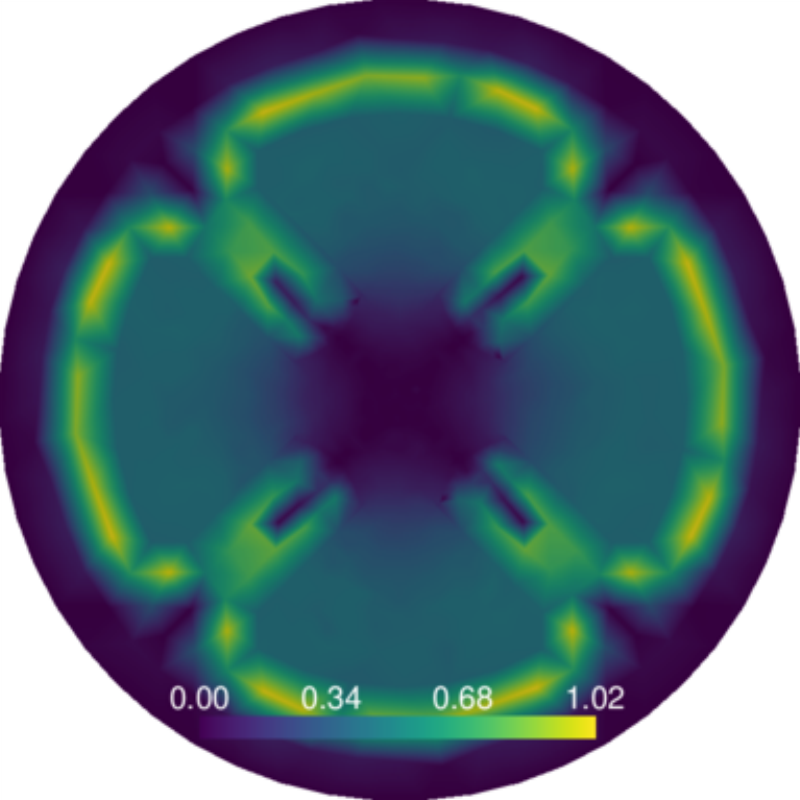}&
		\includegraphics[width=0.25 \linewidth]{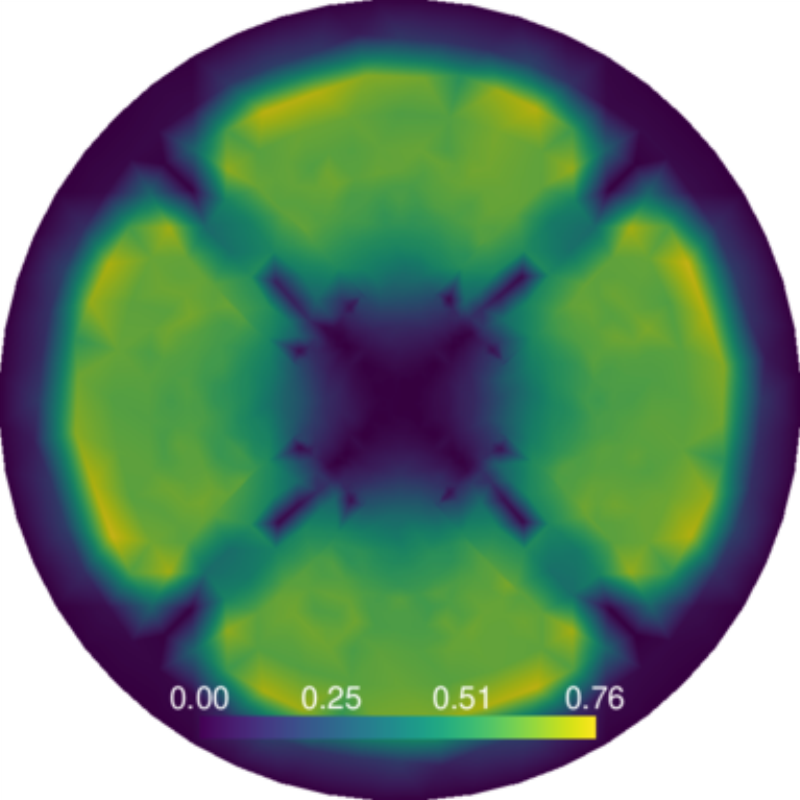}
	\end{tabular}
	\caption{Transient eddy-current simulation  for $k=105,110,115,120$. 
		\\ \indent \textbf{Row 1}: High-fidelity model, parametrized by $I_{\text{hifi}}(t)$. 
		\\ \indent \textbf{Row 2}: Low-fidelity model, parametrized by $I_{\text{lofi}}(t)$ $\left(\Delta_{L^2}\,\mathbf{a}^{\text{lofi}}=39.847\,\%\right)$.
		\\ \indent \textbf{Row 3}: Bias-corrected model $\left(\Delta_{L^2}\,\mathbf{a}^{\text{corr}}=0.613\,\%\right)$.
		\\ \indent \textbf{Row 4}: Absolute values of the discrepancy function coefficients.}
	\label{fig:transient_quadrupole}
\end{figure}

\textcolor{black}{
The potential distribution obtained with the high- and low-fidelity magnet models is depicted in Figure \ref{fig:transient_quadrupole} (rows 1 and 2, respectively).
In both cases, we can observe the eddy current phenomenon occurring in the iron yoke $\partial \Omega_{\text{Fe}}$.
Relative to the high-fidelity model, the error of the low-fidelity model is  $\Delta_{L^2}\,\mathbf{a}^{\text{lofi}}=39.847\,\%$.
The potential distribution of the bias-corrected model is shown in Figure \ref{fig:transient_quadrupole} (row 3). 
The relative error of the bias-corrected model is $\Delta_{L^2}\,\mathbf{a}^{\text{corr}}=0.613\,\%$, which constitutes a major improvement to the low-fidelity model.
In this test case, the correction is also visually obvious, particularly observing the difference in the potential's magnitude, see the corresponding color-bars.
A visualization of the discrepancy between the high- and low-fidelity models is given in Figure \ref{fig:transient_quadrupole} (row 4), where we observe that the discrepancy primarily occurs at the interface between the iron and the air domain.
}

\clearpage

\subsection{Wave propagation in a cavity}
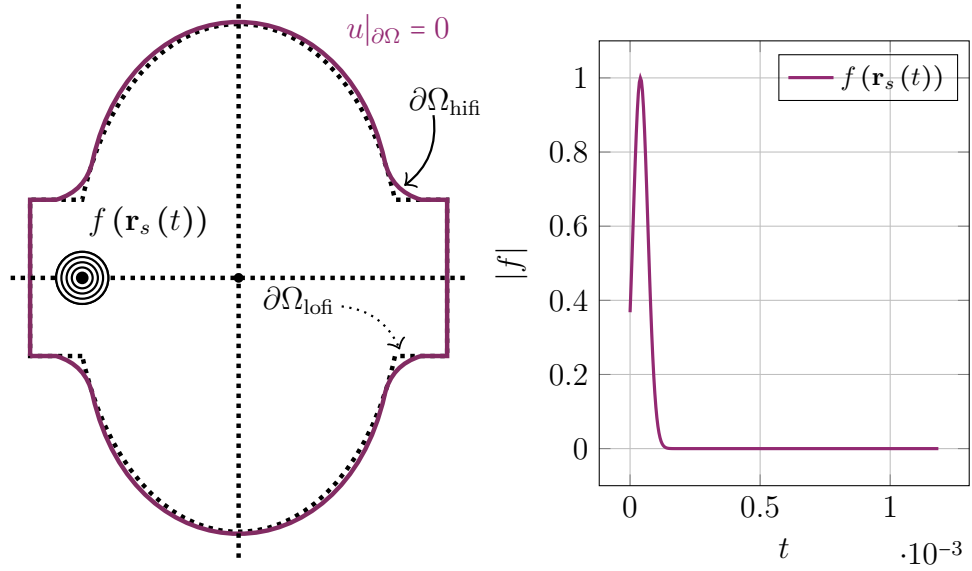
\begin{figure}[t!]
\begin{tikzpicture}
\node at (-2,0){
\includegraphics[scale = 2.1]{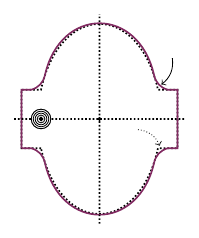}
};
\node at (-3.2,0.75){$f\left(\mathbf{r}_s\left(t\right)\right)$};
\node at (-1.2,-0.3){$\partial \Omega_{\text{lofi}}$};
\node at (0.75,2.35){$\partial \Omega_{\text{hifi}}$};
\node[TUDa-10c] at (0.12,3.3){$u|_{\partial \Omega}=0$};
\node at (4.5,-0.35){
\begin{tikzpicture}
\begin{axis}[scatter/classes={%
    	a={mark=x,draw=green}, b={mark=x,draw=blue}},
		grid=both,
		width = 6.5cm,
		height = 7.5cm,
		xlabel={$t$},
		ylabel={$|f|$},
		legend pos= north east,
		legend style={fill=none}
		]
		\addplot [TUDa-10c, very thick] table [col sep=comma, x=epochs, y = data] {./data/waveguide/exponential.csv};
		\addlegendentry{\footnotesize $f\left(\mathbf{r}_s\left(t\right)\right)$};
\end{axis}
\end{tikzpicture}
};
\end{tikzpicture}
\caption{\textbf{Left:} Schematic of the cavity, where $\partial \Omega_{\text{hifi}}$ indicates the design with round corners (solid purple line) and $\partial \Omega_{\text{lofi}}$ the design with sharp corners (dotted black line). The point of excitation is indicated on the left of the cavity. \textbf{Right:} Excitation function $f\left(\mathbf{r}_s\left(t\right)\right)$ over time.}
\label{fig:waveguide_geo}
\end{figure}

For the third test case, we consider wave propagation in a closed cavity, where one commonly aims at detecting resonance frequencies and eigenmodes.
After excitation, the wave propagates through the cavity until it eventually dissipates and the dominant frequencies are identified.
As to the geometry of the cavity, we assume a ``true'' geometry $\Omega_{\text{hifi}}$, with rounded-off corners between the connectors and the cavity itself, as well as an approximate design, $\Omega_{\text{lofi}}$, with sharp corners at the connectors.
Both geometries are depicted in Figure~\ref{fig:waveguide_geo} (left).
The equation describing the \gls{bvp} reads
\begin{align}
\label{eq:wave}
    \frac{\partial^2 u}{\partial t^2} - v^2\, \Delta u = f \\
    u|_{\partial \Omega} = 0 \nonumber,
\end{align}
where $u$ is the acoustic pressure field, $v$ is the velocity of the propagating wave, $f \in \mathcal{C}^{\infty}(\Omega)$ the excitation function, and $\Omega \in \{ \Omega_{\text{lofi}}, \Omega_{\text{hifi}}\}$ the low and high-fidelity domains.
For boundary conditions, we apply homogeneous Dirichlet boundary conditions on $\partial \Omega$, such that the wave is reflected off the cavity walls and contained therein.
The wave excitation $f$ is a Gaussian impulse, as shown in Figure \ref{fig:waveguide_geo} (right), applied as a point source at $\mathbf{r}_s\in \Omega$ as indicated in Figure \ref{fig:waveguide_geo} (left).

\subsubsection{Finite element modeling}
We express \eqref{eq:wave} in its variational form and spatially discretize with first-order shape and test functions $w_i \in H_0(\text{grad};\,\Omega)$. 
The wave equation solution is approximated via $u = \sum_{j=1}^{N_{\text{dof}}^{\text{lofi}}} \hat{u}_j\,w_j$, where the \gls{dof} $\{\hat{u}_j\}_{j\leq N_{\text{dof}}^{\text{lofi}}}$ lie on the mesh nodes.
As \eqref{eq:wave} is an equation of second order, discretization in time requires a central differences scheme $\frac{\partial^2 u}{\partial t^2} = \frac{u_{t_{k+1}}-2u_{t_k}+u_{t_{k-1}}}{\Delta t^2}$.
The resulting discretized \gls{fe} system reads
\begin{align}
	 \left(\mathbf{M}+v^2\,\Delta t^{2} \,\mathbf{A}\right)\,\mathbf{\hat{u}}_{t_{k+1}} = \Delta t^{2}\,\mathbf{b}(t_{k+1}) + 2\, \mathbf{M}\,\mathbf{\hat{u}}_{t_{k}}-\mathbf{M}\,\mathbf{\hat{u}}_{t_{k-1}},
\end{align}
where $\mathbf{\hat{u}}$ is a vector containing the \gls{dof} of the \gls{fe} basis and the mass and stiffness matrices are given as
\begin{align}
\label{eq:eddy_curr_disc}
    \mathbf{M}_{ij} =
    \int_{\Omega} w_i \, w_j \,\mathrm{d}\Omega \quad \text{ and }\quad 
    \mathbf{A}_{ij} =
    \int_{\Omega} \nabla w_i \cdot \nabla w_j\,\mathrm{d}\Omega,
\end{align}
and the right hand side loading vector $\mathbf{b}$ is given at each time instance $t_k$ as
\begin{align}
\label{eq:eddy_curr_disc}
    \mathbf{b}_{i} \left(t_k \right) =
    \int_{\Omega} f\left(t_k \right) \, w_i \, \mathrm{d}\Omega.
\end{align}

For the simulation, we assume $v=343\,\,\si{ms^{-1}}$, $\Delta t = 4\cdot 10^{-5} \,\si{s}$, and $N_T = 100$ time steps.
The low-fidelity model $\mathbf{\Psi}_{\text{lofi}}\left(\mathbf{\hat{u}}_{t_{k+1}}^{\text{lofi}}, \mathbf{\hat{u}}_{t_{k}}^{\text{lofi}}, \mathbf{\hat{u}}_{t_{k-1}}^{\text{lofi}}, \Delta t\,|\,\Omega_{\text{lofi}}\right)$ uses the approximate cavity geometry $\Omega_{\text{lofi}}$ with $N_{\text{dof}}^{\text{lofi}} = 169$. 
The high-fidelity model $\mathbf{\Psi}_{\text{hifi}}\left(\mathbf{\hat{u}}_{t_{k+1}}^{\text{hifi}}, \mathbf{\hat{u}}_{t_{k}}^{\text{hifi}}, \mathbf{\hat{u}}_{t_{k-1}}^{\text{hifi}}, \Delta t\,|\,\Omega_{\text{hifi}}\right)$ uses the ``true'' cavity geometry $\Omega_{\text{hifi}}$ with $N_{\text{dof}}^{\text{hifi}} = 3997$.
The modeling and discretization errors are here induced by the geometry variation and the difference in mesh discretization.

\subsubsection{Discrepancy function approximation}
\begin{figure}[t!]
    \begin{tikzpicture}[scale = 1]
		\begin{axis}[scatter/classes={%
    	a={mark=x,draw=TUDa-10c, very thick}},
    	ymax = 30,
		grid = both,
		width = \textwidth,
		height = 7cm,
		xlabel={$t$},
		ylabel={$\|\delta_{t}\|_{2}$},
		legend columns = 4,
		legend style={at={(1,1.2)}, fill=none, column sep = 1ex}
		]
		\addplot [TUDa-0d, thick] table [col sep=comma, x=epochs, y=error, x expr = 0.00004*\thisrow{epochs}] {./data/waveguide/l2_errors_ref_waveguide.csv};
		\addlegendentry{\footnotesize reference};
		\addplot [scatter, only marks, draw = TUDa-10c, mark = x, mark size = 4pt, very thick] table [col sep=comma, x=epochs, y=error, x expr = 0.00004*\thisrow{epochs}] {./data/waveguide/training_data.csv};
		\addlegendentry{\footnotesize training data};
		\addplot [TUDa-0c, densely dotted, thick] table [col sep=comma,x=epochs, y=error, x expr = 0.00004*\thisrow{epochs}] {./data/waveguide/error_l2_out_waveguide_overfit.csv};
		\addlegendentry{\footnotesize hybrid model};
		\addplot [TUDa-10c, very thick] table [col sep=comma, x=epochs, y=error, x expr = 0.00004*\thisrow{epochs}] {./data/waveguide/error_l2_out_waveguide_alternative.csv};
		\addlegendentry{\footnotesize hybrid model + up.};
		\end{axis}
		\end{tikzpicture}
	\caption{Spatially integrated discrepancy function $\|\delta_{t}\|_{2}$ for the time $t_0 = 0\, \si{s}$ until $t_{100} = 4\cdot 10^{-3}\, \si{s}$. The reference and the hybrid model with and without artificial upsampling scheme are depicted. The training data is indicated at the respective time steps with \textcolor{TUDa-10c}{$\mathbf{\times}$.}}
	\label{fig:error_waveguide}
	\end{figure}
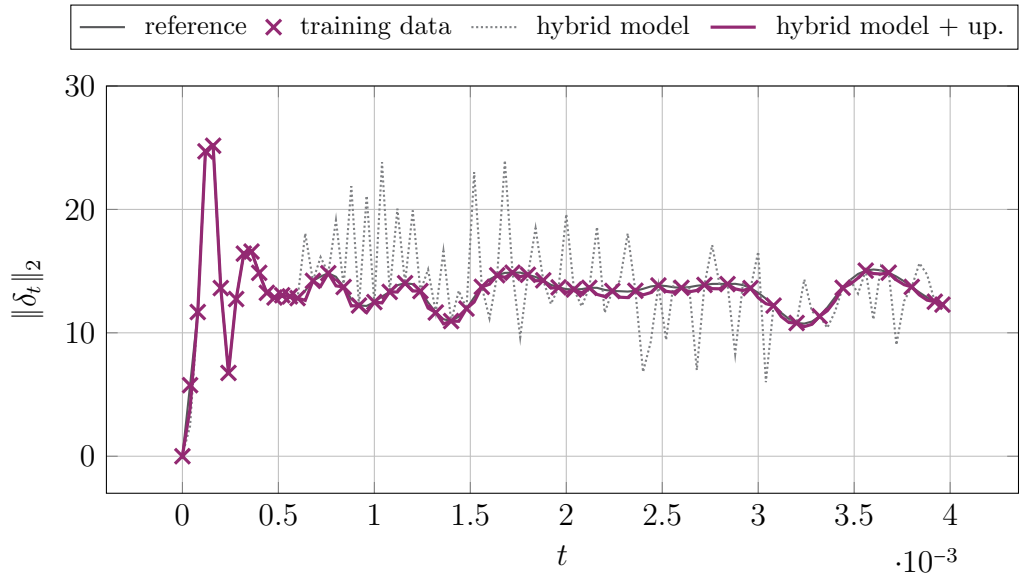

For the discrepancy function approximation, we use 51 out of 100 trajectory instances as training data, depicted as crosses at the respective time steps in Figure \ref{fig:error_waveguide}. The hybrid model is trained according to the parameters in Table \ref{table:waveguide_par}, from which we observe that $2500$ training epochs are required to reduce $\Delta_{L^2} \delta_{\theta}$ to $0.200\%$ for the upsampled model and to $14.797\%$ for the non-upsampled model.
\textcolor{black}{The \gls{rnn} has $N_p=3$ \gls{lstm} cells and $N_{\text{dof}}^{\text{lofi}}=169$ neurons per layer, amounting to $716\,222$ trainable parameters.}

\begin{table}[b!]
	\centering
	\begin{tabular}{l l c c c c}
		\hline\hline
		Description & Symbol & 0-1000 & 1000-2000 & 2000-2500\\ [0.5ex] 
		\hline
		Learning rate    & $\eta$    &  $1\cdot 10^{-3}$ & $5\cdot 10^{-4}$ & $1\cdot 10^{-4}$ \\
		Local weighting factors & $\beta$ & $\times$ & $\times$ & $\checkmark$ \\
		Variance weighting factor            & $\alpha$  &  $\frac{1}{100}$  & $\frac{1}{100}$ & $\frac{1}{100}$   \\
		Error with upsampling & $\Delta_{L^2} \delta_{\theta^{\text{up}}}$  &  $2.587\%$  & $0.436\%$ & $0.200\%$      \\
		Error without upsampling & $\Delta_{L^2} \delta_{\theta}$  &  $20.035\%$  & $17.195\%$ & $14.797\%$ \\[1ex]
		\hline
	\end{tabular}
	\caption{\textcolor{black}{Training parameters and hybrid model errors for the waveguide test case. The \gls{rnn} has $N_p=3$ \gls{lstm} cells, equivalently, $N_p=3$ consecutive time steps are taken into account in each training epoch. Each layer of the \gls{rnn} has $N_{\text{dof}}^{\text{lofi}}=169$ neurons.}}
	\label{table:waveguide_par}
\end{table}

In Figure \ref{fig:error_waveguide}, the spatially integrated discrepancy function $\|\delta_{t}\|_{2}$ is displayed at each time step, for the hybrid model with and without upsampling and the reference data.
Similar to the previous sections, the hybrid model with artificial upsampling generally yields good agreement on the complete data set and the hybrid model without upsampling tends to overfit in regions with sparse data.
Nevertheless, in this case, we require a larger training data set to achieve a good approximation, that is, slightly more than half of the available samples.
This can be attributed to the fact that the underlying dynamical system is of second order and that the underlying discrepancy function exhibits locally non-smooth behavior.
Consequently, higher demands are placed on the choice of linear upsampling scheme resulting in the training data to be sampled more densely.

\subsubsection{Bias correction}
\begin{figure}[h!]
	\vspace{-6em}
	\centering
	\noindent
	\begin{tabular}{cccc}
		$u^{\text{\tiny hifi}}_{t_{25}}$ & 
		$u^{\text{\tiny hifi}}_{t_{50}}$ & 
		$u^{\text{\tiny hifi}}_{t_{60}}$ &  
		$u^{\text{\tiny hifi}}_{t_{70}}$
		\\
		\includegraphics[width=0.16\textwidth]{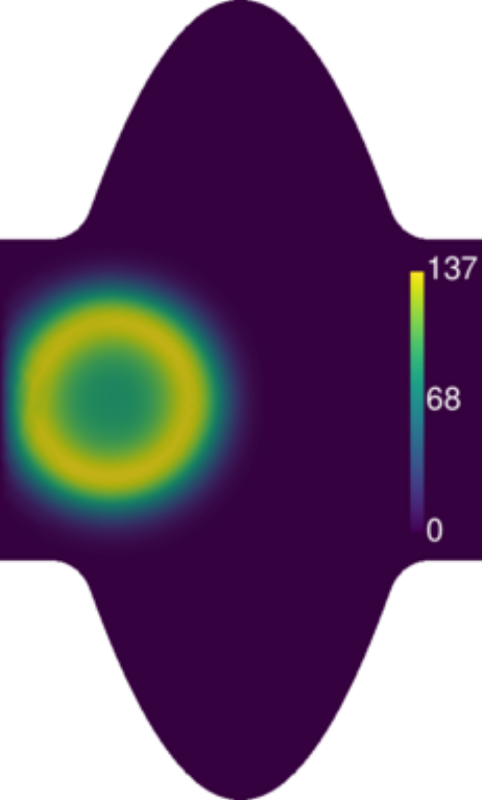}&
		\includegraphics[width=0.16\textwidth]{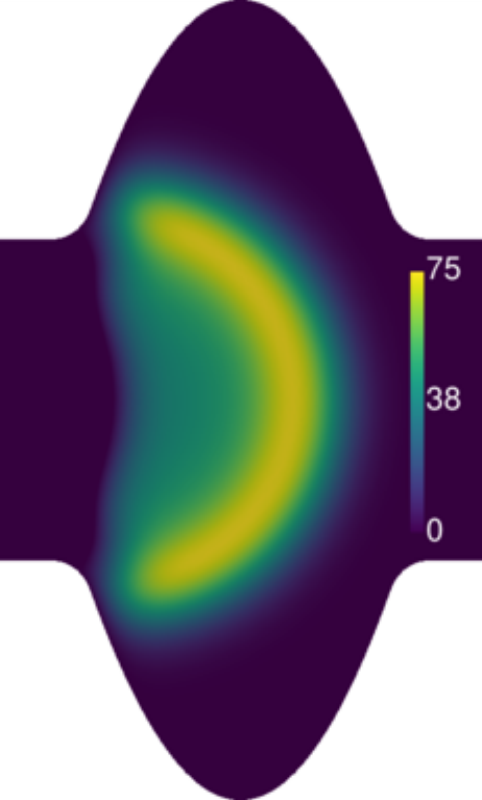}&
		\includegraphics[width=0.16\textwidth]{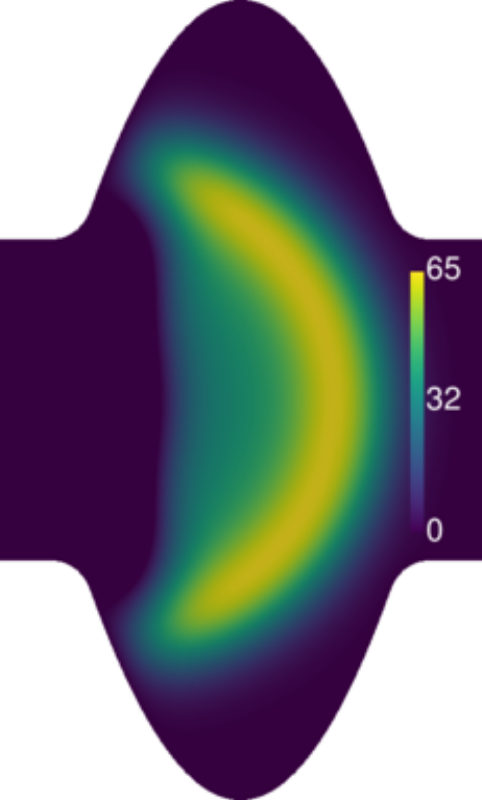}&
		\includegraphics[width=0.16\textwidth]{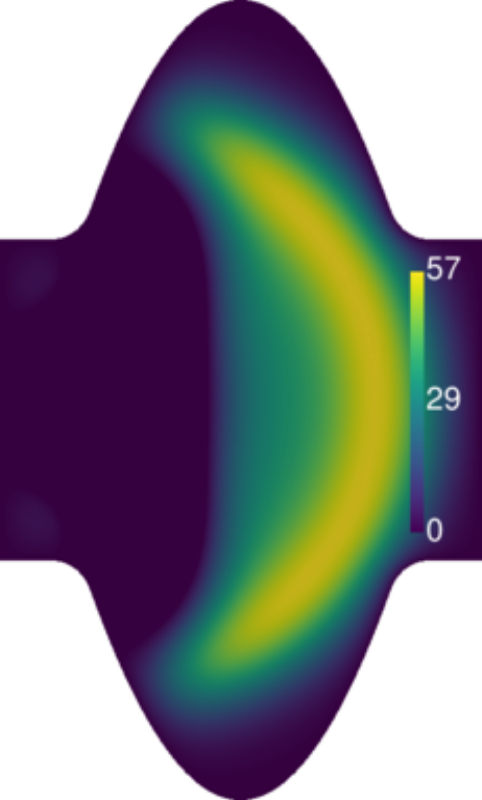} 
		\\
		$u^{\text{\tiny lofi}}_{t_{25}}$ & 
		$u^{\text{\tiny lofi}}_{t_{50}}$ & 
		$u^{\text{\tiny lofi}}_{t_{60}}$ &  
		$u^{\text{\tiny lofi}}_{t_{70}}$
		\\
		\includegraphics[width=0.16\textwidth]{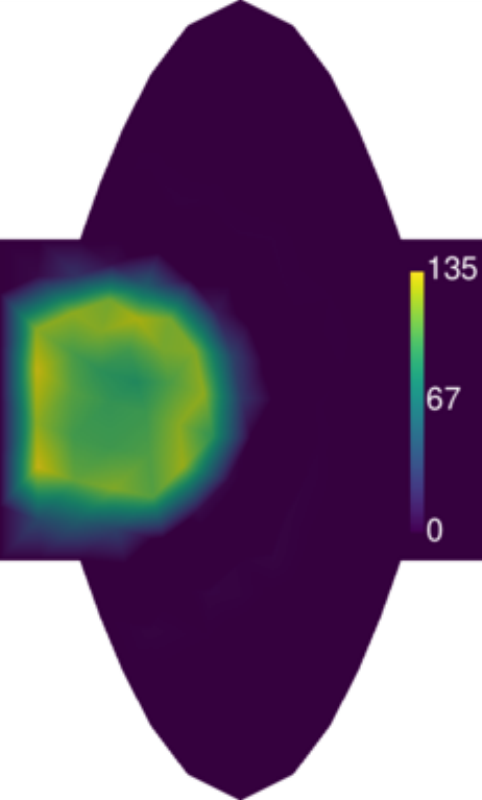}&
		\includegraphics[width=0.16\textwidth]{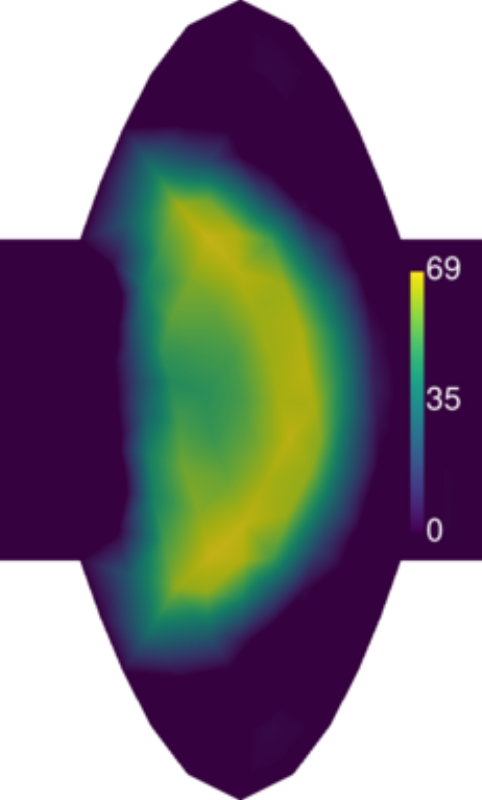}&
		\includegraphics[width=0.16\textwidth]{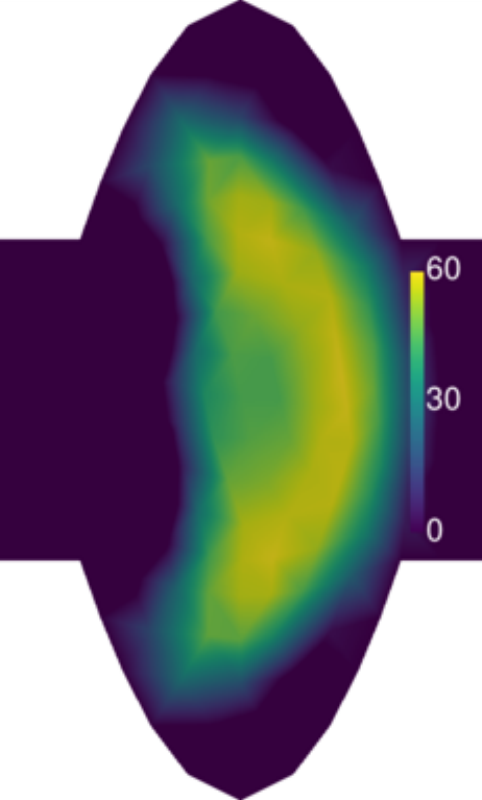}&
		\includegraphics[width=0.16\textwidth]{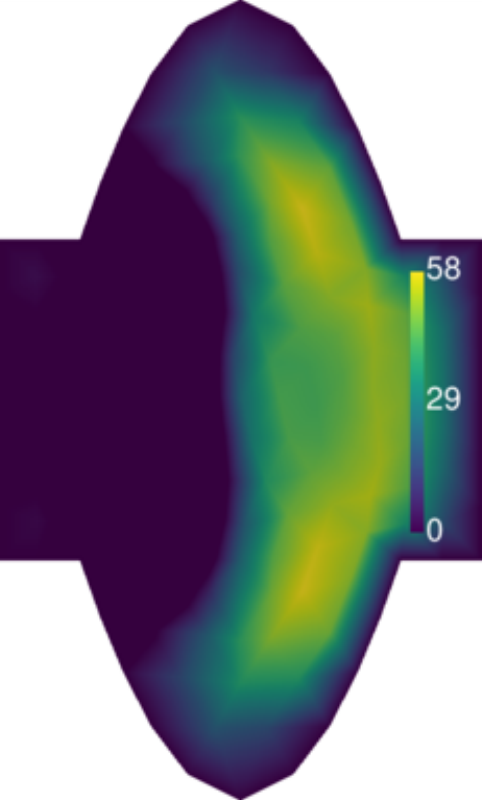} 
		\\
		$u^{\text{\tiny corr}}_{t_{25}}$ & 
		$u^{\text{\tiny corr}}_{t_{50}}$ & 
		$u^{\text{\tiny corr}}_{t_{60}}$ &  
		$u^{\text{\tiny corr}}_{t_{70}}$  
		\\
		\includegraphics[width=0.16\textwidth]{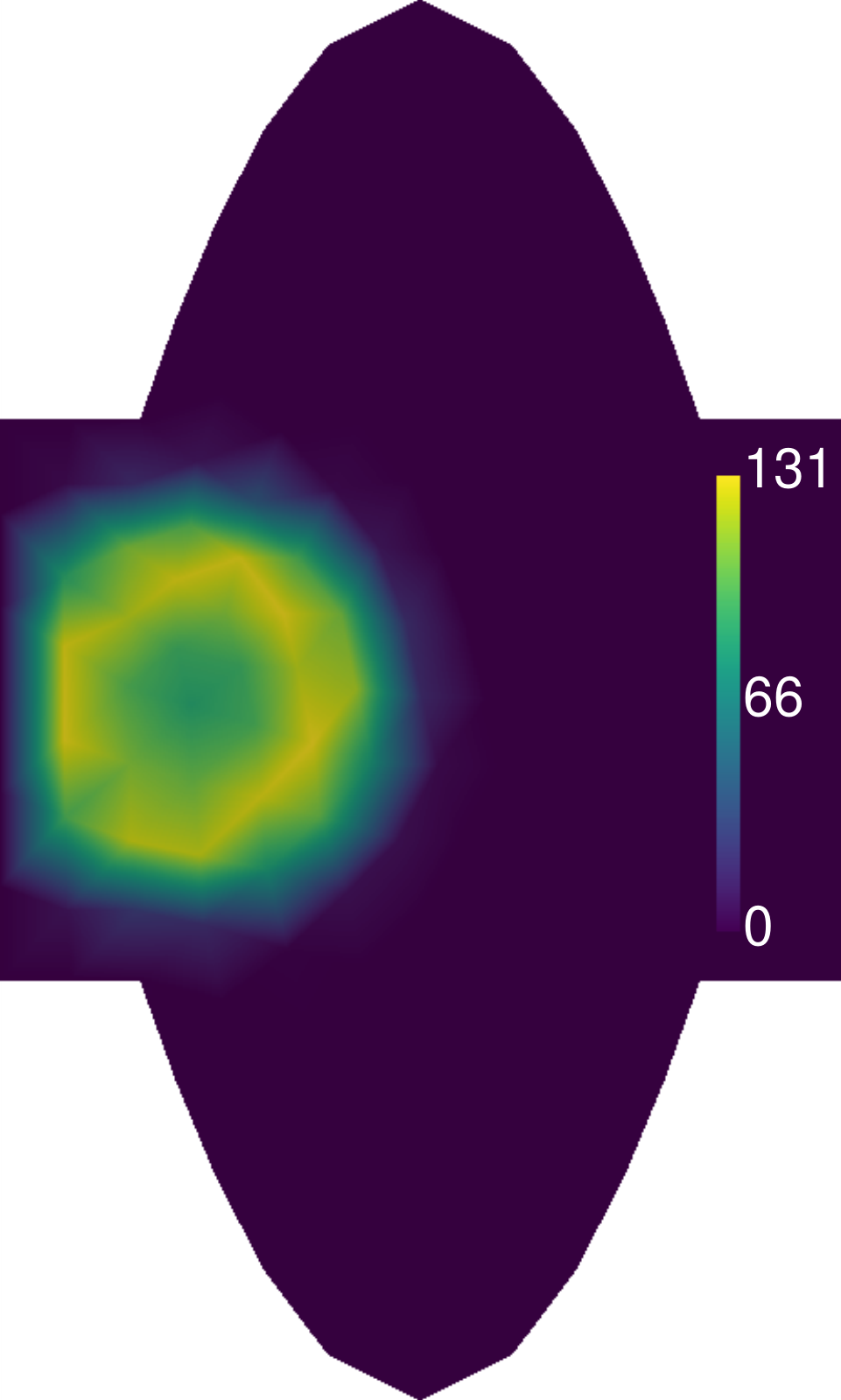}&
		\includegraphics[width=0.16\textwidth]{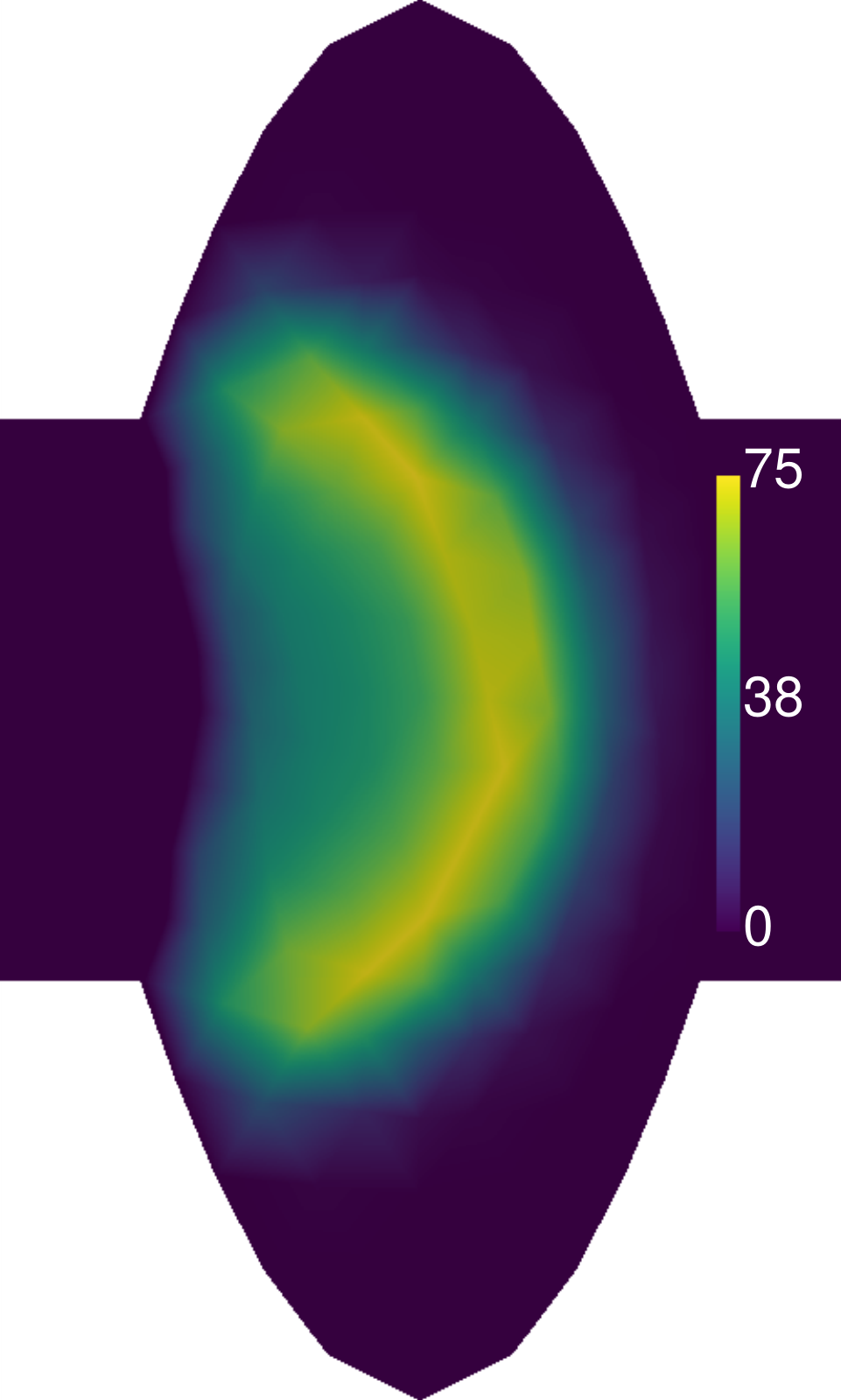}&
		\includegraphics[width=0.16\textwidth]{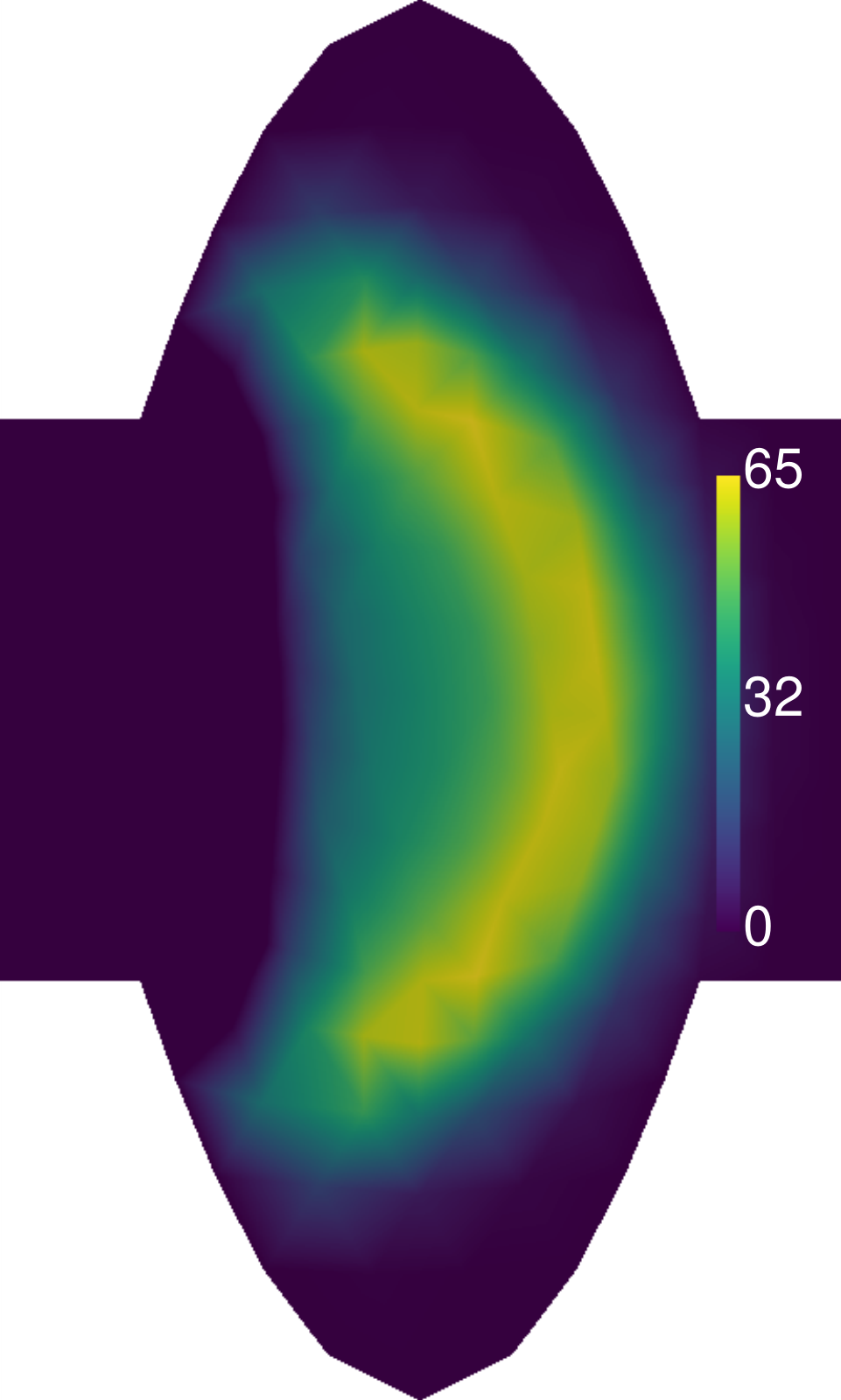}&
		\includegraphics[width=0.16\textwidth]{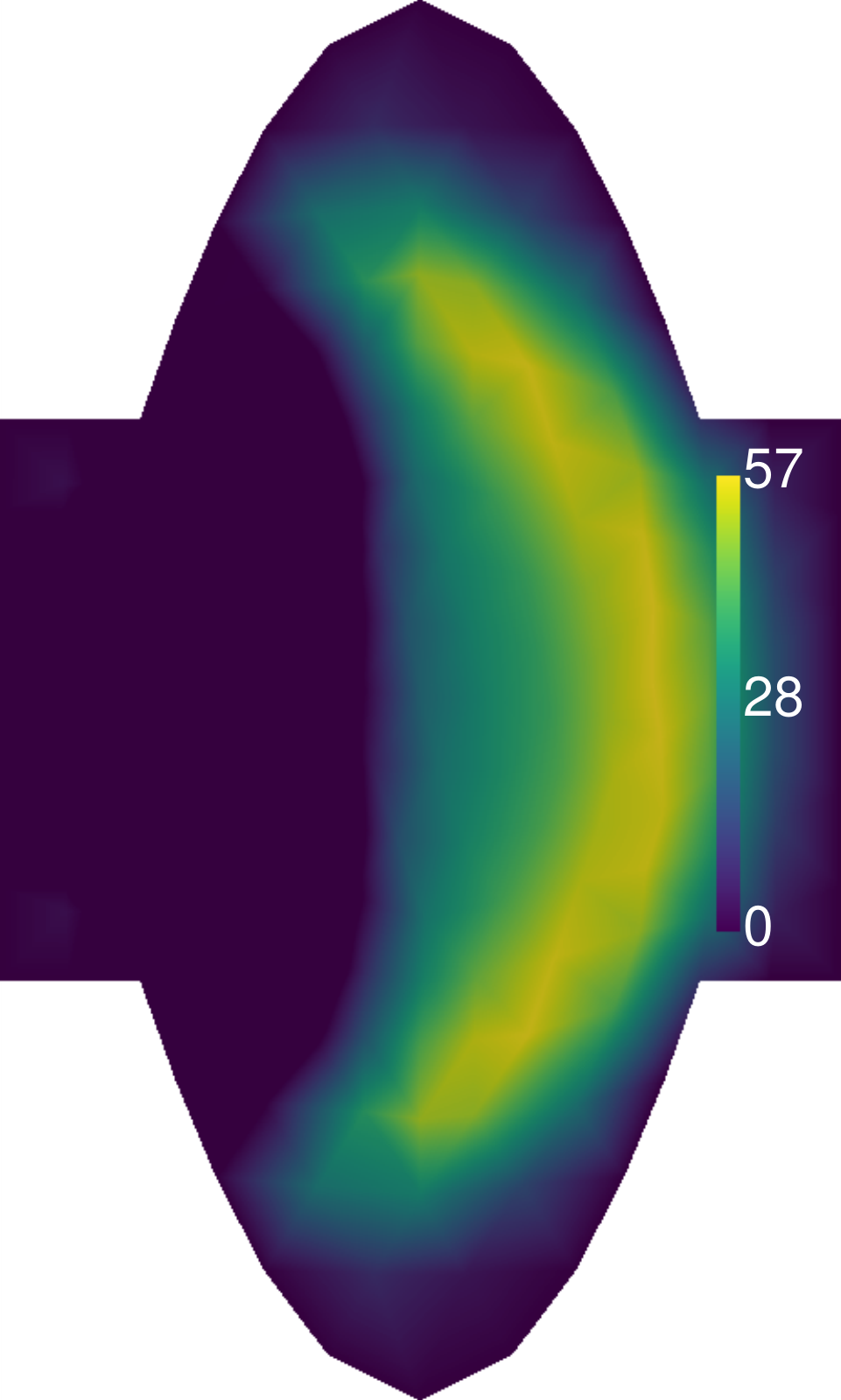}
		\\
		\scriptsize $\sum_{i=1}^{N^\text{lofi}_{\text{dof}}} \,\left|\delta^{\text{\tiny NN}}_{i,t_{25}}\,\right|\phi_i(\mathbf{r})$ & 
		\scriptsize $\sum_{i=1}^{N^\text{lofi}_{\text{dof}}}\,\left|\delta^{\text{\tiny NN}}_{i,t_{50}}\,\right|\phi_i(\mathbf{r})$ & 
		\scriptsize $\sum_{i=1}^{N^\text{lofi}_{\text{dof}}}\,\left|\delta^{\text{\tiny NN}}_{i,t_{60}}\,\right|\phi_i(\mathbf{r})$ &  
		\scriptsize $\sum_{i=1}^{N^\text{lofi}_{\text{dof}}}\,\left|\delta^{\text{\tiny NN}}_{i,t_{70}}\,\right|\phi_i(\mathbf{r})$
		\\
		\includegraphics[width=0.16\textwidth]{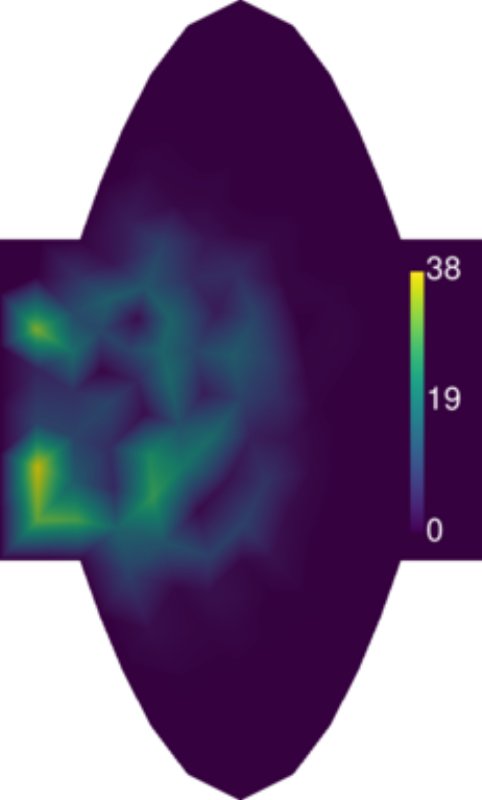}&
		\includegraphics[width=0.16\textwidth]{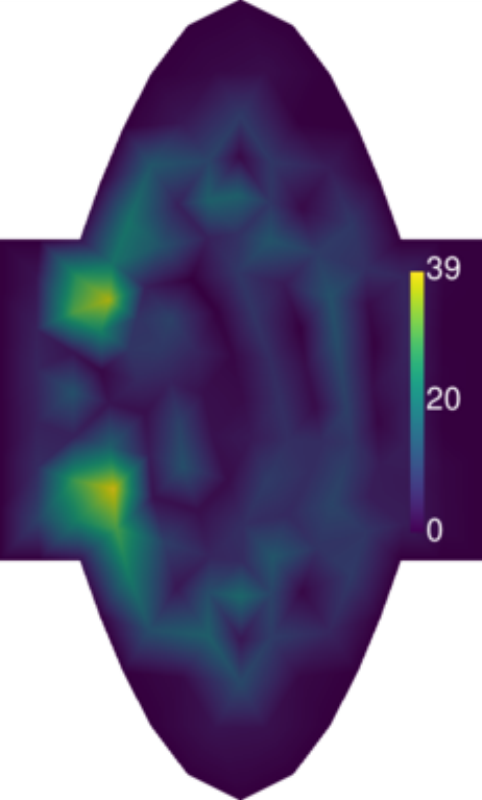}&
		\includegraphics[width=0.16\textwidth]{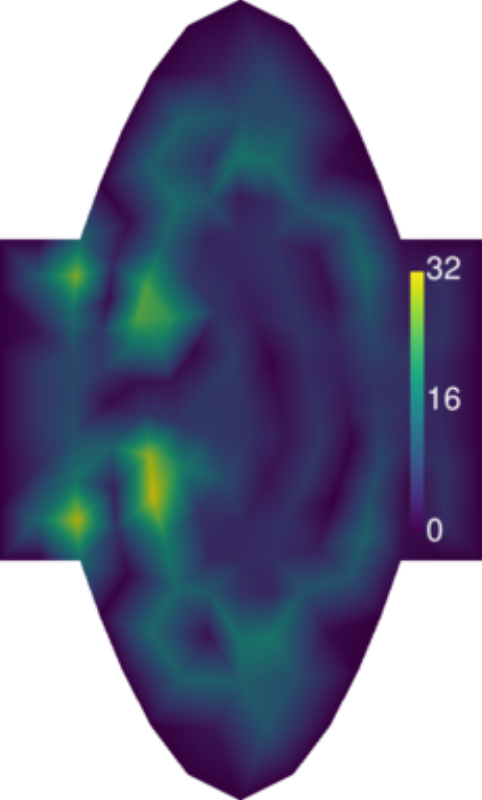}&
		\includegraphics[width=0.16\textwidth]{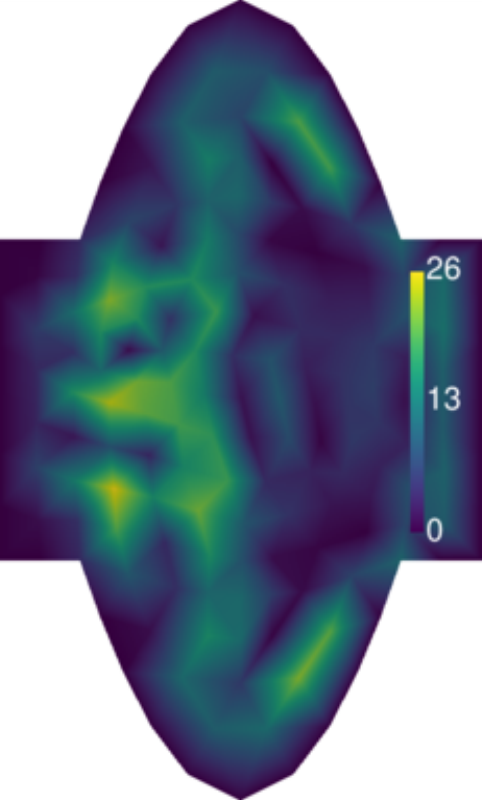} 
	\end{tabular}
	\caption{Wave propagation for $k=25,50,60,70$. 
		\\ \indent \textbf{Row 1:} High-fidelity model, parametrized by $\Omega_{\text{hifi.}}$. 
		\\ \indent \textbf{Row 2:} Low-fidelity model, parametrized by $\Omega_{\text{lofi.}}$ $\left(\Delta_{L^2}\,u^{\text{lofi}}=27.53\,\%\right)$.
		\\ \indent \textbf{Row 3:} Bias-corrected model $\left(\Delta_{L^2}\,u^{\text{corr}}=9.99\cdot 10^{-2}\,\%\right)$.
		\\ \indent \textbf{Row 4:} Absolute values of the discrepancy function coefficients.}
	\label{fig:transient_waveguide}
\end{figure}

\textcolor{black}{A visualization of the wave propagation using the high- and low-fidelity models is given in Figure \ref{fig:transient_waveguide} (rows 1 and 2, respectively).
We observe that both waves propagate simultaneously, however, the low-fidelity solution is much coarser and its magnitude is significantly off.
Relative to the high-fidelity model, the error of the low-fidelity model is $\Delta_{L^2}\,u^{\text{lofi}}=27.53\,\%$.
In Figure~\ref{fig:transient_waveguide} (row 4), the solution of the bias-corrected model is shown, the relative error of which is $\Delta_{L^2}\,u^{\text{corr}}=9.99\cdot 10^{-2}\,\%$. 
Once again, a significant correction to the low-fidelity model has been achieved, which can be visually verified by the magnitudes of the pressure fields corresponding to each model, see the corresponding color-bars.
A visualization of the discrepancy function is given in Figure \ref{fig:transient_waveguide} (row 4), where we observe in $t_{50}=2\cdot 10^{-3}\,\si{s}$ and $t_{60}=2.4\cdot 10^{-3}\,\si{s}$ that a significant portion of the error occurs near the variation in the geometry.
}

\clearpage

\section{Conclusion and Outlook}
\label{sec:conclusion}

This work presented a hybrid modeling framework for the approximation and correction of model bias in the setting of multi-fidelity modeling and simulation of dynamic/transient systems.
The proposed hybrid modeling approach combines standard \gls{fe} approximation with an \gls{rnn}, where the former accounts for the spatial effects of the approximation and latter for the temporal dynamics.
The hybrid model first approximates the discrepancy between models of varying fidelity and is subsequently used to correct the low-fidelity models and increase their accuracy.

The presented numerical results show that the proposed hybrid modeling method is capable of yielding accurate approximations of the discrepancy function and, accordingly, significantly improved bias-corrected models, for a variety of dynamical \gls{fe} engineering simulations. 
In all considered test cases, the discrepancy function is approximated with an error below $2\%$, even if it displays a locally non-smooth behavior.
Furthermore, the hybrid model remains accurate even if sparse training data sets are employed, given that a suitable upsampling scheme is provided, which controls the interpolation behavior in regions with only sparse data.
Local weighting factors further ensure that the \gls{rnn} provides a good fit of the training data and avoids over-smoothing.

In all considered test cases, the \glspl{rnn}, consequently, the hybrid model is trained to sufficient accuracy after 1000 - 2500 epochs, thus resulting in quite modest training times, given the dimensionality of input and output.
This is accomplished due to the nature of the proposed hybrid model, which splits spatial and temporal dependencies.
The model already provides a spatial basis in the form of the low-fidelity \gls{fe} basis functions, hence, the \gls{rnn} focuses on approximating the temporal dependencies only, which is a task very much suited to \glspl{rnn} in general.

Despite the promising results shown in this work, there are limitations to the proposed hybrid modeling method worth mentioning.
First, the accuracy of the hybrid model depends significantly on the upsampling scheme, especially within the domains with sparse data.
The choice of upsampling scheme is thus problem dependent and requires assumptions on the underlying dynamical systems and discrepancy functions.
In addition, the training data cannot be chosen completely at random, as samples within specific regions of the trajectory contain more information about the underlying dynamics than others.
Introducing an optimal experimental design approach, e.g., knowing a priori which time steps to sample as to achieve a good approximation, would be advisable.
Finally, the projection operator is problem dependent as well, depending primarily on the choice of basis functions used to represent the discrepancy function.
In the context of this work, this choice comes naturally due to the use of the \gls{fe} method, however, the generalization to other basis representations should be explored.

To provide an outlook, there are numerous possible extensions to the presented methodology.
Apart from considering more sophisticated \gls{rnn} architectures, a possible avenue would be to explore localized Gaussian Processes with physics-inspired priors to upsample the physical data in a better way \cite{hanuka2021physics,williams2006gaussian}. 
Another possibility would be to include physics inspired loss functions, similar to the concept of \glspl{pinn} \cite{cai2021physics, karniadakis2021physics, raissi2019physics}, as a substitute or complement to data upsampling.
Last, the performance of hybrid models trained with real-world observations, for example, collected from measurements or experiments, could be investigated.
In such cases, one would face the additional task noise treatment, exemplarily, separating noise from the underlying trajectory data.

\section*{Acknowledgements}
The authors would like to thank Armin Galetzka for thoroughly reading this manuscript and for providing helpful insights for improvements. 
Moritz von Tresckow acknowledges the support of the German Federal Ministry for Education and Research (BMBF) via the research contract 05K19RDB. 
Dimitrios Loukrezis and Herbert De Gersem acknowledge the support of the Deutsche Forschungsgemeinschaft (DFG, German Research Foundation), Project-ID 492661287 -- TRR 361.

\bibliographystyle{elsarticle-num}
\biboptions{sort&compress}
\bibliography{mybib.bib}
\end{document}